\documentclass[a4paper,11pt]{article}
\pdfoutput=1
\usepackage{jheppub}
\usepackage[T1]{fontenc}
\usepackage{orcidlink}
\usepackage{amsmath, amsthm, amssymb, amsfonts, mathrsfs}
\usepackage{mathtools}
\usepackage{braket, slashed, bm}
\usepackage{hyperref}
\usepackage{booktabs}
\usepackage{scalefnt}
\usepackage{enumitem}

\usepackage{bbm}
\usepackage{lmodern}
\usepackage{amsmath, amsthm, amssymb, amsfonts, mathrsfs}
\usepackage{scalefnt}
\usepackage{braket, slashed, bm}
\usepackage{bbold}
\usepackage{listings}
\usepackage{siunitx}
\usepackage{booktabs}
\usepackage{soul}
\usepackage{setspace}
\usepackage{color}
\usepackage[dvipsnames]{xcolor}

\title{\boldmath Ensemble Learning with Unsupervised and Semi-Supervised Models for New Physics Searches at Colliders}

\author[1]{Alexandre Alves\,\orcidlink{0000-0002-2454-7874}}
\author[2]{Eduardo da Silva Almeida\,\orcidlink{0000-0003-1397-0244}}
\author[1]{Douglas Roberto de Matos Pimentel\,\orcidlink{0009-0005-2421-4331}}

\affiliation[1]{Departamento de F\'isica, Universidade Federal de S\~ao Paulo, UNIFESP, 09972-270, Diadema-SP, Brazil}
\affiliation[2]{Departamento de Física do Estado Sólido, Universidade Federal da Bahia,\\
UFBA, 40170-115, Salvador-BA, Brazil }

\emailAdd{aalves@unifesp.br}
\emailAdd{almeidae@ufba.br}
\emailAdd{douglas.pimentel@unifesp.br}

\abstract{
New physics beyond the Standard Model is widely believed to exist motivated by key limitations of the current theory. However, in the absence of undisputed experimental signals, there is little guidance on what type of extended physics model should be built. To address the challenge of training machine learning classifiers to spot on unknown phenomena, many algorithms have been applied to the task of either identifying anomalous events relative to background expectations or cluster collision events together in classes of similarity. However, as in the supervised case, the no free lunch theorem assures that the performance of unsupervised models depends on the specific signal hypothesis and their inductive biases, and the successful identification of a given signal by a particular algorithm does not guarantee comparable performance for other signals. In this study, we take a step toward a more model-independent unsupervised and semi-supervised search strategy by combining classifiers into ensembles. Our goal is addressing two complementary aspects of unsupervised ensemble learning -- overall performance and robustness of the classification task from a thorough assessment of ensemble models. We show that, despite sacrificing some performance relative to the best individual model in the majority of cases, the ensembles provide robust and reliable classification predictions across several signal hypotheses, increasing confidence that genuinely unknown new physics will not pass unnoticed.
}

\begin{document}

\maketitle
\flushbottom


\newpage
\section{Introduction}

 The Standard Model (SM) is perhaps the most successful physical theory ever built and has passed decades of precise experimental tests. Notwithstanding, if one believes it is the ultimate theory of fundamental interactions, it should be delicately fine-tuned to explain many phenomena. Disregarding possible fine-tunings, however, if the existence of dark matter is confirmed, a new Standard Model would be needed. Dark matter is not the only possible silver bullet for the Standard Model, though. The strong-CP problem, neutrino masses, the observed baryon asymmetry, among other problems, can lead to diverse scenarios of new interactions and particles. These new physics scenarios are not guaranteed to ever be thought of by us. Even restricting ourselves to the space of models that we might design to solve all the SM problems, the size of that space is overwhelming. After decades of dedicated efforts to search for new physics, only a tiny fraction of possibilities has actually been scanned~\cite{Craig:2016rqv,Kim:2019rhy}. 

 Despite innumerable phenomenological analyzes performed over the years using machine learning (ML) or not, new physics might go unnoticed because these analyzes are tailored to strong model assumptions. The smoking gun signature -- a narrow resonance with pairs of visible particles -- like the one that thrilled the particle physics community back in 2017, where a 750 GeV resonance-like signature was observed with high significance both in ATLAS and CMS data, would be a dream for bump hunters, but much less identifiable signals could arise, for example, as smooth excesses on the tails of mass or transverse momentum distributions. Conditional on the particles and interactions we suppose to observe, one always tries to optimize the signal significance by judiciously choosing the final state topology, representations (observables), jet parameters when they are present, and kinematic cuts. The statistical significance can also depend strongly on the tools used for simulations but also, and more importantly, on the order of the perturbation theory of the scattering amplitudes of the simulation, the renormalization/factorization scales, the efficiencies and acceptances encoded in detector simulators and parameters of hadronization tools and jet clustering algorithms.

  Ultimately, the ML algorithm employed to enhance the separability of the signal adds new dimensions of assumptions through architectures, internal parameters, and training/test/validation methodology hyperparameters. Within the supervised framework, it is well known that Boosted Decision Trees (BDT) present excellent performance on tabular data and have been extensively used by experimental collaborations for years. Convolutional neural networks (CNN) excel in disentangling jet images. More recently, transformers and graph neural networks have surpassed other algorithms in many particle physics tasks. All these physics and ML model assumptions could be well accommodated in an extended no free lunch theorem -- {\it when averaged on all possible new physics searches, all joint physical/ML frameworks perform equally well}~\cite{585893}. In other words, no statistical learning model is superior across every possible dataset or new physics search. 

  The strong dependence on physical and ML model assumptions poses hard obstacles to minimal guaranties in finding a new signal. This acknowledgment has led ML research in particle physics  into the semi-supervised and unsupervised domains. Semi-supervised algorithms learn to recognize background events from background simulations, from a training set, and assign to test events a measure of {\it backgroundness}. A low {\it backgroundness} thus flags an anomalous event collision. Unsupervised models try to separate classes of events in the test set without learning a general rule beforehand but relying on measures of similarity. A paradigm is the clustering algorithm K-nearest neighbors (KNN). In KNN, the algorithm identifies clusters of events in the dataset and the only hyperparameter is the number of clusters it should search for.  It is clear, from this discussion, that unsupervised algorithms make fewer assumptions than semi-supervised algorithms which, by their turn, make less assumptions than supervised models.

 However, it does not mean that the performance of semi-supervised and unsupervised models can not depend strongly on the specific type of new physics signal, representation of the dataset, and physical parameters of simulations, for example. It has been reported that the performance of variational autoencoders (VAE) to identify outliers can strongly depend on its hyperparameters~\cite{Fraser:2021lxm} where the optimum hyperparameter settings of a VAE to target a type of signal change for another signal. 

 What do we do when we do not trust the answer from a particular expert? A good strategy is to ask other experts and decide on the pool of answers we gather from that mix of experts. For supervised algorithms, prediction stabilization and even enhancement of performance can be achieved with various ensemble strategies. Bagging, boosting, and random forests can be classified as homogeneous strategies, while stacking, voting, and committees as heterogeneous, and they are all very effective types of combination~\cite{hansen1990neural,schapire1990strength,wolpert1992stacked,krogh1995neural,breiman1996bagging,breiman2001random,freund1997decision,ho1998random,opitz1999popular,dietterich2000ensemble,zhou2012ensemble}. In particular, boosting~\cite{schapire1990strength} is observed to dramatically increase the performance of the meta-learner compared to weak learners that compose the ensemble in case of decision trees. As we are going to show in the next section, with a proper choice of diverse algorithms, the meta-learner error can be made smaller than the average error of the individual models.
 
 Unsupervised ensemble learning is a more subtle instance of meta-learning. Contrary to the supervised case, evaluating the performance of the meta-learner compared to the base models without data labels is difficult~\cite{aggarwal2013outlier,aggarwal2015theoretical} preventing us from obtaining the same relative guaranties of enhanced performance compared to the supervised case. Notwithstanding, many unsupervised ensemble strategies are expected to increase the robustness of the predictions by suppressing underperforming models~\cite{chiang2017anomaly}. This is an important aspect of unsupervised classification in particle physics that has been neglected so far. 

 In this work, we advocate the importance of robustness and stabilization of predictions in unsupervised searches for new physics in colliders. Looking for optimal classifiers is certainly an essential target in particle phenomenology, but in the space of all possible new physics models, we cannot know previously which one will perform best for a given signal. Of course, we could experiment with a large number of algorithms and decide for the best one. Yet, nature might surprise us with something our models never learned to identify. In this respect, our best strategy is the mix-of-experts. By combining virtues and compensating for weaknesses, the ensemble might present better prospects for finding new phenomena in particle physics data with diminished chances to miss signals. 

 In particle physics, homogeneous ensemble strategies in supervised learning as BDT, gradient boosting, and random forests are ubiquitous. Heterogeneous ensemble applications are much rarer but important to cite. Tevatron constructs a {\it super classifier} in single-top observation~\cite{CDF:2010eor}. The stacking technique is used in~\cite{alves2017stacking} to identify Higgs bosons in the $\tau^+\tau^-$ channel, and to search for a new scalar decaying to $W^+W^-\to\ell^+\ell^-+\!\!\!\not E_T$~\cite{alves2022reconstruction}. Other examples of ensembles include top-quark~\cite{Araz:2021wqm} and charm-quark tagging~\cite{Alves:2019ppy}.

 Ensemble learning with anomaly detection algorithms has been proposed in Ref.~\cite{vanBeekveld:2020txa} in the search for supersymmetry at the LHC. The method employed in this work was to combine the normalized anomaly scores of four baseline algorithms. For a given event, they compute the scores, $s_i$, from a Variational Autoencoder, an Autoencoder (AE), an Isolation Forest (IF), and a Gaussian Mixture Model (GMM) trained on the latent space representation of the events obtained from a VAE. The best combination is obtained from the logical AND operator, $s^{\text{AND}}=min(s_i)$. They show that this combination improves the signal significance compared to the individual algorithms. 

 Another landmark work on anomaly detectors is the one that reports on the Dark Machines Challenge Initiative conducted back in 2019~\cite{darkmachinechallenge}. In that challenge, the objective was to investigate the performance of various anomaly detectors in a variety of new physics scenarios. There are some clear lessons from this work to emphasize: (1) no individual algorithm performs best for all physics scenarios, (2) ensembles of anomaly detectors can surpass the performance of the individual algorithms, (3) the performance rank of the algorithms depends on the metric, in this case, in terms of AUC and signal efficiency for a fixed background efficiency. 

 In this work, we explore an important but overlooked aspect of unsupervised search for new physics -- the robustness and stabilization of predictions across diverse new physics models. We perform two studies: (1) with a fixed background hypothesis and seven different new physics scenarios with the same final state, namely, a $t\bar{t}$ pair, and (2) with varying signal and respective background final states from the Dark Machine Unsupervised Challenge~\cite{darkmachinechallenge}. If, on one hand, the enhancement of performance in identifying new signals using ensembles has been demonstrated in previous works, on the other hand, a phenomenological study of the variation of the predictions is lacking. This study is necessary as it complements and reaffirms the importance of combining algorithms to find new physics. Beyond the combination schemes presented in~\cite{vanBeekveld:2020txa,darkmachinechallenge}, we explore other types of ensembles found in the literature. We conduct a thorough comparison between all the combinations in order to establish a clear picture of performance enhancement and stabilization.

 We organized this work as follows. In section~\ref{section:building_ensembles}, we describe in detail the ensemble models analyzed in the work; section~\ref{section:signals-backg} is devoted to the characterization of signal and backgrounds with the fixed $t\bar{t}$ final state; the simulations and representation of data for $t\bar{t}$ processes are presented in section~\ref{section:simulations}; in section~\ref{section:train-test}, the methodology for training and testing the machine learning models is described; the results for $t\bar{t}$ final states can be found in section~\ref{section:results}; the results for Dark Machines Challenge final states can be found in section~\ref{section:darkmachines}; we conclude the work in section~\ref{section:conclusions}.

\section{Building Ensembles of Unsupervised Classifiers}
\label{section:building_ensembles}

\subsection{Preliminaries}

 In this section, we present the unsupervised ensemble classifiers that will be used to combine the various unsupervised and semi-supervised algorithms. Before delving into detailed explanations of the algorithms, we now present some technical preliminaries necessary for a better understanding of key concepts. The classification task targeted in this work is the binary distinction between signal and background classes. The ensembles to be constructed will be of parametric and non-parametric types, as we will explain in the next sections. 
 
 The performance of the individual classifiers and ensembles will be compared using two metrics : (1) the area under the receiver operating characteristic curve (ROC-AUC or AUC for short), and (2) the background rejection for a fixed signal acceptance. 
 
 To compute AUC, we need to estimate the true and false positive rates (TPR and FPR).
Suppose $\hat{y}=f(X)$ is the model prediction for the data instance $X$. The prediction is made based on a threshold of the output score of the algorithm -- it predicts $\hat{y}=1$ when $s(X)\geq s_c$ and $\hat{y}=-1$ when $s(X)< s_c$. For a given operating score point $s_c$ the true and false positive rates can be interpreted as signal ($\epsilon_S)$ and background ($\epsilon_B$) efficiencies. They are calculated as follows
 \begin{eqnarray}
     \epsilon_S = \text{TPR}(s_c) &=& \int_{s_c}^\infty p(s|\hat{y}=+1)\; ds \\
     \epsilon_B = \text{FPR}(s_c) &=& \int_{s_c}^\infty p(s|\hat{y}=-1)\; ds\; .
 \end{eqnarray}
This is a $s$-parametrized curve in the $\epsilon_S$ {\it versus} $\epsilon_B$ plane limited in the region $[0,1]^2$.  The area can be expressed as a line integral through the contour composed by the lines $\gamma_1: \epsilon_S=0,0\leq \epsilon_B\leq 1$, $\gamma_2: \epsilon_B=1,0\leq \epsilon_S\leq 1$, and the curve $\Gamma: \epsilon_S=\epsilon_S(\epsilon_B)$. Now, it is possible to show that
\begin{equation}
\text{AUC} = \frac{1}{2} +\frac{1}{2}\int_0^1\left[-\epsilon_S(s) \epsilon_B^\prime(s)+\epsilon_B(s)\epsilon_S^\prime(s)\right]\; ds\; .
\end{equation}
It is easy to see that a random classifier where $\epsilon_S=\epsilon_B$ for all $s$ has $\text{AUC}=1/2$ while a perfect classifier is such that $\epsilon_S=1, 0\leq\epsilon_B\leq 1$ with $\text{AUC}=1$.

It is also important to note that new physics signals can only be safely identified when the false positive rate is low. For this reason, it is also useful to consider the signal efficiency at fixed background efficiency:  $\epsilon_S@\epsilon_B$ -- in addition to AUC. ROC curves with the same area can have different shapes and favor higher background rejection for a fixed signal acceptance. For the majority of new physics scenarios, algorithms that are capable of rejecting backgrounds for the same signal acceptance are preferable.

To compare the ensembles with the individual classifiers that compose them, we rank all algorithms according to their AUC and $\epsilon_S@\epsilon_B$ across seven classification tasks. These tasks involve different signal hypotheses leading to the same final state as fully leptonic $t\bar{t}$ production in $pp$ collisions at the 14 TeV LHC. We also compute the standard deviation of the ranks across the seven tasks. The most desirable algorithms are therefore those with both a high average rank\footnote{By best rank, we mean the algorithm with rank=1.} and a small rank dispersion, indicating that they consistently achieve strong performance across different signal hypotheses. 

A suitable figure-of-merit for AUC comparisons is the normalized euclidean distance of the $i$-th algorithm in the plane spanned by the mean and standard deviation of ranks (performance {\it versus} stability space). First, we apply a min-max transformation on the mean ranks and standard deviations evaluated in seven datasets and then calculate the distance
\begin{equation}
    d_{rank,i} = \sqrt{\langle rank\rangle^2_i+\sigma_{rank,i}^2}\; .
    \label{eq:distance}
\end{equation}
The best algorithm is the one with the lower $d_{rank}$, reinforcing that the rank-1 algorithm is the one with the larger AUC.

One important aspect of ensemble learning is the relationship between performance and diversity.
For majority voting in supervised learning~\cite{brown2010goodbad,jiang2017generalized}, the ensemble error, $E_{vote}$, can be reduced compared to the generalization error $E_D$ according to
\begin{equation}
    E_{vote} = E_D\underbrace{-D_{good}+D_{bad}}_{\text{Diversity}}\; ,
\end{equation}
where $D_{good}$($D_{bad}$) is the disagreement among the algorithms, in which the ensemble provides right(wrong) predictions. Let us demonstrate this result.

Suppose we train $N$ models, $h_t(x)$, to classify a dataset $x\sim p(x)$ labeled as $y(x)\in \{-1,+1\}$. The majority vote ensemble of the $N$ classifiers is given by
\begin{equation}
    H(x) = \text{sign}\left(\frac{1}{N}\sum_{t=1}^N h_t(x)\right),\;\; H(x)\in\{-1,+1\}\; .
\end{equation}
We want to decompose the error of the ensemble in classifying the events. For that aim, we calculate the individual errors of the models
\begin{equation}
    e_t(x) = \frac{1}{2}[1-y(x)h_t(x)]=\left\{\begin{array}{c} 0, h_t(x)=y(x) \\ 1, h_t(x)\neq y(x)\end{array}\right.
\end{equation}
and the disagreement between the individual models and the ensemble
\begin{equation}
    \delta_t(x) = \frac{1}{2}[1-h_t(x)H(x)]=\left\{\begin{array}{c} 0, h_t(x)=H(x) \\ 1, h_t(x)\neq H(x)\; .\end{array}\right.
\end{equation}

The difference between the ensemble loss and the average loss of the individual models can now be computed as follows
\begin{eqnarray}
    \Delta &=& e_{vote}-\bar{e}_{ind} \\
    &=& \frac{1}{2}[1-y(x)H(x)]-\frac{1}{N}\sum_{t=1}^N e_t(x) \nonumber \\
    &=& \frac{1}{2}[1-y(x)H(x)]-\frac{1}{N}\sum_{t=1}^N \frac{1}{2}[1-y(x)h_t(x)] \nonumber \\
    &=& -y(x)H(x) \frac{1}{N}\sum_{t=1}^N \frac{1}{2}\left(1-\frac{h_t(x)}{H(x)}\right) \nonumber \\
    &=& -\frac{1}{N} y(x) H(x)\sum_{t=1}^N\frac{1}{2}[1-h_t(x)H(x)],\;\; \text{noting that }\; \frac{1}{H(x)}=H(x) \nonumber \\
    &=& -\frac{1}{N} y(x) H(x)\sum_{t=1}^N\delta_t(x),\;\; \text{finally, we have}\nonumber \\
    e_{vote} &=& \bar{e}_{ind}-\frac{1}{N} y(x) H(x)\sum_{t=1}^N\delta_t(x)
\end{eqnarray}

To obtain the error over the entire dataset, we integrate over the distribution $p(x)$,
\begin{eqnarray}
   && \int dx\; p(x)e_{vote}(x) =  \int dx\; p(x)\bar{e}_{ind}(x)-\int dx\; p(x)y(x)H(x)\frac{1}{N}\sum_{t=1}^N \delta_t(x) \nonumber \\
   && E_{vote} = \overline{E}-\int_{x_+} p(x)\overline{\delta}(x)dx+\int_{x_-} p(x)\overline{\delta}(x)dx \nonumber \\
   && E_{vote} = \overline{E}-D_{good}+D_{bad}\; ,
\end{eqnarray}
where $D_{good}(D_{bad})$ is the disagreement between the individual models and the ensemble in those events where the ensemble provides a correct(incorrect) prediction, $x_+(x_-)$. 

This decomposition allows us to understand how ensemble learning might work in reducing the loss. For a given event, $x$, if $D_{good}$ is low, many votes are wasted as the agreement about the predictions is high and essentially no new information is gained from the experts, so more disagreement means fewer errors. On the other hand, $D_{bad}$ must be low and the experts have to agree about the events where the ensemble gives a wrong prediction. This way, if $D_{bad}<D_{good}$, the ensemble might perform better than the individual models. Disagreement is promoted by diversity. That is why combining many different models is observed to work in supervised ensemble learning. Unfortunately, in unsupervised learning, we do not have access to the diversity terms without labels, so we cannot have the same supervised guaranties. Yet, we can rely on the heuristics of observations and experiments and learn from experience. It is important to notice that, as observed in the literature of unsupervised learning~\cite{chiang2017anomaly}, there are no guaranties of performance gains with unsupervised ensembles, and even in the supervised case, as we demonstrated, it depends on the balance between the diversity as given by $D_{bad}-D_{good}$.

Wrapping up all these concepts into an analysis framework, we will show how diversifying an ensemble of
unsupervised and semi-supervised algorithms can make predictions more reliable, that is, reducing $d_{rank}$ in terms of AUC across several distinct classification tasks in the search for new physics at the LHC, demonstrating how to reduce ML model dependence in unsupervised learning. Now, we proceed to describe the meta-learners used in this work.

\subsection{Improved Spectral Meta-Learner (i-SML)}
\label{section:isml}

The first unsupervised ensemble considered in this work is the improved Spectral Meta-Learner (i-SML) framework developed by Jaffe, Nadler and Kluger in \cite{jaffe2015estimating}, which is an extension of previous works \cite{parisi2014ranking, dawid1979maximum}. The method estimates the performance of multiple classifiers and combines their predictions without access to labeled data. Its key observation is that, under the assumption that the classifiers are conditionally independent given the true label, the off-diagonal entries of the population covariance matrix of the classifier outputs exhibit a rank-one structure. The leading eigenvector associated with this structure contains information about the relative balanced accuracies of the classifiers, up to a global sign. This property allows the performance of the individual classifiers to be estimated directly from their predictions on unlabeled data and subsequently used to construct an ensemble classifier.

Consider an instance space $\mathcal{X}$ and a binary output space $\mathcal{Y} = \{-1,1\}$. Let $(X,Y) \in \mathcal{X} \times \mathcal{Y}$ be a random pair drawn from an unknown joint distribution. An important quantity in the model is the class imbalance $b$, defined as
\begin{equation}
b = \Pr(Y=1)-\Pr(Y=-1).
\end{equation}
Since the class probabilities sum to one, they can be expressed in terms of $b$ as
\begin{equation}
\Pr(Y=1)=\frac{1+b}{2},
\qquad
\Pr(Y=-1)=\frac{1-b}{2}.
\end{equation}
For each classifier $f_i$, let $\psi_i$ denote its sensitivity and $\eta_i$ its specificity,

\begin{equation}
\psi_i = \Pr\left(f_i(X)=1 \mid Y=1\right),\qquad \eta_i = \Pr\left(f_i(X)=-1 \mid Y=-1\right).
\end{equation}
Using the definition of balanced accuracy of the classifier $i$, $\pi_i = \frac{\psi_i+\eta_i}{2}$, and also conveniently defining $\delta_i=\frac{\psi_i-\eta_i}{2}$, the following expressions are obtained
\begin{equation}
\psi_i=\pi_i+\delta_i,
\qquad
\eta_i=\pi_i-\delta_i.
\end{equation}
Let $Z$ be the $M\times N$ matrix containing the predictions of the $M$ classifiers over $N$ unlabeled instances, with $Z_{ij}=f_i(x_j)$ and $i=1,\ldots,M$, $j=1,\ldots,N$. No prior knowledge of the classifier performances is assumed; in particular, the sensitivities $\psi_i$ and specificities $\eta_i$ are unknown.

The i-SML framework relies on three main assumptions:
\begin{itemize}
\item[(i)] The instances $\{x_j\}_{j=1}^N$ are independent and identically distributed samples from the marginal distribution $p_X(x)$.

\item[(ii)] The classifiers are conditionally independent given the true label. That is, for any pair $i\neq j$, any $a_i,a_j\in\{-1,1\}$, and $y\in\{-1,1\}$,
\begin{equation}
\Pr(f_i=a_i,f_j=a_j\mid Y=y) = \Pr(f_i=a_i\mid Y=y)\Pr(f_j=a_j\mid Y=y).
\end{equation}

\item[(iii)] Most classifiers perform better than random guessing, i.e., more than half of the classifiers satisfy $\pi_i>0.5$.
\end{itemize}

The derivation initially treats the class imbalance $b$ as known. Under this condition, the classifier performance parameters can be recovered from the first- and second-order moments of the prediction matrix. Since $b$ is unknown in the unsupervised setting, it is subsequently estimated from the observed predictions using a restricted likelihood procedure.


Let $\mu_i=\mathbb{E}[f_i(X)]$ denote the population mean of the predictions of the classifier $i$, and let $r_{ij}=
\mathbb{E}\left[
(f_i(X)-\mu_i)(f_j(X)-\mu_j)
\right]$ denote the elements of the corresponding covariance matrix $R_{\text{cov}}$. Since $f_i(X) \in \{-1,1\}$,
\begin{equation}
\mu_i=\sum_i y_i {\Pr}_i = \Pr(f_i=1)-\Pr(f_i=-1).
\end{equation}
Using the definitions of $b$, $\pi_i$, and $\delta_i$, this expectation can be written as
\begin{equation}
\mu_i=2\delta_i+b(2\pi_i-1).
\label{eq:mu}
\end{equation}
Under the conditional independence assumption, the covariance between two classifiers has the form

\begin{equation}
r_{ij} =
\begin{cases}
1 - \mu_i^2, & \text{if } i=j, \\
(2\pi_i-1)(2\pi_j-1)(1-b^2), & \text{if } i\neq j.
\end{cases}
\label{eq:cov}
\end{equation}
Thus, the off-diagonal covariance entries admit the factorization
\begin{equation}
r_{ij}=v_i v_j,\qquad i\neq j,
\end{equation}
where
\begin{equation}
v_i=\sqrt{1-b^2}(2\pi_i-1).
\end{equation}
Consequently, the leading eigenvector associated with the off-diagonal covariance structure provides an estimate of the vector $\vec v$. This vector is identifiable only up to a global sign, since both $\vec v$ and $-\vec v$ generate the same pairwise products. Assumption (iii) resolves this ambiguity by choosing the sign for which the majority of the components satisfy $v_i>0$.
For a given value of $b$, the balanced accuracies can therefore be obtained from
\begin{equation}
2\pi_i-1=\frac{v_i}{\sqrt{1-b^2}}.
\end{equation}
Combining this relation with Eq.~\eqref{eq:mu} and the definitions of $\pi_i$ and $\delta_i$ gives the sensitivity and specificity as
\begin{equation}
\psi_i =\frac{1}{2}\left(1+\mu_i + v_i\sqrt{\frac{1-b}{1+b}}\right),
\qquad
\eta_i = \frac{1}{2}\left(1-\mu_i + v_i\sqrt{\frac{1+b}{1-b}}\right).
\label{eq:psi-eta}
\end{equation}
Therefore, once $b$ is known, the first- and second-order moments of the unlabeled classifier predictions provide the quantities required to characterize the individual classifiers.
In practice, the population quantities $\mu_i$ and $r_{ij}$ are unknown and are replaced by their empirical counterparts,
\begin{equation}
\hat{\mu}_i=\frac{1}{N}\sum_{j=1}^{N} Z_{ij},
\qquad
\hat{r}_{ij}=\frac{1}{N-1}\sum_{k=1}^{N}(Z_{ik}-\hat{\mu}_i)(Z_{jk}-\hat{\mu}_j).
\end{equation}
The estimated covariance structure is then used to obtain $\hat v_i$, which, for a given value of $b$, yields estimates $\hat\psi_i$ and $\hat\eta_i$ through Eq.~\eqref{eq:psi-eta}.

Having estimated the classifier performance parameters, the individual predictions can be combined using the likelihood-ratio test. Under the conditional independence assumption, the likelihood ratio factorizes over the classifiers, giving, for a given instance $k$
\begin{equation}
t_k=\sum_i\ln\left[\frac{\Pr(f_i(x_k)\mid Y=+1)}{\Pr(f_i(x_k)\mid Y=-1)}\right].
\end{equation}
For $f_i=+1$, the contribution of the classifier $i$ to the log-likelihood ratio is $\ln\left(\frac{\psi_i}{1-\eta_i}\right)$,
whereas for $f_i=-1$ it is $\ln\left(\frac{1-\psi_i}{\eta_i}\right)$. These two cases can be combined as
\begin{align}
t_k
&=\sum_i\left[\left(\frac{1+f_i}{2}\right)\ln\left(\frac{\psi_i}{1-\eta_i}\right) + \left(\frac{1-f_i}{2}\right)\ln\left(\frac{1-\psi_i}{\eta_i}\right)\right]
\nonumber\ \\
&=
\frac{1}{2}\sum_i\ln\beta_i + \frac{1}{2}\sum_i f_i\ln\alpha_i,
\end{align}
where
\begin{equation}
\alpha_i=
\frac{\psi_i\eta_i}
{(1-\psi_i)(1-\eta_i)},
\qquad
\beta_i=
\frac{\psi_i(1-\psi_i)}
{\eta_i(1-\eta_i)}.
\end{equation}
Since the factor $1/2$ does not affect the sign of the test statistic, the resulting SML ensemble classifier is
\begin{equation}
\hat y_k^{(SML)}=\operatorname{sign}\left[\sum_i f_i(x_k)\ln\alpha_i + \sum_i\ln\beta_i\right].
\end{equation}
Thus, the ensemble combines the individual predictions through weights determined entirely by the estimated sensitivities and specificities of the classifiers.

The remaining unknown quantity is the class imbalance $b$. To estimate it without labeled data, the i-SML framework uses a restricted likelihood approach. For each candidate value $\tilde b$ in a suitable interval, the estimates $\hat\psi_i(\tilde b)$ and $\hat\eta_i(\tilde b)$ are obtained from Eq.~\eqref{eq:psi-eta}. These parameters define the probability of observing a vector of classifier predictions $\vec f$ through
\begin{equation}
\Pr(\vec f)=\sum_{y\in\{-1,1\}}\Pr(Y=y)\prod_i\Pr(f_i\mid Y=y),
\end{equation}
where the conditional independence assumption is used in the factorization of the likelihood.
The corresponding empirical log-likelihood is then defined as
\begin{equation}
\hat{G}_N(Z\mid\tilde b)=\frac{1}{N}\sum_{j=1}^{N}\ln\Pr\left(f(x_j)\mid\hat\psi(\tilde b),\hat\eta(\tilde b),\tilde b\right).
\end{equation}
The estimated class imbalance is obtained by maximizing this functional,
\begin{equation}
\hat b =\underset{\tilde b\in[-1+\delta,1-\delta]}{\operatorname{argmax}} \hat{G}_N(Z\mid\tilde b),
\end{equation}
where $\delta>0$ excludes degenerate cases $b=\pm1$.

\subsection{Latent Spectral Meta Learner (L-SML)}
\label{section:lsml}

Unsupervised ensemble methods, such as the i-SML presented in Section \ref{section:isml}, rely fundamentally on the conditional independence assumption of the classifiers given the ground truth $Y$.


Under conditional independence, high pairwise correlation between two classifiers serves as strong evidence of high accuracy, as purely random noise models exhibit zero expected correlation. However, in real-world scenarios, individual classifiers frequently share underlying training subsets, feature spaces, or architectural priors. Consequently, their error profiles exhibit strong conditional dependencies, violating this assumption.

Neglecting conditional dependencies introduces severe vulnerabilities. For example, high correlation no longer uniquely implies accuracy; it may instead reflect shared algorithmic bias or redundancy. Another issue is that the optimal decision boundary becomes non-linear in the classifier predictions. Thus, linear aggregation strategies (such as the linear meta-classifier derived under the i-SML model) become strictly suboptimal.

To relax the strict conditional independence requirement, Jaffe et al. \cite{jaffe2016unsupervised} introduced a three-layer hierarchical Bayesian generative model by incorporating an intermediate layer of $K \le M$ unobserved binary latent variables $\{\alpha_k\}_{k=1}^K$, where $\alpha_k \in \{-1, +1\}$ and $M$ is the total number of individual classifiers used. The model establishes conditional independence across two decoupled levels:

\textbf{Latent-to-Label Independence}: The $K$ latent variables $\alpha_1, \dots, \alpha_K$ are conditionally independent given the ground truth $Y$:

\begin{equation}
\Pr(\alpha_1 = a_1, \dots, \alpha_K = a_K \mid Y) = \prod_{k=1}^K \Pr(\alpha_k = a_k \mid Y)   
\end{equation}

\textbf{Classifier-to-Latent Independence}: Each classifier $f_i$ is mapped to a single latent group $k \in {1, \dots, K}$ via an assignment function $c: [M] \to [K]$. Given its assigned latent variable $\alpha_{c(i)}$, the classifier $f_i$ is conditionally independent of all other classifiers and of $Y$:

\begin{equation}
\Pr(f_i = a_i \mid \alpha_{c(i)}, Y) = \Pr(f_i = a_i \mid \alpha_{c(i)}).
\end{equation}

Thus, classifiers belonging to distinct groups ($c(i) \neq c(j)$) remain conditionally independent given $Y$, while classifiers within the same group ($c(i) = c(j)$) are conditionally independent given $\alpha_{c(i)}$, but conditionally dependent given $Y$.

To operationalize this model without labeled data, one must estimate both the number of latent groups $K$ and the assignment function $c$. This is achieved by exploiting the structural rank properties of the classifier covariance matrix $R_{\text{cov}} \in \mathbb{R}^{M \times M}$.


Under the 3-layer model, the population covariance matrix $R_{\text{cov}}$ (off-diagonal) decomposes into two rank-one matrices: $$r_{ij} = \mathbf{1}_c(i,j) , v_i^{\text{on}} v_j^{\text{on}} + \left(1 - \mathbf{1}_c(i,j)\right) v_i^{\text{off}} v_j^{\text{off}}, \quad \forall i \neq j$$ where $\mathbf{1}_c(i,j) = 1$ if $c(i) = c(j)$ and $0$ otherwise, while $v^{\text{on}}, v^{\text{off}} \in \mathbb{R}^m$ represent state vectors for on-block (intra-group) and off-block (inter-group) covariances, respectively.

In \cite{jaffe2016unsupervised} the authors explore the matrix $R_{\text{cov}}$ to build a score that identifies conditional dependencies (or shared latent groups) between pairs of individual models. For any quadruplet of classifiers $(i,j,k,l)$, consider the determinant of the corresponding $2\times2$ covariance submatrix,

\begin{equation}
M_{ijkl}=\det\begin{pmatrix}r_{ij} & r_{il}\\r_{kj} & r_{kl}
\end{pmatrix}=r_{ij}r_{kl}-r_{il}r_{kj}.
\end{equation}
The two rank-one decomposition above implies that these determinants exhibit a characteristic behavior depending on the group membership of the classifiers. Suppose first that $c(i)\neq c(j)$. For the vast majority of choices of $(k,l)$ the classifiers belong to external groups. In these cases, the four covariance terms in $M_{ijkl}$ are governed by the off-block component such that

\begin{equation}
\begin{aligned}
r_{ij} &= v_i^{\mathrm{off}}v_j^{\mathrm{off}}, \\
r_{il} &= v_i^{\mathrm{off}}v_l^{\mathrm{off}},\\
r_{kj} &= v_k^{\mathrm{off}}v_j^{\mathrm{off}},\\
r_{kl} &= v_k^{\mathrm{off}}v_l^{\mathrm{off}}.
\end{aligned}
\end{equation}
Consequently,

\begin{equation}
M_{ijkl} = v_i^{\mathrm{off}}v_j^{\mathrm{off}}v_k^{\mathrm{off}}v_l^{\mathrm{off}}-v_i^{\mathrm{off}}v_l^{\mathrm{off}}v_k^{\mathrm{off}}v_j^{\mathrm{off}}=0.
\end{equation}
Thus, at the population level, the dominant contribution to a score based on these determinants for classifiers belonging to different groups vanishes. In contrast, if $c(i)=c(j)$, the intra-group covariance is governed by the on-block component,

\begin{equation}
r_{ij}=v_i^{\mathrm{on}}v_j^{\mathrm{on}},
\end{equation}
while the remaining terms involving classifiers from external groups are predominantly governed by the off-block component. In this case, the two products entering the determinant generally no longer coincide, and therefore

\begin{equation}
M_{ijkl}\neq0.
\end{equation}

Hence, the aggregation of these determinants provides a mechanism for distinguishing intra-group from inter-group classifier pairs: pairs belonging to different groups tend to produce a small score, whereas pairs belonging to the same group produce a substantially larger score. The formal separation between these two regimes is established by Lemma 4 in \cite{jaffe2016unsupervised}.

In practice, the population covariance matrix $R_{\text{cov}}$ is unknown and must be estimated from a finite sample. Let $\hat{R}_{\text{cov}}$ denote the sample covariance matrix. Statistical fluctuations imply that determinants that vanish at the population level are generally nonzero in the sample. We therefore define the pairwise Score Matrix $\hat{S}\in\mathbb{R}^{M\times M}$ as

\begin{equation}
\hat{s}_{ij}=\sum_{\substack{k\neq i,j \ l\neq i,j}}
\left|\hat{r}_{ij}\hat{r}_{kl}-\hat{r}_{il}\hat{r}_{kj}\right|,
\qquad
\hat{s}_{ii}=0.
\end{equation}
The score $\hat{s}_{ij}$ thus measures the aggregate deviation from the rank-one covariance structure across quadruplets. Low values indicate that $f_i$ and $f_j$ are consistent with belonging to different latent groups, whereas high values indicate that they are more likely to belong to the same group.

Using $\hat{S}$ as an affinity matrix, Spectral Clustering is used to obtain a candidate assignment $\hat{c}$ of the $M$ classifiers into $K$ latent groups. Since $K$ is unknown, the clustering procedure is evaluated for candidate values of $K$. For each candidate partition, the corresponding on-block and off-block covariance components are estimated, and the residual of the two rank-one covariance model is computed. The number of latent groups is then selected as the value of $K$ that minimizes this residual, following the model-selection procedure described in \cite{jaffe2016unsupervised}.

Once the assignment function $\hat{c}$ and the number of latent groups $\hat{K}$ have been established, parameter estimation and label inference proceed through the latent structure in two stages:

\begin{itemize}
\item \textbf{Stage 1 (Sub-Ensemble Latent Consensus):} For each latent group $k\in{1,\dots,\hat{K}}$, the classifiers assigned to that group are conditionally independent given the latent variable $\alpha_k$. The i-SML is therefore applied to each sub-ensemble to estimate the classifier parameters and infer the corresponding latent predictions $\hat{\alpha}_k\in\{-1,+1\}^N$.

\item \textbf{Stage 2 (Global Latent Aggregation):} The latent variables $\alpha_1,\dots,\alpha_{\hat{K}}$ are conditionally independent given the ground-truth label $Y$. The inferred latent predictions $\hat{\alpha}_1,\dots,\hat{\alpha}_{\hat{K}}$ can therefore be combined using the same conditional-independence meta-learning framework to estimate the parameters relating the latent variables to $Y$ and infer the final ensemble prediction $\hat{Y}\in\{-1,+1\}^N$.

\end{itemize}

\subsection{Strategy for Unsupervised Multiple Method Aggregation (SUMMA)}
\label{section:summa}

The Strategy for Unsupervised Multiple Method Aggregation (SUMMA) proposed by Ahsen, Vogel and Stolovitzky \cite{ahsen2019unsupervised} extends the spectral approach of i-SML to classifiers that produce real-valued scores. Instead of binarizing these outputs, SUMMA converts the scores into rankings and uses the statistical dependence between the rankings produced by different classifiers to estimate their relative performance. In this way, the method is insensitive to the scale of the original classifier outputs while retaining the ordering information relevant for classification. Under the assumption of conditional independence of the rank predictions given the true class, the off-diagonal entries of the covariance matrix of the ranks exhibit a rank-one structure whose factors are proportional to the performance of the individual classifiers. This structure is then used to construct a weighted aggregation of the rankings.

As in the previous sections, consider $N$ unlabeled instances and $M$ base classifiers. SUMMA uses the convention that lower ranks correspond to stronger evidence for the positive class. Let $\rho=N_1/N$ denote the prevalence of the positive class, where $N_1$ is the number of positive instances. For classifier $i$, let $g_i(X)=\left(g_i(x_1),\ldots,g_i(x_N)\right)^\top$ be its vector of real-valued scores. Collecting the score vectors from all $M$ classifiers yields the score matrix $S \in \mathbb{R}^{M \times N}$. Each score vector is then converted into a rank vector $r_i=(r_{i1},\ldots,r_{iN})^\top$, where the ranks are a permutation of ${1,\ldots,N}$. Ties between scores are broken uniformly at random. Thus, the absolute scale of the original classifier outputs is discarded, while their relative ordering is preserved.


For a sample $k$, let $\Pr(R_i = r\mid Y=y_k)$ denote the probability that classifier $i$ assigns rank $r \in \{1,\ldots,N\}$ to a sample belonging to class $y_k\in\{-1,+1\}$\footnote{This label notation convention differs from the original paper \cite{ahsen2019unsupervised}, which uses $\{0,1\}$. This adaptation is adopted for consistency with the previously presented ensemble methods (i-SML and L-SML), ensuring uniform notation throughout this manuscript. This choice has no impact on the theoretical results and entails only minor algebraic adjustments, such as using the $\operatorname{sign}(\cdot)$ function instead of the Heaviside step function in the final decision rule.}, and let the joint probability $\Pr(R_1 = r_{1k}, \dots, R_M = r_{Mk} \mid Y= y_k)$ as the probability that each classifier \(i\) assigns rank \(r_{ik}\) to sample \(k\) with label \(y_k\). The performance of classifier $i$ is characterized by the difference between its mean ranks for the two classes,
\begin{equation}
\Delta_i=\mathbb{E}[R_i\mid Y=-1]-\mathbb{E}[R_i\mid Y=+1].
\label{eq:delta_summa}
\end{equation}
With the adopted ranking convention, $\Delta_i>0$ indicates that the classifier tends to assign lower ranks to positive instances and is therefore informative, while $\Delta_i=0$ corresponds to a random classifier. A negative value indicates that the classifier ranks the classes in the opposite direction. For a perfect classifier, the positive instances occupy the first $N_1$ ranks and the negative instances the remaining ranks,

\begin{equation}
 \underbrace{1,\ldots,N_1}_{\text{positive}} \qquad \underbrace{N_1+1,\ldots,N}_{\text{negative}},   
\end{equation}
which gives
\begin{equation}
\mathbb{E}[R_i\mid Y=+1] = \frac{N_1+1}{2}, \qquad \mathbb{E}[R_i\mid Y=-1] = \frac{N+N_1+1}{2},    
\end{equation}
and consequently
\begin{equation}
\Delta_i=\frac{N}{2}.
\end{equation}
More generally, this performance measure has a direct relation to the AUC. As demonstrated in \cite{ahsen2019unsupervised}, given a set of rank predictions with known ground-truth labels
\begin{equation}
\mathrm{AUC}_i=\frac{1}{2}+\frac{\Delta_i}{N}.
\label{eq:auc_delta_summa}
\end{equation}
Thus, $\Delta_i$ is an alternative representation of the classifier performance, with the advantage that it can be inferred without access to the class labels.



Ahsen et al. \cite{ahsen2019unsupervised} demonstrated that higher-order central moments of rank predictions encode information about individual classifier performances. Nevertheless, for constructing the ensemble, the second-order covariance matrix $\Sigma_2$\footnote{Unlike the previous sections (\ref{section:isml} and \ref{section:lsml}), where the covariance matrix was denoted by $R_{\text{cov}}$, this section adopts the original notation from \cite{ahsen2019unsupervised}, using $\Sigma_2$ for the covariance matrix of the rankings. This choice avoids ambiguity with the symbol $r_{ik}$, which is reserved for the ordinal rank position.} is sufficient. Its elements are given by:
\begin{equation}
\Sigma_2(i,j)=
\begin{cases}
\dfrac{N^2-1}{12}, & i=j,\\
C(\rho)\Delta_i\Delta_j, & i\neq j.
\end{cases}
\label{eq:summa_cov}
\end{equation}
The diagonal term follows from the fact that each rank vector is a permutation of ${1,\ldots,N}$ and therefore has variance $(N^2-1)/12$. More importantly, the off-diagonal terms decompose into the product of the performance measures of the two classifiers.
Defining
\begin{equation}
\lambda_c=C(\rho)\|\Delta\|^2,
\qquad
v_i=\frac{\Delta_i}{\|\Delta\|},
\end{equation}
where $\Delta=(\Delta_1,\ldots,\Delta_M)^\top$ and $\|\Delta\|$ its euclidean norm, the covariance matrix can be decomposed as
\begin{equation}
\Sigma_2=\lambda_c vv^\top-\operatorname{diag}(\lambda_c vv^\top)+\frac{N^2-1}{12}I.
\end{equation}
Hence, the off-diagonal covariance structure is rank-one and its leading singular vector provides an estimate $\hat v_i$ proportional to the performance of each classifier. Importantly, only the direction of $\Delta$ is required to construct the ensemble; its absolute magnitude, and consequently the exact value of the AUC, is not necessary at this stage. In practice, the empirical covariance matrix is obtained directly from the observed ranks as
\begin{equation}
\widehat{\Sigma}_2(i,j)=\frac{1}{N}\sum_{k=1}^{N}\left(r_{ik}-\frac{N+1}{2}\right)\left(r_{jk}-\frac{N+1}{2}\right).
\label{eq:summa_emp_cov}
\end{equation}
A rank-one decomposition of its off-diagonal structure then provides the estimator $\hat v$.

Once the relative classifier performances have been estimated, SUMMA constructs the ensemble by maximum likelihood estimation of the latent class of each sample. For a given sample $k$, the likelihood under a candidate class $y_k$ is
\begin{equation}
\Pr(R_{1}=r_{1k},\ldots,R_M=r_{Mk}\mid Y=y_k)=
\prod_{i=1}^{M}{\Pr}(R_i=r_{ik}\mid y_k),
\end{equation}
where conditional independence has again been used. Therefore, the maximum likelihood decision can be written in terms of the log-likelihood ratio,
\begin{equation}
t_k=\sum_{i=1}^{M}\ln\left[\frac{\Pr(R_i=r_{ik}\mid Y=+1)}{\Pr(R_i=r_{ik}\mid Y=-1)}\right].
\label{eq:summa_llr}
\end{equation}
Noting that $\Pr(r)=1/N$, $\Pr(Y=+1)=\rho$, and $\Pr(Y=-1)=1-\rho$, the Bayes theorem yields
\begin{equation}
t_k=\sum_{i=1}^{M}\ln\left[\frac{\Pr(Y=+1\mid R_i=r_{ik})}{\Pr(Y=-1\mid R_i=r_{ik})}\right]+M\ln\left(\frac{1-\rho}{\rho}\right).
\label{eq:summa_bayes}
\end{equation}
The remaining problem is therefore to determine the conditional probability $\Pr(Y=+1\mid R_i=r)$ from the unlabeled rank data.
Rather than imposing a parametric form for this probability, SUMMA uses the maximum entropy principle. The idea is to choose the distribution that satisfies the statistical constraints available from the data while making no additional assumptions about the unknown quantities. In this case, the constraints are the prevalence $\rho$ and the conditional mean-rank difference $\Delta_i$. Under these constraints, the maximum entropy distribution is approximated by
\begin{equation}
{\Pr}(Y=+1\mid R_i=r)=
\left[1+\exp\left[
\frac{12\Delta_i}{N^2-1}
\left(
r-\frac{N+1}{2}
\right)
+
\ln\left(
\frac{N-N_1}{N_1}
\right)
\right]
\right]^{-1}.
\label{eq:summa_maxent}
\end{equation}
This expression shows that the posterior probability of the positive class decreases as the assigned rank increases, consistent with the convention that lower ranks correspond to stronger evidence for the positive class. The derivation of this expression uses a weakly predictive approximation in the original formulation, although the authors report that the resulting expression remains a good approximation beyond this regime. Substituting Eq.~\eqref{eq:summa_maxent} into the likelihood-ratio expression and using
\begin{equation}
\Delta_i=
\sqrt{\frac{\lambda_c}{\rho(1-\rho)}}v_i,
\end{equation}
the terms independent of the sample rank do not affect the final decision. The maximum likelihood estimate of the class label can therefore be written as
\begin{equation}
\hat y_k^{SUMMA}=
\operatorname{sign}
\left(
\sum_{i=1}^{M}
\hat v_i
\left(
\frac{N+1}{2}-r_{ik}
\right)
\right).
\label{eq:summa_classifier}
\end{equation}

Equation~\eqref{eq:summa_classifier} shows the main idea of SUMMA: each classifier contributes its rank to the sample $k$, but the contribution is weighted by $\hat v_i$, which is proportional to its estimated performance. Thus, unlike unweighted rank aggregation (Borda Count Method \cite{dwork2001rank}, which is described in Section~\ref{section:borda_count}), more informative classifiers have a greater influence on the final decision.
Because $\sum_{i=1}^{M} \hat v_i\frac{N+1}{2}$ is constant for every sample, it does not affect the ordering induced by the ensemble. Consequently, SUMMA defines the continuous ensemble score
\begin{equation}
c_k^{SUMMA}=
-\sum_{i=1}^{M}
\hat v_i r_{ik}.
\label{eq:summa_score}
\end{equation}
Samples with larger values of $c_k^{SUMMA}$ are therefore ranked as more likely to belong to the positive class. The resulting procedure can be interpreted as an unsupervised weighted rank aggregation, where the weights are inferred directly from the covariance structure of the unlabeled predictions.




\subsection{Non-Parametric and Heuristic Ensemble Combination Baselines}

Non-parametric and heuristic ensemble methods provide robust baselines by directly aggregating continuous predictions or ordinal ranks from individual base classifiers without requiring hyperparameter tuning or model fitting. Below, we detail the non-parametric combination rules evaluated in this work.

\subsubsection{Borda Count Method}
\label{section:borda_count}

Similarly to SUMMA, Borda Count operates on the continuous score matrix $S$ produced by the base classifiers. To ensure scale-invariant aggregation, each detector's score for an instance $x_j$ is converted into its ordinal rank $r_{ij} \in \{1, \dots, N\}$ (ordered in ascending score magnitude). The final consensus score for an instance $x_j$ is computed as the unweighted mean rank across all $M$ classifiers, corresponding to the classic unsupervised Borda Count consensus rule \cite{dwork2001rank, aslam2001models}:

\begin{equation}
c_{\text{Borda}}(x_j) = \frac{1}{M} \sum_{i=1}^M r_{ij}.
\end{equation}

\subsubsection{Score Combination Rules}
Operating on standard-scaled score matrix $\tilde{S}$, three classic heuristic combination rules from the outlier ensemble literature \cite{aggarwal2015theoretical, zhao2019pyod} are evaluated:
\begin{itemize}
    \item \textbf{Median Combination:} Computes the element-wise median across the base classifiers. It is highly robust to isolated detector failures or extreme outlier scores:
    \begin{equation}
    c_{\text{Median}}(x_j) = \text{median}\left(\tilde{s}_{1j}, \tilde{s}_{2j}, \dots, \tilde{s}_{Mj}\right)
    \end{equation}
    
    \item \textbf{Maximum Combination (Max):} Takes the maximum normalized score among all classifiers, maximizing sensitivity to subtle anomalies detected by any single model:
    \begin{equation}
    c_{\text{Max}}(x_j) = \max_{1 \le i \le M} \tilde{s}_{ij}
    \label{eq:max}
    \end{equation}
    
    \item \textbf{Average-of-Maximum (AOM):} Partitions the $M$ base classifiers randomly into $B$ disjoint sub-ensembles $\mathcal{G}_1, \dots, \mathcal{G}_B$ \cite{aggarwal2015theoretical}. It computes the maximum score within each subgroup and averages across subgroups, balancing sensitivity and variance reduction:
    \begin{equation}
    c_{\text{AOM}}(x_j) = \frac{1}{B} \sum_{b=1}^B \max_{i \in \mathcal{G}_b} \tilde{s}_{ij}.
    \end{equation}
\end{itemize}

\subsubsection{Score Combinations Rules (from Dark Machines Challenge)}

In order to complement the previous models and adopt the benchmark implementations from \cite{darkmachinechallenge} for direct comparison in Section~\ref{section:results}, we evaluate four additional heuristic strategies:

\begin{itemize}
    \item \textbf{OR}:
    \begin{equation}
        c_{\text{OR}}(x_j) =
        \max_{1 \leq i \leq M} s'_{ij}
        \label{eq:or}
    \end{equation}

    \item \textbf{AND}:
    \begin{equation}
        c_{\text{AND}}(x_j) =
        \min_{1 \leq i \leq M} s'_{ij}
    \end{equation}

    \item \textbf{Average (AVG)}:
    \begin{equation}
        c_{\text{AVG}}(x_j) =
        \frac{1}{M}\sum_{i=1}^{M} s'_{ij}
    \end{equation}

    \item \textbf{Product (PROD)}:
    \begin{equation}
        c_{\text{PROD}}(x_j) =
        \prod_{i=1}^{M} s'_{ij}.
    \end{equation}
\end{itemize}
Although \textbf{OR} (Eq.~(\ref{eq:or})) and \textbf{Max} (Eq.~(\ref{eq:max})) share the same mathematical operator ($\max$), they differ fundamentally in their input calibration: while \textbf{Max} operates on standard-scaled scores $\tilde{s}_{ij}$ (measuring dispersion in units of standard deviation $\sigma$), the Dark Machines combinations apply min-max normalization $s'_{ij} \in [0, 1]$. Consequently, \textbf{OR} acts as a fuzzy logic union over bounded pseudo-probabilities, making its sensitivity to extreme outliers and heavy-tailed score distributions distinct from \textbf{Max}.

\section{Signals and Backgrounds with a Fixed $t\bar{t}$ Final State}
\label{section:signals-backg}

 To illustrate the variation in predictions across unsupervised models and its stabilization through combinations in ensembles, we first fix the final state topology to two $b$-tagged jets, two opposite-sign leptons, and missing energy: $pp\to b\bar{b}\ell^+\ell^-+\not\!\!\! {E_T},\; \ell=e,\mu$ where the missing energy is due to escaping neutrinos. This final state is predicted in the production and decay of many interesting new physics models. In the SM, it represents one of the most promising channels to measure the trilinear Higgs self-interaction in double Higgs production via gluon fusion. By fixing the background hypothesis, we factor out a key aspect of unsupervised searches, making the measurement of the performance of algorithms more straightforward. In section~\ref{section:darkmachines}, we perform a study with varying signal and background final states.

\subsection{Dominant Background}
\begin{figure}[t!]
    \centering
    \includegraphics[width=0.7\linewidth]{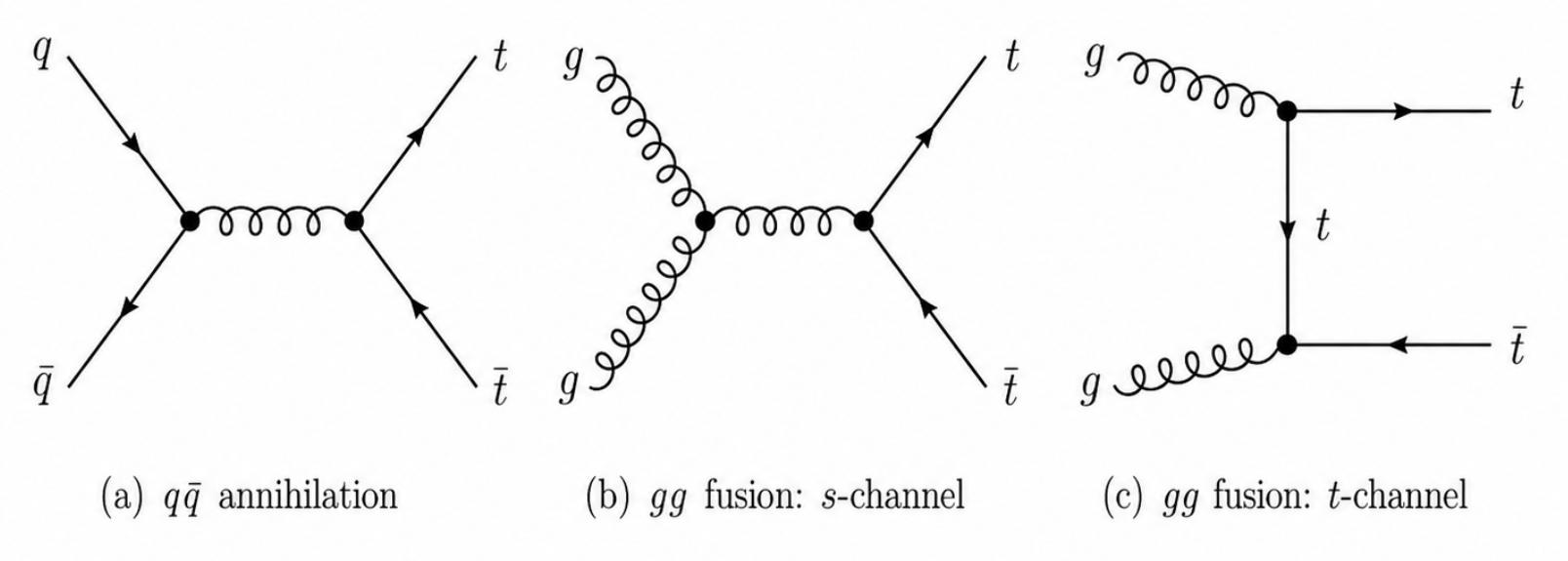}
    \caption{QCD production of top quark pairs at Leading Order. The diagrams depict the (a) $q\bar{q}$-initiated and (b,c) $gg$-initiated processes.}
    \label{fig:sm-tt}
\end{figure}

 At the LHC, the dominant background to signals of this type is $t\bar{t}$ production, where both top quarks decay leptonically with a NNLO+NNLL QCD cross section of around 44 pb after $W$-bosons decay to electrons and muons for a center-of-mass energy of 14 TeV.

 There are two contributions at Leading Order to this process. The $gg\to t\bar{t}$ and $q\bar{q}\to t\bar{t}$ where all quarks $q$ in the proton contribute to this channel. The Feynman diagrams are shown in Fig.~\ref{fig:sm-tt}.

\subsection{Non-Resonant and Resonant Double Higgs production}

Many interesting models predict distinct signals in double Higgs production, both resonantly with the presence of a new heavy scalar, for example, and non-resonantly as modifications of the SM Higgs sector parameters. The double Higgs production is also of prime interest for measuring the trilinear Higgs self-interaction. Consider the case of an extended scalar sector with a new SM singlet real scalar, $H$. The scalar potential might contain the following interactions, after spontaneous symmetry breaking,
\begin{equation}
V(h,H) \supset \kappa_3{\color{red}{\lambda_{hhh}}} vh^3+\kappa_4{\color{red}{\lambda_{hhh}}} h^4 + {\color{blue}{\lambda_{Hhh}}}h^2 H\; .
\label{eq:VhH}
\end{equation}

A fruitful model for an extended Higgs sector is the SM-singlet real scalar -- the xSM~\cite{Profumo:2007wc,Profumo:2014opa,Huang:2017jws}, an interesting laboratory for studying electroweak phase transitions that can be tested in gravitational waves and collider experiments. Even when $H$ is too heavy, deviations from the SM can be found in the trilinear and quartic interactions: $\kappa_3{\color{blue}\lambda_{hhh}} v$ and $\kappa_4{\color{red}{\lambda_{hhh}}}$, respectively . Yet, the smoking gun signature would be the production of the heavy scalar $H$ followed by its decay into $hh$. In many portions of the xSM parameter space, this is the preferred decay channel of H~\cite{Huang:2017jws} once all the other channels are proportional to $\sin^2\theta$, the $h$--$H$ mixing angle, which has been constrained by collider data as $|\theta|<0.30$ at 95\% CL~\cite{Bosse:2026bdk}.

Many decay channels of the Higgs boson can be used to detect signals from an extended scalar sector affecting double Higgs production. A compromise between signal dilution and background contamination is the decay channel $hh\to b\bar{b}W^+W^-\to b\bar{b}\ell^+\ell^-+\not\!\!\! {E_T}$ with a combined branching ratio of 1.8\%. This figure compares favorably with $b\bar{b}\gamma\gamma$ whose branching ratio is 0.26\%. However, the $b\bar{b}W^+W^-$ mode competes with the large QCD $t\bar{t}$ background for statistical significance. Yet, $b\bar{b}W^+W^-$ has been established as the best channel for Higgs pair production searches~\cite{Alves:2025slf,Bosse:2026bdk}

\begin{equation}
     pp\to hh\to b\bar{b}W^+W^-\to b\bar{b}\ell^+\ell^-+\not\!\!{E_T}
 \end{equation}
\begin{figure}[t!]
    \centering
    \includegraphics[width=0.8\linewidth]{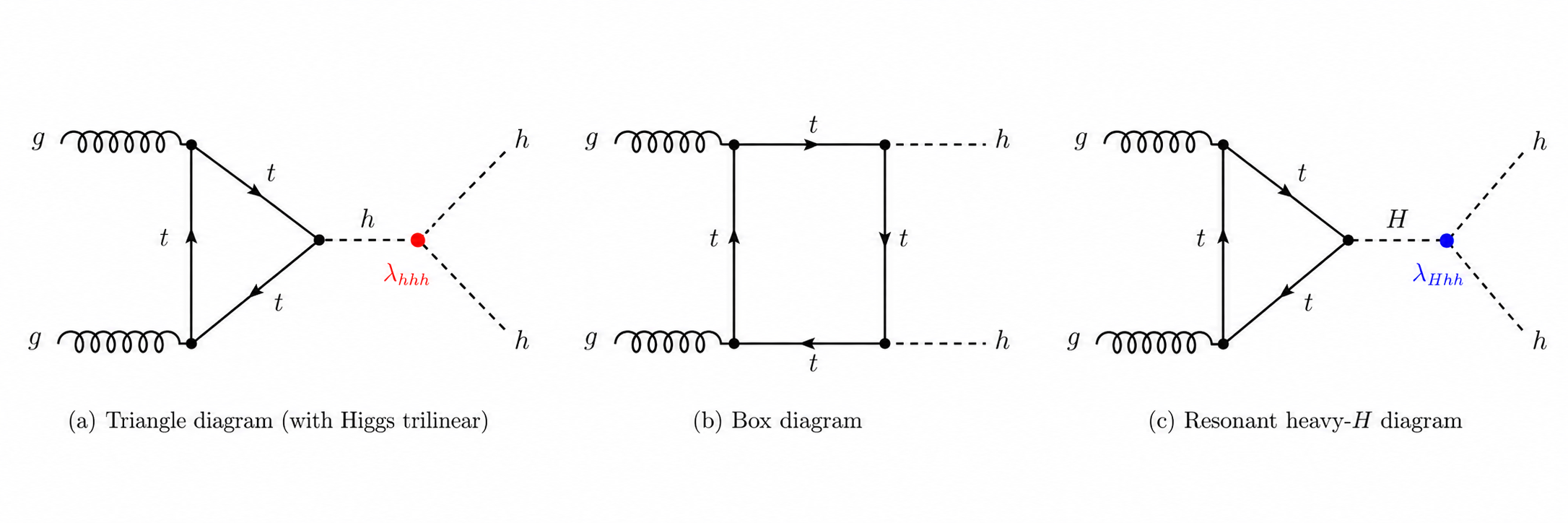}
    \caption{Feynman diagrams for double Higgs production. Diagrams (a) and (b) represent the non-resonant $hh$ production, and diagram (c) a resonant production through a new scalar, $H$.}
    \label{fig:hh}
\end{figure}

The non-resonant $hh$ production receives two contributions, (a) and (b), depicted in Figure~\ref{fig:hh}. Because the interference of the triangle and box diagrams is destructive, the $\kappa_3$ modifier and $\theta$ affect the kinematics of the produced particles in significant ways that can be used to distinguish between SM and new physics. When there is a new contribution from a heavy scalar, diagram (c) in Figure~\ref{fig:hh}, the $b$-jets, leptons, and missing energy tend to be harder, facilitating the distinction. Despite all diagrams of Figure~\ref{fig:hh} contributing to $hh$ production, when $H\to hh$ is present, we can easily isolate its contribution with kinematic cuts. If the $H$ width is sufficiently narrow, no interference is expected between $H\to hh$ and the SM.  

 \subsection{Resonant Top Quark Pair production in $H\to t\bar{t}$}

  In the xSM, the heavy Higgs boson decays to all SM particles, but with branching ratios proportional to $\sin^2\theta$ as discussed above. Contrary to the SM Higgs boson, however, if $m_H>2m_t$, the heavy Higgs decays to top quarks, whose interactions are strengthened by the top mass. Again, the heavy Higgs boosts its decay yields, making them harder and more distinguishable from the background, which is composed mainly of QCD $t\bar{t}$. Moreover, in certain regions of the parameter space, double Higgs blind spots arise, depleting the $hh$ channel~\cite{Alves:2020bpi}.

\begin{figure}[t!]
    \centering
    \includegraphics[width=0.4\linewidth]{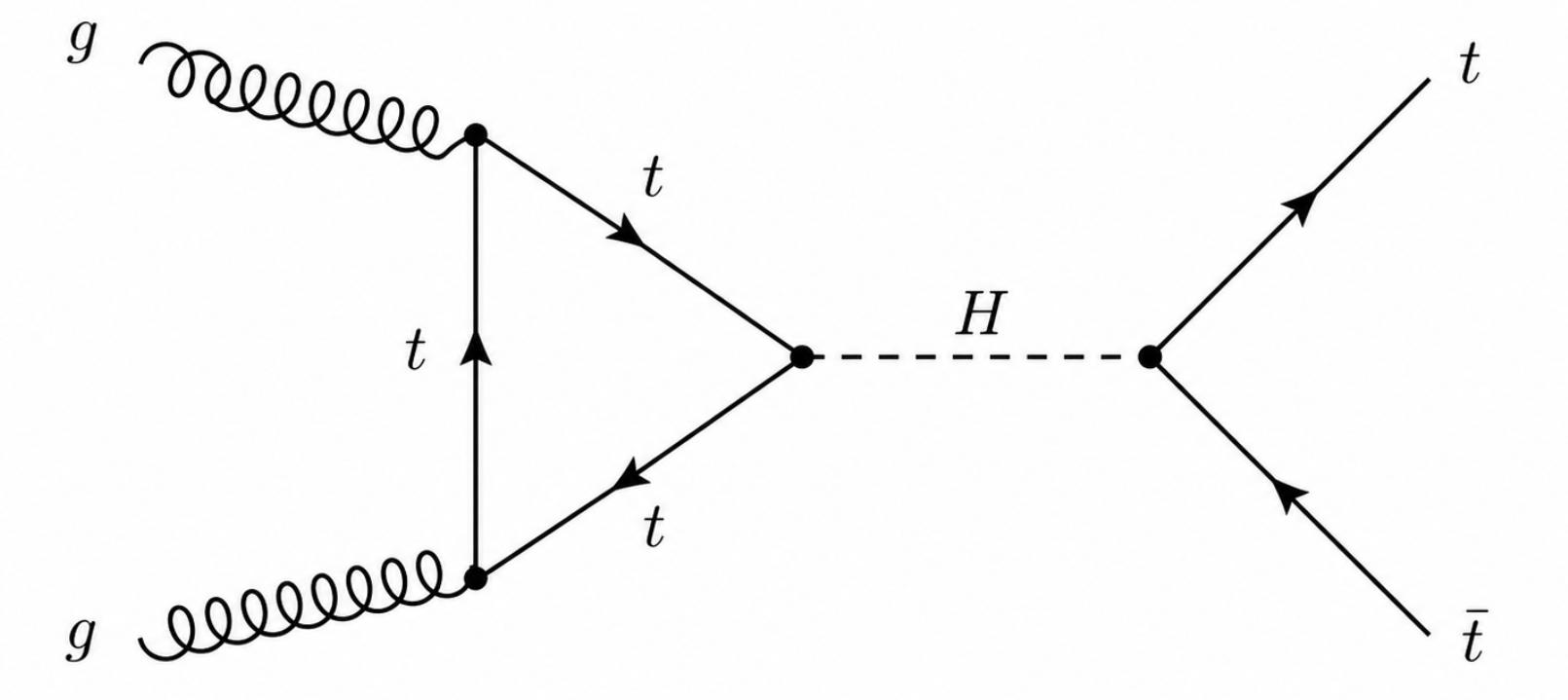}
    \caption{Feynman diagram for heavy Higgs production and decay to top quark pair.}
    \label{fig:Htt}
\end{figure}

  The Feynman diagram for $H$ production and decay to top pairs is shown in Figure~\ref{fig:Htt}. Note that the production cross section also depends on $\sin^2\theta$.
  
\subsection{Dimension-6 Operators in Top Pair Production}

 Suppose a new heavy Higgs boson or $Z^\prime$ is produced in gluon-fusion and $q\bar{q}$ annihilation, respectively, and then decays into a top quark pair. If the new particles are too massive to be produced on their mass-shell, we cannot resolve the particles and interactions that are driving the new physics even though they can produce excesses of events in comparison to SM expectations and also influence the kinematic distributions of the top quark decays. 

 In those cases, we can use an effective field theory (EFT) formulation of the interactions. In the case of $gg\to t\bar{t}$ and $q\bar{q}\to t\bar{t}$ interactions that respect the SM symmetries, many effective operators could contribute, but restricting ourselves to dimension-6 operators, we can pick, for example, the following ones
 \begin{equation}
     {\cal L}_{EFT} \supset {\color{blue}\frac{f_{gg}}{\Lambda^2}} f^{abc}\tilde{G}^{a,\mu}_\nu G^{b,\nu}_{\rho}G^{c,\rho}_{\mu}+{\color{blue}{\frac{f_{qq}}{\Lambda^2}}}(\bar{q}_L\gamma^\mu q_L)(\bar{t}_L\gamma_\mu t_L)\; ,
 \end{equation}
where $G^a_{\mu\nu}=\partial_\mu A^a_\nu-\partial_\nu A^a_\mu$ is the gluon field strength tensor and $f^{abc}$ are the $SU(3)_C$ structure constants. The effective couplings are given by $f_{gg}/\Lambda^2$ and $f_{qq}/\Lambda^2$, where $\Lambda$ is the mass scale at which new physics manifests.

\begin{figure}[t!]
    \centering
    \includegraphics[width=0.6\linewidth]{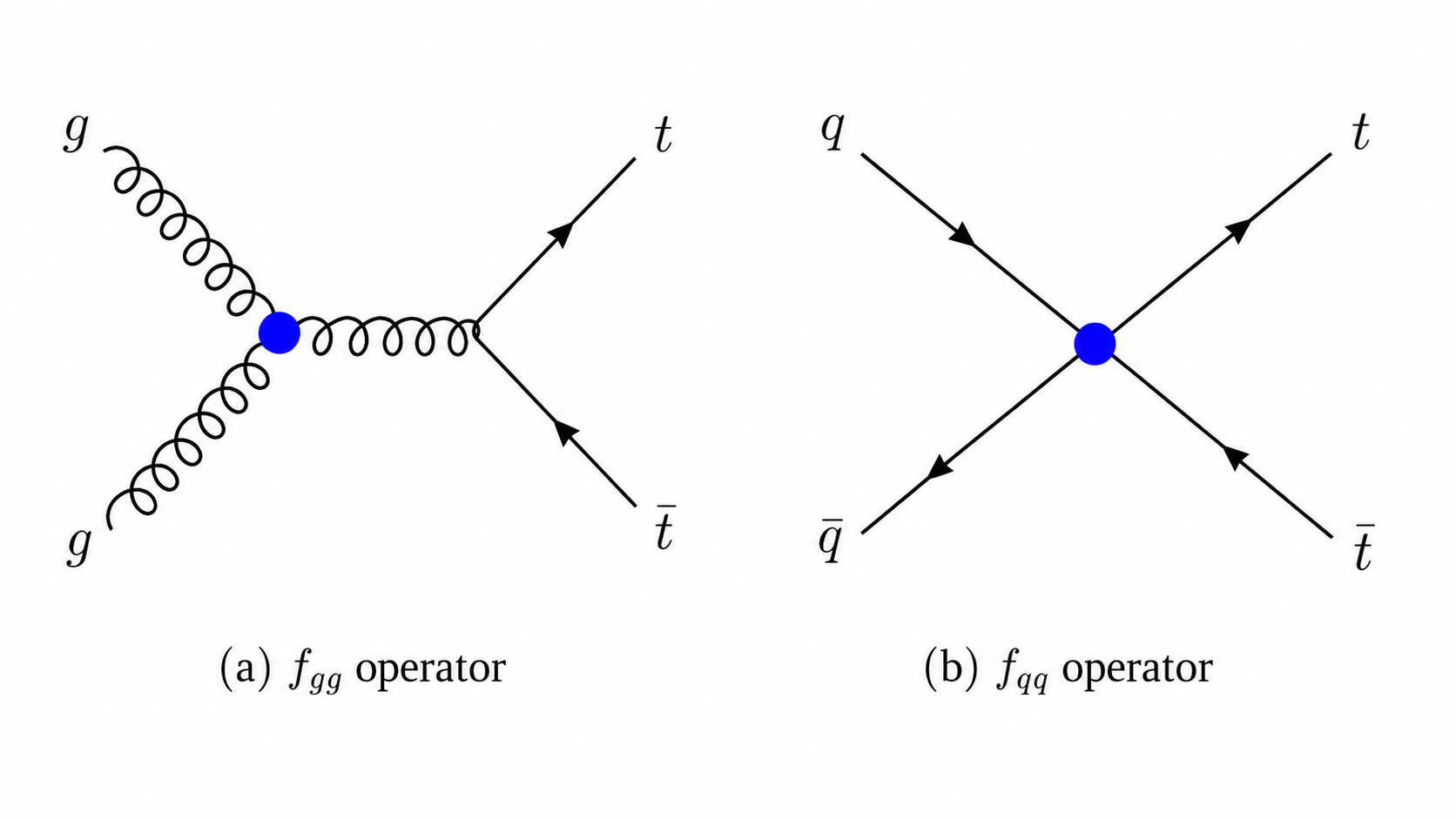}
    \caption{Feynman diagrams for top quark pair production from EFT interactions. In panel (a) we show the effective triple-gluon interaction, and in panel (b), the four-fermion interaction $q\bar{q}t\bar{t}$.}
    \label{fig:fop}
\end{figure}

 In our simulations, we take just one operator at a time, setting either (a) $(f_{gg},f_{qq})=(1,0)$ or (b) $(f_{gg},f_{qq})=(0,1)$. The Feynman diagrams corresponding to (a) $gg\to t\bar{t}$ and (b) $q\bar{q}\to t\bar{t}$ operators are depicted in Figure~\ref{fig:fop}. Note that in the case of gluon-initiated production, the new interaction occurs in the triple-gluon interaction, but not in the $gt\bar{t}$ vertex. This could correspond to new heavy quarks running in a loop coupled to three gluons, for example. In the case of quark-initiated production, the effective interaction is a typical four-fermion interaction hiding a heavy force carrier particle, for example. We set $f_{gg}$ and $f_{qq}$ to unity and $\Lambda=1$ TeV in the simulations. As we will see, identifying these operators in action is challenging as they resemble and interfere with the $t\bar{t}$ backgrounds.

\section{Simulations and Representation of Data}
\label{section:simulations}

 All parton-level simulations are performed with \texttt{MadGraph5}~\cite{Alwall:2014hca} for the 14 TeV LHC at Leading Order QCD. We chose
the NNPDF2.3QED~\cite{Ball:2013hta} for the parton distribution functions. The renormalization($\mu_R$)/factorization($\mu_F$) scales are chosen as the Higgs boson mass, $m_h=125$ GeV, the new heavy Higgs boson $m_H=1$ TeV, and the top quark mass $m_t=172$ GeV, for the $hh$ and $H\to hh$ and $H\to t\bar{t}$ signals, and for the $t\bar{t}$ background, respectively. For EFT operators, we set the scales dynamically as $\mu_R=\mu_F=H_T=\sum\limits_{i\in \mathrm{event}} p_{T_i}$. Hadronization of quarks and gluons is performed with \texttt{Pythia8}~\cite{Sjostrand:2007gs}. Detector effects are simulated with \texttt{Delphes3}~\cite{deFavereau:2013fsa} with default cards. \texttt{Fastjet3}~\cite{Cacciari:2011ma} is used for jet clustering with anti-$k_t$ algorithm for jet cone of size $\Delta R=0.4$.

 Selection criteria for the basic identification and triggering  of two opposite-charge leptons, two tagged-bottom jets, and missing energy are the following:
 \begin{eqnarray}
&& p_T(b) > 30\;\hbox{GeV},\;\; |\eta_b|<2.5,\\
&& p_T(\ell) > 20\;\hbox{GeV},\;\; |\eta_\ell|<2.5,\\
&& \not\!\! E_T > 20\;\hbox{GeV}\; .
\label{eq:cuts}
\end{eqnarray}

The representation of events for all algorithms is composed of nine kinematic variables. These variables are described below:
\begin{itemize}
    \item invariant masses: $M_{bb}$, $M_{\ell\ell}$, $M_{\ell\ell\nu\nu}$, and $M_{bb\ell\ell\nu\nu}$. They capture the resonant and non-resonant behavior of intermediate particles in the process;
    \item transverse momenta: $p_T^{(bb)}$ and $p_T^{(\ell\ell)}$. Harder spectra occur when heavy particles are present. Deviations from the QCD tails are expected when additional new physics contributions interfere with the SM amplitudes;
    \item distance in the azimuth-rapidity plane: $\Delta R_{bb}$ and $\Delta R_{\ell\ell}$. They also capture resonant structures;
    \item $\text{RIV}_t=\not\!\!{M}/M_{vis}$ where $M_{inv}$ is the invariant mass of the visible particles in the event and $\not\!\! {M}=\sqrt{p_T(\nu_{t,1})^2+p_T(\nu_{t,2})^2+M^2(\nu_{t,1},\nu_{t,2})}$. 
\end{itemize}

\begin{figure}[t!]
    \centering
    \includegraphics[width=0.41\linewidth]{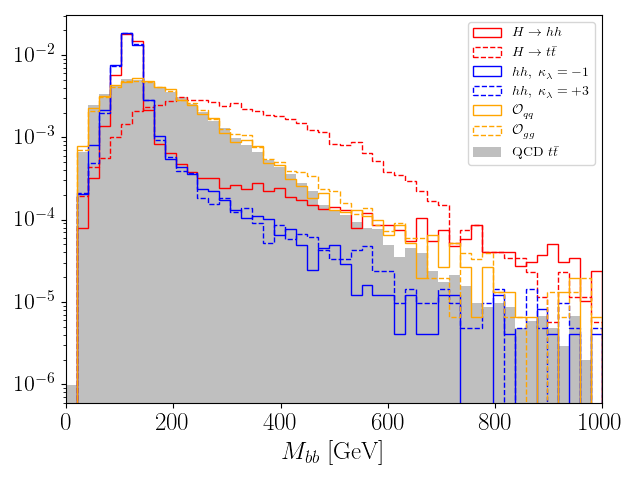}
    \includegraphics[width=0.41\linewidth]{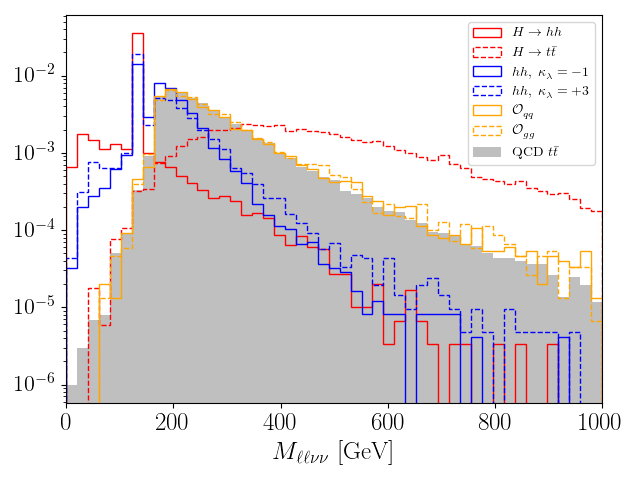} \\
    \includegraphics[width=0.41\linewidth]{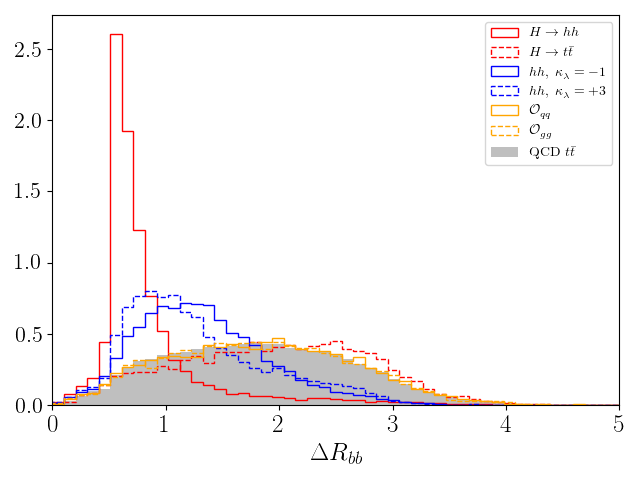}
    \includegraphics[width=0.41\linewidth]{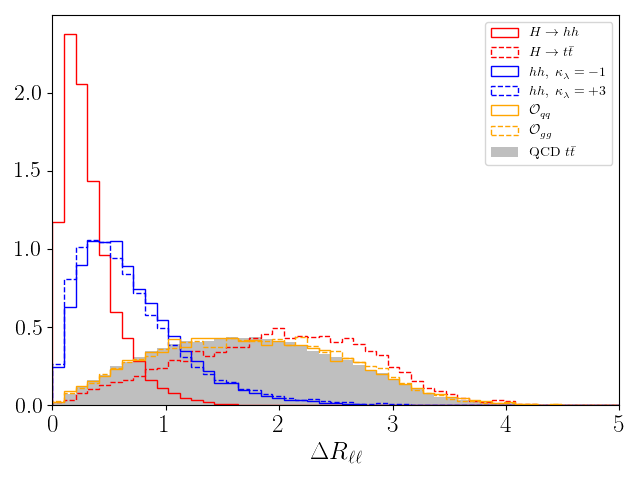} \\
    \includegraphics[width=0.41\linewidth]{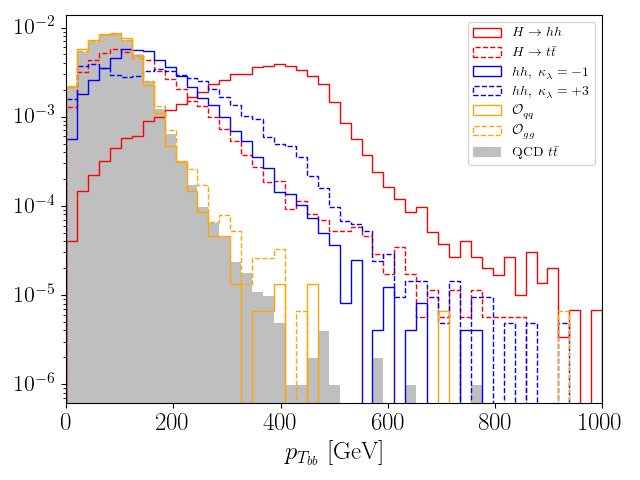}
    \includegraphics[width=0.41\linewidth]{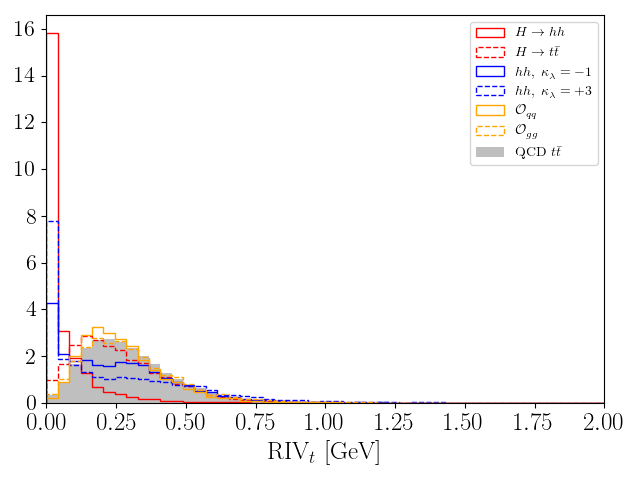}
    \caption{In the top row, we show the $M_{bb}$ distributions at left, and $M_{\ell\ell\nu\nu}$ at right panel. The middle row depicts the $\Delta R_{bb}$ at the left panel, and $\Delta R_{\ell\ell}$ at right. In the bottom row, the $p_{T_{bb}}$ distributions is shown at the left panel, while the RIV$_t$ distribution at right.}
    \label{fig:dist}
\end{figure}

In these variables, the neutrino momenta are reconstructed with the computation of the {\it Higgsness} and {\it Topness} variables~\cite{Kim:2018cxf}, which basically choose the neutrino momenta in order to minimize the mass constraints of double Higgs and top quark pair events. In particular, $p(\nu_{t,1})$ and $p(\nu_{t,2})$ correspond to solutions that minimize top quark pair production constraints. The variable $\text{RIV}_t$, by its turn, is designed for optimum separation between $hh$ and $t\bar{t}$ events~\cite{Alves:2025slf}. We show, in Figure~\ref{fig:dist}, the plots of some selected kinematic distributions of the $t\bar{t}$ background and the new physics signals considered in this work. We can expect, from these distributions, that scenarios with heavy Higgs production will be easier to identify than non-resonant $hh$ production, and that effective interactions will be very challenging scenarios.

Now, let us discuss the training and testing procedures and the baseline algorithms used to construct the ensembles.

\section{Training and Testing the Algorithms}
\label{section:train-test}
\subsection{Base Algorithms}

Firstly, we describe the base algorithms that will be combined in ensembles for the task of unsupervised and semi-supervised new physics classification. They can be categorized into two groups: the clustering models and the outlier detectors.

\subsubsection{Clusterings Models}

The clustering models evaluated in this work are unsupervised learning methods applied in a transductive setting. Rather than learning a decision function from a separate training dataset and subsequently applying it to unseen data, the models are fitted directly to the dataset of interest, without using event labels. For anomaly detection, the data are partitioned into $k=2$ clusters, with the resulting partition interpreted as a separation between standard background events and potential signal-like anomalies.

The evaluated clustering algorithms are categorized into the following paradigms:
\begin{itemize}
    \item \textbf{Centroid-based Models:} \textbf{K-Means} \cite{mcqueen1967some} is partitioning algorithm that assigns data points to $k=2$ clusters by iteratively minimizing the Within-Cluster Sum of Squares (WCSS) to determine optimal cluster centroids. \textbf{MiniBatch K-Means} \cite{sculley2010web} is a optimization variant of K-Means that updates centroid positions using small, random mini-batches of data, significantly reducing computational overhead while introducing stochastic diversity beneficial for ensemble combinations.
    
    \item \textbf{Probabilistic and Mixture Models:}  \textbf{Gaussian Mixture Models (GMM)} \cite{bishop2006pattern} is a parametric density estimation method that models data as a linear superposition of $k=2$ Gaussian components, providing soft cluster assignments through posterior probabilities. An extension of GMM is the \textbf{Bayesian Gaussian Mixture Models (BGMM)} \cite{attias1999variational} which is fitted via Variational Inference (VI) over Dirichlet process priors. By optimizing the Evidence Lower Bound (ELBO), BGMM infers posterior parameter distributions and provides natural regularization against overfitting.
    
    \item \textbf{Hierarchical Model:} \textbf{Agglomerative Clustering} \cite{ward1963hierarchical} is a bottom-up hierarchical method that initializes each data point as an individual cluster and iteratively merges proximity-based cluster pairs until obtaining $k=2$ final partitions.

    \item \textbf{Tree-based Subcluster Model:} \textbf{BIRCH} \cite{zhang1996birch} is a incremental tree-based algorithm that dynamically compresses data into a compact hierarchy of subclusters stored in a Clustering Feature Tree (CF-Tree), followed by a global clustering step that partitions subcluster centroids into $k=2$ final clusters.

\end{itemize}

\subsubsection{Outlier Detections Models}

The outlier detection algorithms in this study are deployed in an inductive setting, where models are trained exclusively on a pure, controlled background sample. This yields a generalized decision function capable of scoring unseen events. Physically, events exhibiting high anomaly scores deviate significantly from the expected background distribution, representing potential candidates for Beyond the Standard Model (BSM) new physics.

The evaluated detectors leverage diverse algorithmic paradigms:
\begin{itemize}
    \item \textbf{Isolation-based Models:} Isolation Forest (\textbf{IForest}) \cite{liu2008isolation} and Isolation Nearest Neighbors Ensemble (\textbf{INNE}) \cite{bandaragoda2018isolation} assign anomaly scores based on the ease of isolating instances. While IForest uses recursive axis-aligned random splits (where anomalies yield shorter path lengths), INNE isolates events using hyperspherical boundaries derived from nearest-neighbor ensembles.
    
    \item \textbf{Density-based Models:} Kernel Density Estimation (\textbf{KDE}) \cite{parzen1962estimation, latecki2007outlier}, Histogram-based Outlier Score (\textbf{HBOS}) \cite{goldstein2012histogram}, and Local Outlier Factor (\textbf{LOF}) \cite{breunig2000lof} estimate event density across feature space. Events residing in low-density regions (or exhibiting lower local reachability density than their $k$-nearest neighbors) receive high anomaly scores.
    
    \item \textbf{Distance-based Models:} $k$-Nearest Neighbors (\textbf{KNN}) \cite{ramaswamy2000efficient} evaluates isolation by measuring the distance of an event to its $k$-th nearest background neighbor; larger distances indicate anomalous behavior.
    
    \item \textbf{Statistical and Distribution-based Models:} \textbf{COPOD} \cite{li2020copod} and \textbf{ECOD} \cite{li2022ecod} use non-parametric empirical cumulative distribution functions (marginal copulas and ECDFs, respectively) to compute joint tail probabilities without parametric assumptions. Complementarily, Minimum Covariance Determinant (\textbf{MCD}) \cite{estimator1999fast} estimates a robust location and covariance matrix of the background, scoring outliers via robust Mahalanobis distance.
    
    \item \textbf{Reconstruction-based Models:} Principal Component Analysis (\textbf{PCA}) \cite{shyu2003novel}, Autoencoders (\textbf{AE}) \cite{sakurada2014anomaly}, and Variational Autoencoders (\textbf{VAE}) \cite{an2015variational} project features into a lower-dimensional latent space and reconstruct the input. The anomaly score is defined by the reconstruction residual (or reconstruction probability), where high reconstruction errors highlight events poorly captured by the background model.
    
    \item \textbf{Boundary-based Models:} One-Class Support Vector Machine (\textbf{OCSVM}) \cite{scholkopf2001estimating} maps features into a high-dimensional Hilbert space via kernel functions to construct a maximum-margin hyper-plane separating dense background regions from the origin.
    
    \item \textbf{Projection Ensembles:} Lightweight On-line Detector of Anomalies (\textbf{LODA}) \cite{pevny2016loda} projects data onto multiple sparse, random 1D sub-spaces, constructing 1D histograms to compute the final anomaly score as the average negative log-likelihood across sub-spaces.
    
    \item \textbf{Deep Representation Models:} Deep Support Vector Data Description (\textbf{DeepSVDD}) \cite{ruff2018deep} maps background events into a minimal-volume latent hypersphere through a deep neural network; the distance of an event to the center of this hypersphere serves as its anomaly score.
\end{itemize}

\subsection{Train-Test Methodology}

Now we describe the training and inference procedures adopted to generate the inputs required by the ensemble methods presented in Section \ref{section:building_ensembles}. Since the base models belong to two distinct learning paradigms, namely transductive clustering and inductive semi-supervised outlier detection, their training and inference procedures are described separately.

The evaluation is performed on a mixed test dataset with a signal-to-background ratio of $S/B = 0.005$, reproducing highly imbalanced conditions expected in realistic scenarios. Because the considered kinematic variables span different numerical scales, all features are standardized using statistics computed exclusively from a pure background sample. This normalization assigns equal importance to all variables while preventing information leakage from the signal. Moreover, by estimating the normalization parameters solely from background events, anomalous events are naturally expected to lie farther from the normalized background distribution, which is advantageous for anomaly detection methods.

The first family comprises clustering algorithms. These methods follow a transductive learning paradigm, meaning that no general predictive model is learned prior to inference. Instead, each clustering algorithm is applied directly to the mixed test dataset, partitioning the events into $k=2$ clusters corresponding to the expected signal and background populations. Since clustering algorithms do not assign semantic labels to the resulting clusters, the cluster containing the largest number of events is identified as the background cluster, reflecting the overwhelming dominance of $t\bar{t}$ events in the considered datasets.

To provide a common interface for the ensemble methods, each clustering algorithm is associated with a scoring function that reflects its confidence that an event belongs to the signal class. Specifically, centroid-based algorithms (KMeans and MiniBatchKMeans) use the Euclidean distance to the background centroid, Gaussian mixture models (GMM and BayesianGMM) use the posterior probability of the signal component, whereas hierarchical methods (Agglomerative and Birch) use the cluster assignments directly.

The second family consists of semi-supervised outlier detection models implemented in the \texttt{PyOD} library. Unlike clustering algorithms, these methods follow an inductive learning paradigm. Each detector is trained exclusively on a pure background sample to learn a generalized anomaly scoring function, which is subsequently applied to the mixed test dataset. The same background-based normalization described above is employed during both training and inference. To determine the decision threshold without introducing data leakage, an independent background calibration sample is used. For each detector $j$, the threshold $\tau_j$ is defined as the $99.5$ percentile of the calibration scores, and any event satisfying $S_{j,i}\ge\tau_j$ is classified as anomalous.

Finally, the outputs produced by both model families are converted into the score matrix $S$ and the binary decision matrix $Z$, providing the common representation required by the ensemble methods described in Section \ref{section:building_ensembles}.

\section{Results for $t\bar{t}$ Final State Scenarios}
\label{section:results}

This section presents the performance of the methods considered in this study across seven different physics scenarios, with particular emphasis on both discrimination performance and stability across repeated runs. The physics scenarios, as described in Section~\ref{section:signals-backg}, comprise:
\begin{itemize}
\item non-resonant double Higgs production with $\kappa_3=-1,1,+3$ corresponding to two new physics couplings plus the SM case, this is the only SM scenario considered in this work,
\item resonant $H\to hh$ production with $m_H=1$ TeV and $\Gamma_H=10$ GeV,
\item resonant $H\to t\bar{t}$ production with $m_H=1$ TeV and $\Gamma_H=10$ GeV,
\item EFT operators with $(f_{gg},f_{qq})=(1,0)$, and $(f_{gg},f_{qq})=(0,1)$, and $\Lambda=1$ TeV
\end{itemize}

The resonant scenarios are expected to be the easiest to identify since some kinematic distributions tend to be much harder compared to the background yields. The non-resonant cases should present an intermediate degree of difficulty for classification, while the EFT new physics should be challenging.

Because the physics scenarios may differ substantially in their intrinsic difficulty, directly comparing their absolute AUC values can obscure the relative behavior of the methods. To mitigate this effect, we complement the AUC analysis with a ranking-based comparison. For each scenario, the methods are ranked according to their mean AUC, with the best-performing method receiving the highest rank. The average rank across all scenarios is then used as an aggregate measure of relative performance, assigning equal importance to each physics scenario regardless of its intrinsic difficulty.

\subsection{Fully unsupervised models}

The first experiment considers fully unsupervised models, in which no information about the signal or background labels is used during model fitting. For each physics scenario, each method is independently trained and evaluated 50 times using a repeated sub-sampling procedure. In each run, a new subset of events is randomly sampled from the corresponding dataset, and the method is trained and evaluated on that sample. This procedure results in a distribution of AUC values for each model and physics scenario.

The mean AUC over the repetitions is used to characterize the average discriminative performance of each method, while the corresponding standard deviation provides a measure of its stability under changes in the sampled data. This distinction is particularly relevant for the present analysis, since a high average AUC accompanied by a large dispersion may indicate that the observed performance is sensitive to the specific sample used during the procedure.
\begin{figure}
\begin{minipage}{0.48\linewidth}
\centering
\includegraphics[width=\linewidth]{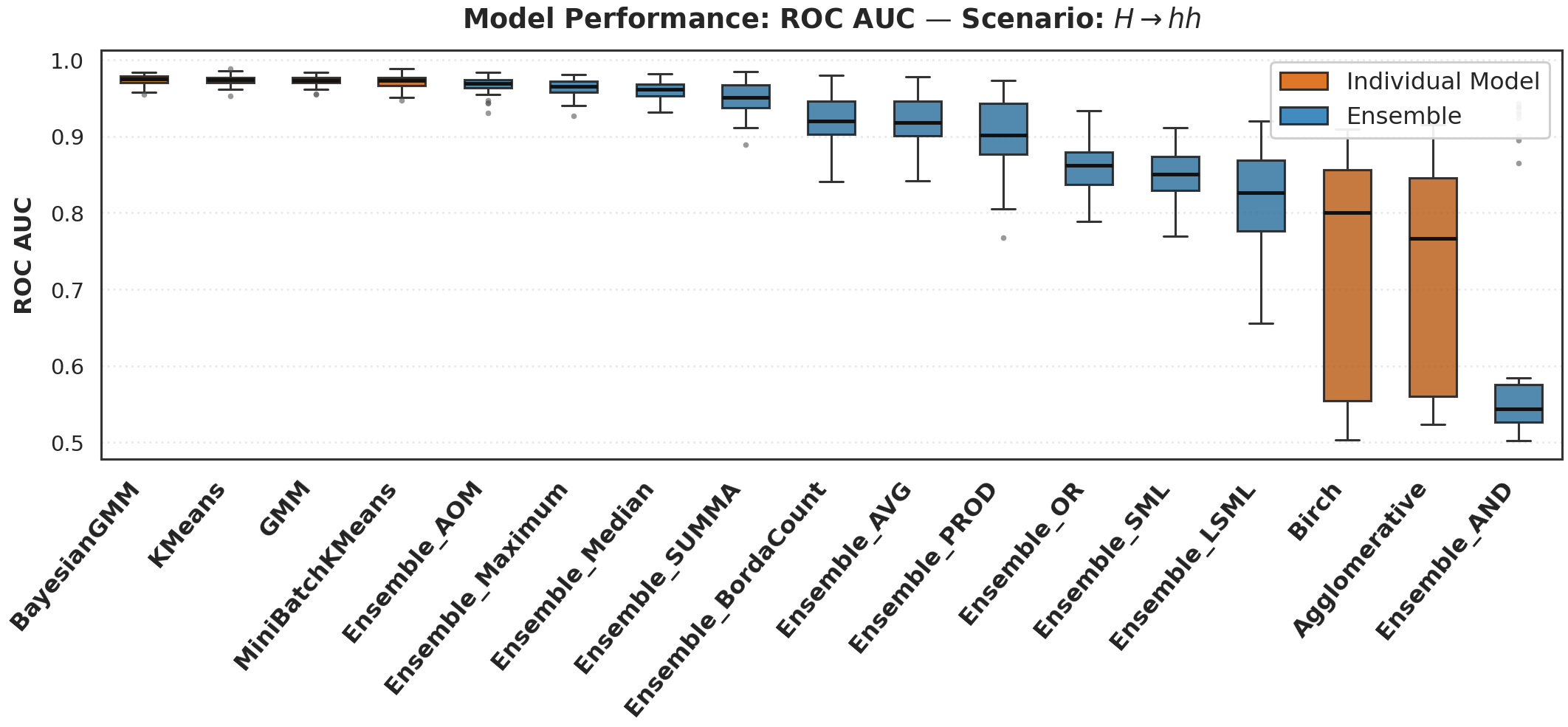}
\end{minipage}\hfill
\begin{minipage}{0.48\linewidth}
\centering
\includegraphics[width=\linewidth]{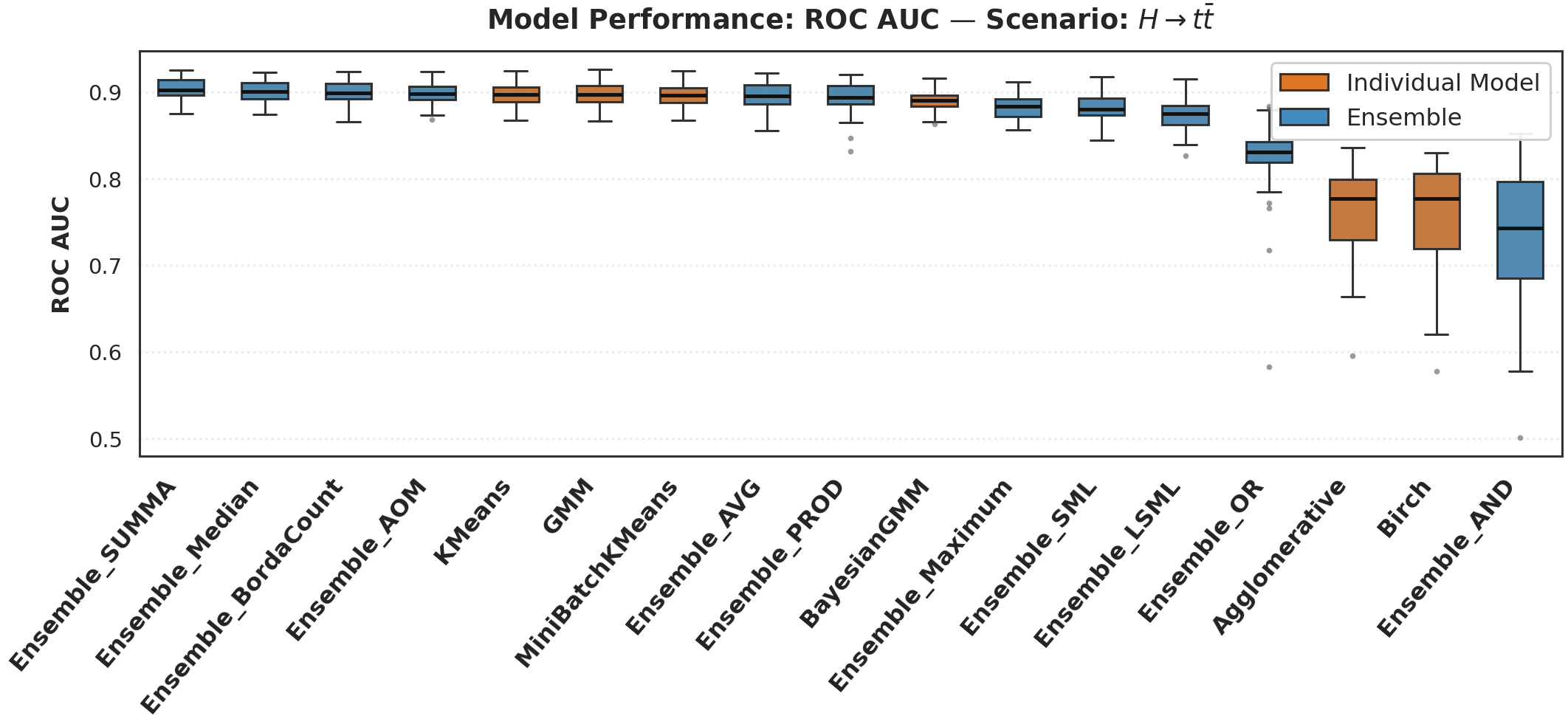}
\end{minipage}

\begin{minipage}{0.48\linewidth}
    \centering
    \includegraphics[width=\linewidth]{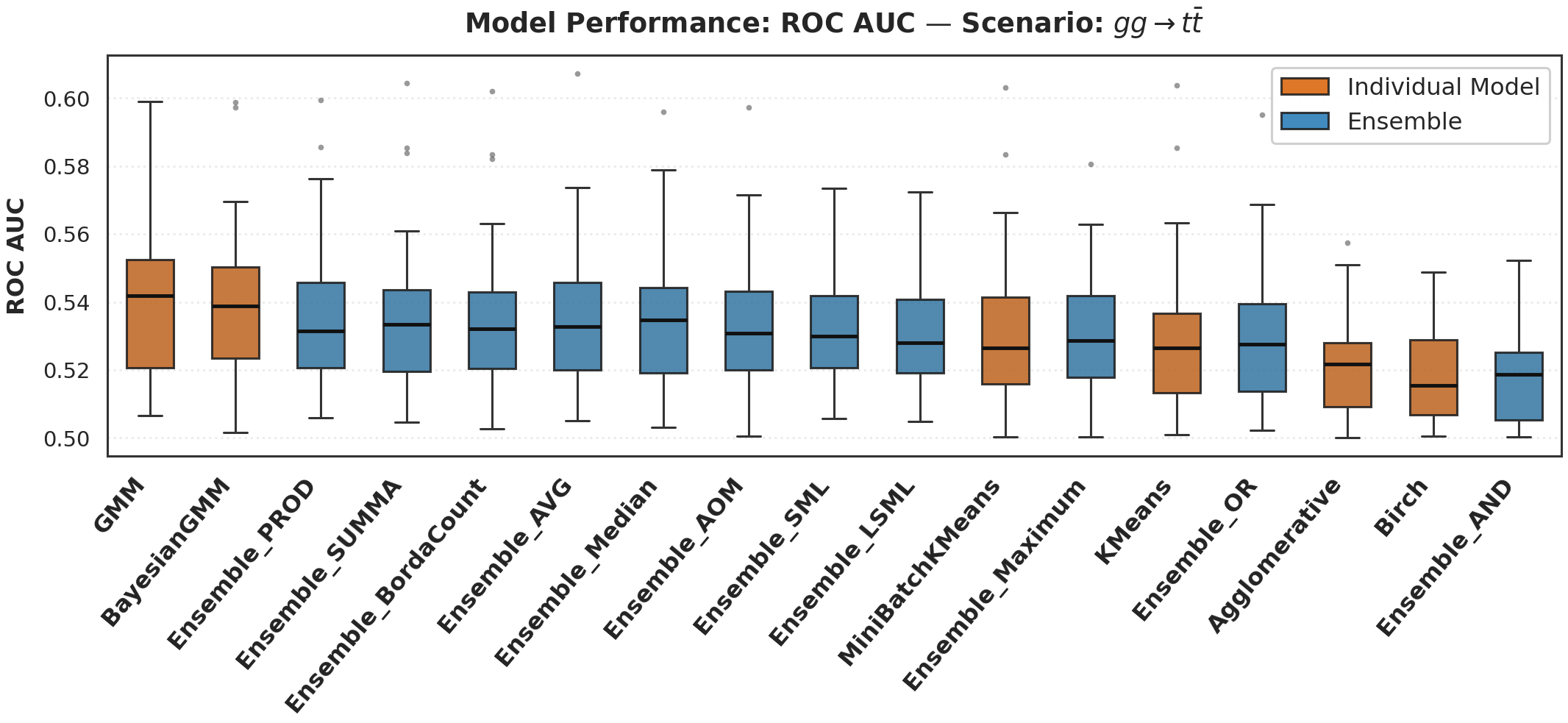}
\end{minipage}\hfill
\begin{minipage}{0.48\linewidth}
    \centering
    \includegraphics[width=\linewidth]{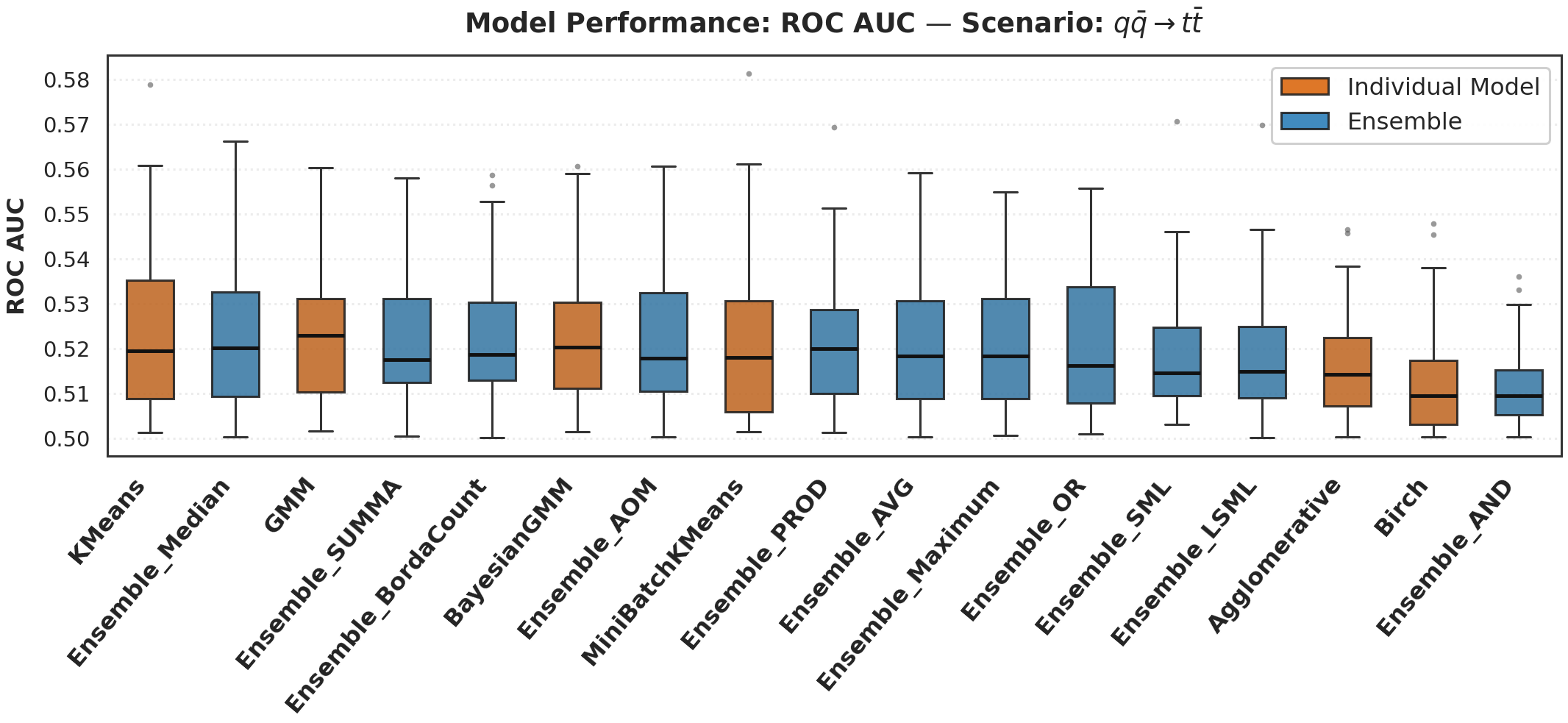}
\end{minipage}

\begin{minipage}{0.48\linewidth}
    \centering
    \includegraphics[width=\linewidth]{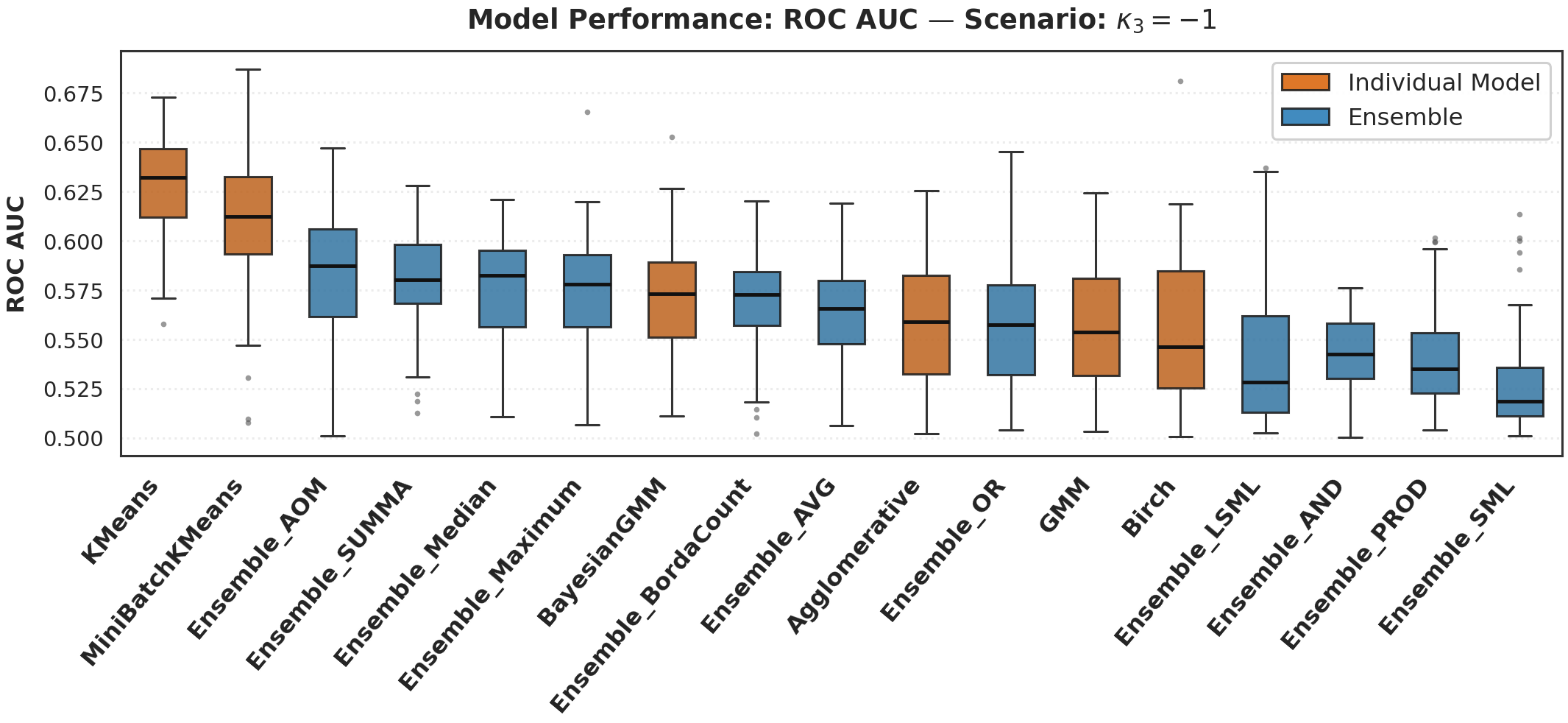}
\end{minipage}\hfill
\begin{minipage}{0.48\linewidth}
    \centering
    \includegraphics[width=\linewidth]{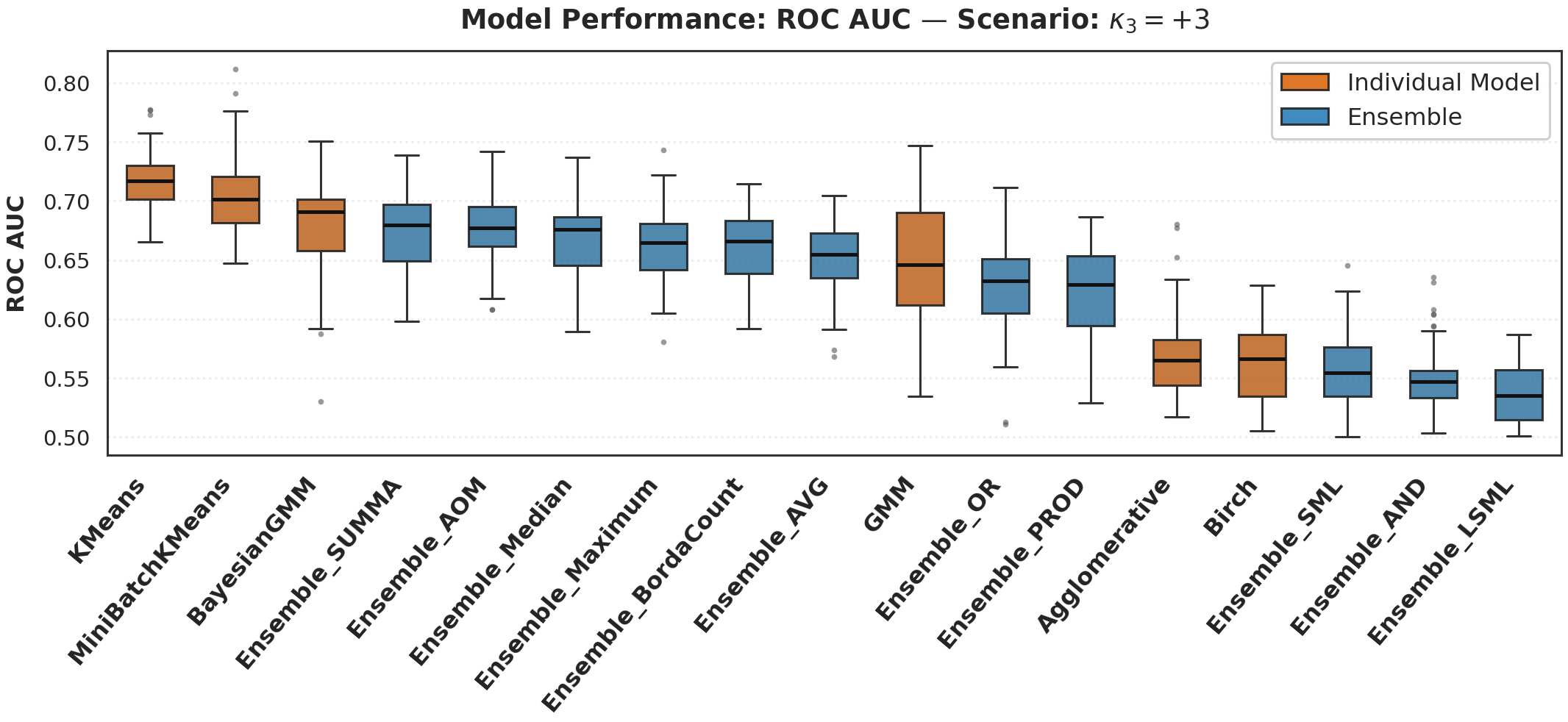}
\end{minipage}


\caption{AUC distributions obtained for the unsupervised and ensemble methods across the six new physics physics scenarios considered in this work. Each subplot corresponds to a specific physics scenario, with the distributions obtained from ten independent repeated sub-sampling runs. We do not show, here and hereafter, the SM $hh$ case but the final ranking of performances take it into account as well.}
\label{fig:boxplots_unsupervised}
\end{figure}
Figure \ref{fig:boxplots_unsupervised} shows the resulting AUC distributions for all methods across the physics scenarios considered here. The results indicate that no single method consistently dominates across all scenarios. Instead, the relative performance of the methods depends on the underlying physics process. Among the individual unsupervised models, K-Means and its variant exhibit competitive performance across a large fraction of the scenarios, frequently appearing among the highest-performing methods.

The effective-operator scenarios represent a substantially more challenging regime. In these cases, most methods remain close to the random-classification baseline, with AUC values around 0.5, indicating limited discriminative information captured by the corresponding unsupervised representations.

The ensemble methods, in particular, Borda Count and SUMMA repeatedly appear among the best-performing approaches across the different scenarios. Their recurrence across distinct physics configurations suggests that combining the information provided by multiple unsupervised models can yield a more robust characterization of the signal-background separation than relying on a single model. In particular, four ensemble models rank among the best AUC for the $H\to t\bar{t}$ events.

\begin{figure}
    \centering
    \includegraphics[width=0.95\linewidth]{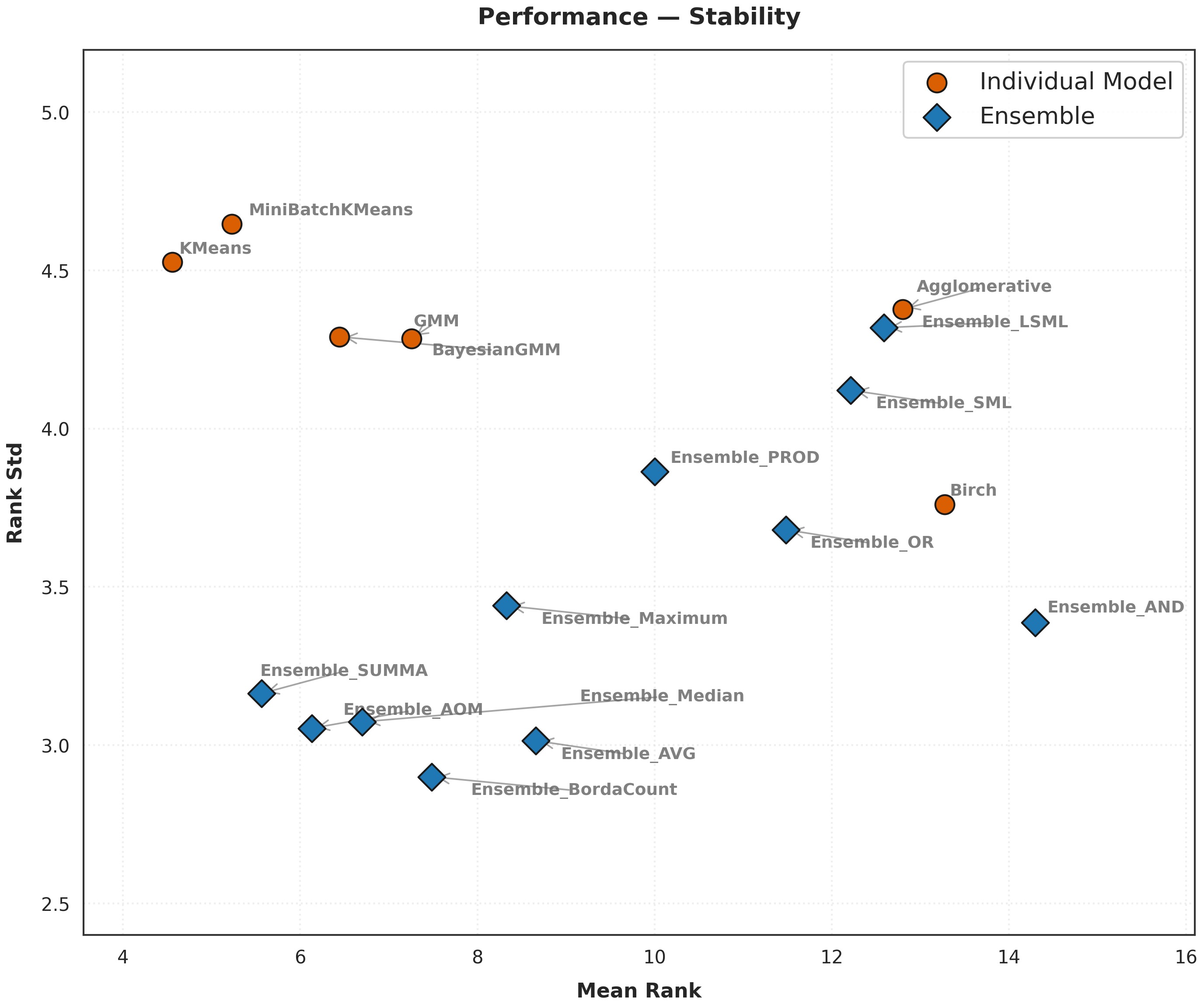}
    \caption{Performance {\it versus} stability space comparing ensemble methods and individual base models in the clustering experiment. The $x$-axis shows the mean rank, $\mu_{\text{rank}}$, of each method across all runs and physics scenarios considered, while the $y$-axis represents the standard deviation of ranks, $\sigma_{\text{rank}}$, as a measure of ranking variability across runs. Orange circles represent individual models, while blue diamonds represent ensembles. The most favorable methods are located in the bottom-left region of the plot, with the ideal point at $(\mu_{\text{rank}}, \sigma_{\text{rank}}) = (1, 0)$, corresponding to the best possible mean rank and zero variability across runs.}
    \label{fig:pareto_unsupervised}
\end{figure}

Although Figure \ref{fig:boxplots_unsupervised} provides a detailed view of the AUC distributions within each individual physics scenario, a consolidated analysis is required to identify methods that remain competitive across different scenarios and independent runs. Figure \ref{fig:pareto_unsupervised} summarizes this behavior by jointly considering the average ranking and its corresponding standard deviation over all experimental runs.

For each independent run, the methods are ranked according to their AUC within the corresponding physics scenario. The horizontal axis, therefore, represents the average ranking ($\mu_{\text{rank}}$) of each method over all scenarios and repetitions, while the vertical axis represents the standard deviation of these rankings ($\sigma_{\text{rank}}$). Lower values on both axes indicate more desirable behavior: a lower average rank corresponds to consistently stronger relative performance, while a lower ranking standard deviation indicates that the method preserves a similar competitive position despite variations in both the physics scenario and the randomly generated datasets. Using rankings instead of absolute AUC values reduces the influence of the intrinsic difficulty of each physics scenario, ensuring that all scenarios contribute equally to the overall assessment.

As shown in Figure \ref{fig:pareto_unsupervised}, SUMMA, AOM and Median ensemble models occupy the most favorable region of the performance {\it versus} stability space, combining high average rankings with remarkable ranking stability across the experimental runs. The remaining methods are generally located farther from the lower-left region of the plot, indicating either lower average performance, higher ranking variability, or both. Table~\ref{tab:top5_clustering} quantitatively summarizes these results by highlighting the top five performing models, ranked according to their normalized Euclidean distance, Eq.~\eqref{eq:distance}, to the ideal performance point ($\mu_{\text{rank}}=1$, $\sigma_{\text{rank}}=0$). As we can see, the ensemble methods dominate this rank, exhibiting the best trade-off in the performance {\it versus} stability space.


\begin{table}[b!]
\centering
\begin{tabular}{cccc}
\hline
\textbf{Ranking} & \textbf{Model} & \textbf{Type} & \textbf{Normalized Distance} \\
\hline
1 & AOM      & Ensemble & 0.7621 \\
2 & SUMMA & Ensemble & 0.7624 \\
3 & Median        & Ensemble & 0.7883 \\
4 & BordaCount     & Ensemble & 0.7926 \\
5 & AVG     & Ensemble & 0.8676 \\
\hline
\end{tabular}
\caption{Ranking of the top five methods in the clustering experiment based on the normalized distance in the performance {\it versus} stability space.}
\label{tab:top5_clustering}
\end{table}

\subsection{Semi-supervised models}

The second experimental setup investigates the behavior of semi-supervised models. In this configuration, the models have access to a sample of simulated background events during the learning procedure, while the signal information remains unavailable. This additional background information allows us to investigate whether incorporating partial knowledge of the background distribution leads to more consistent signal-background discrimination.

\begin{figure}
\centering
\begin{minipage}{0.48\linewidth}
\centering
\includegraphics[width=\linewidth]{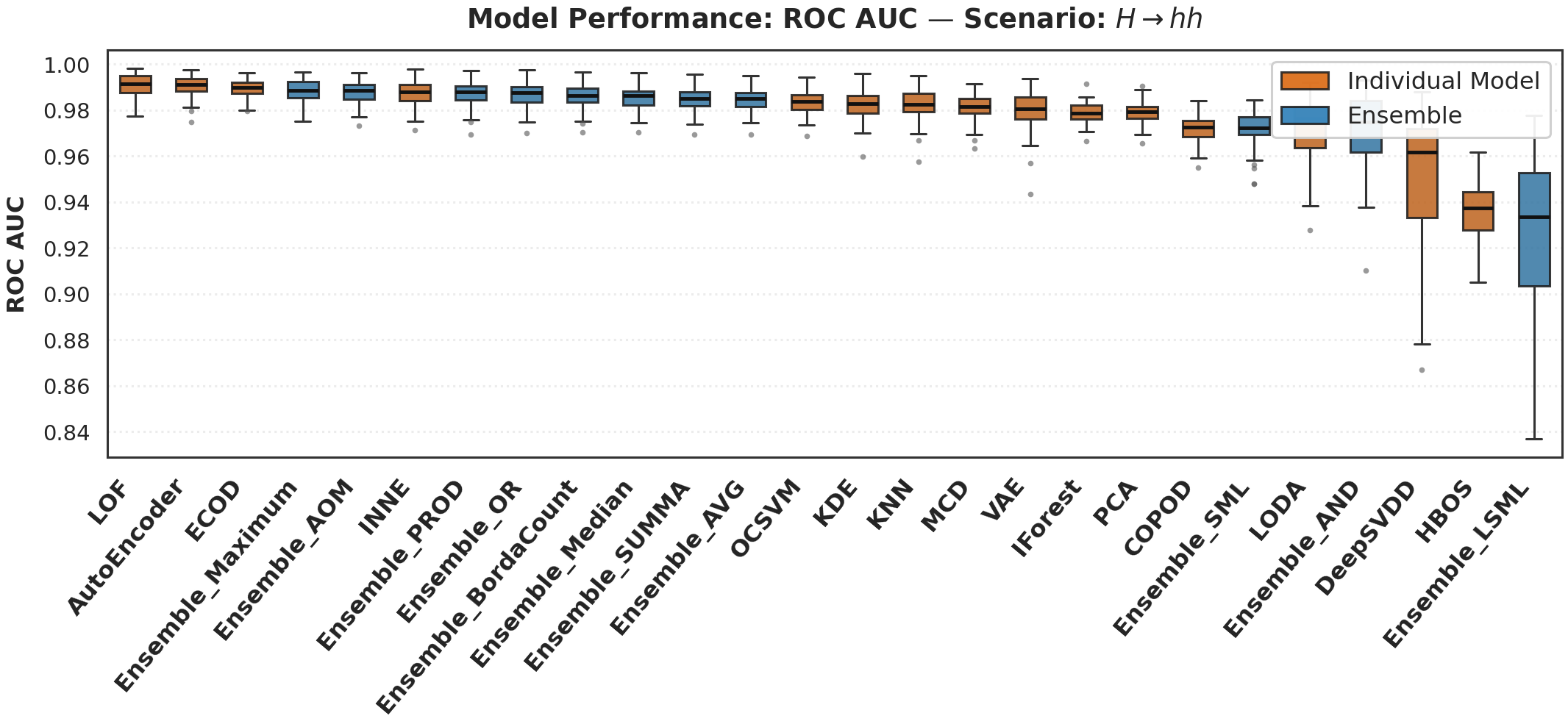}
\end{minipage}\hfill
\begin{minipage}{0.48\linewidth}
\centering
\includegraphics[width=\linewidth]{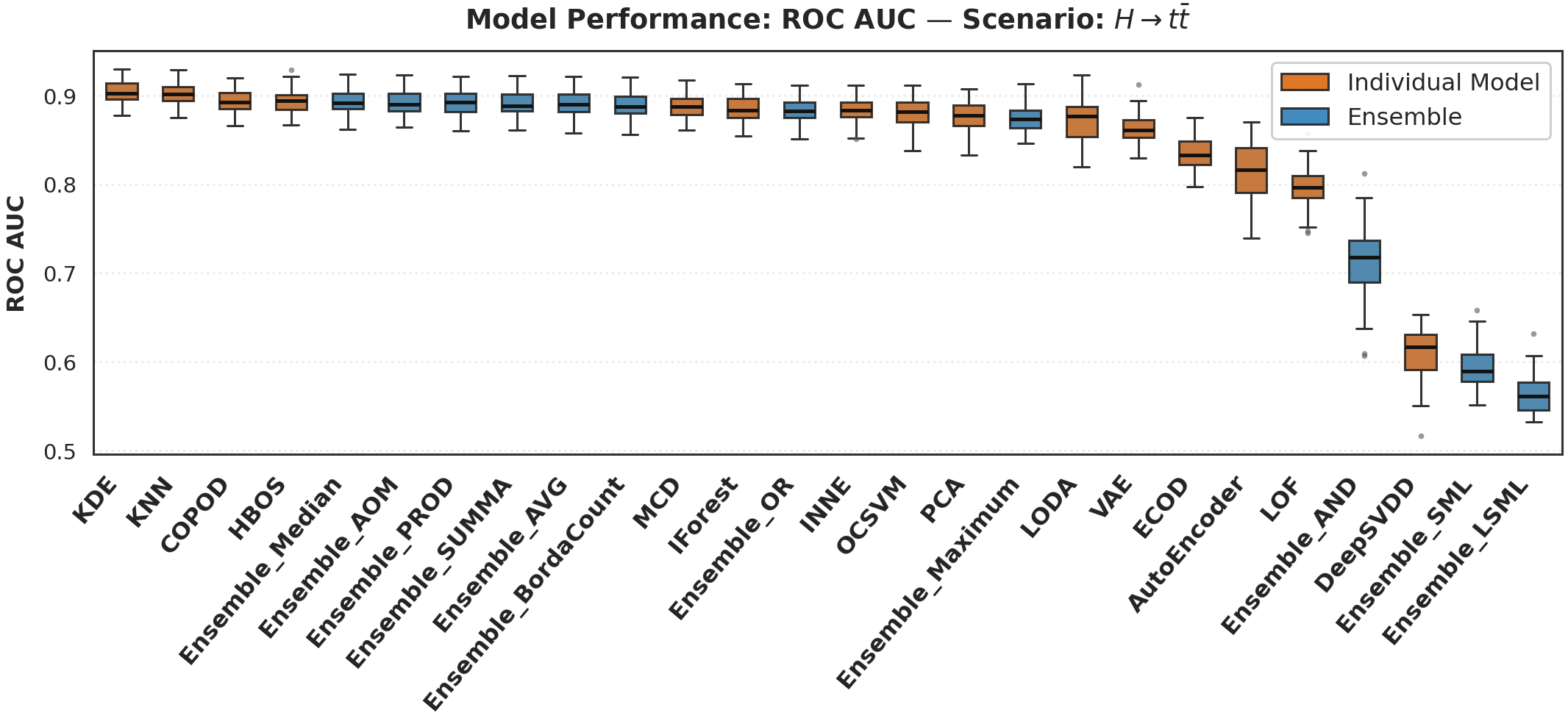}
\end{minipage}

\begin{minipage}{0.48\linewidth}
    \centering
    \includegraphics[width=\linewidth]{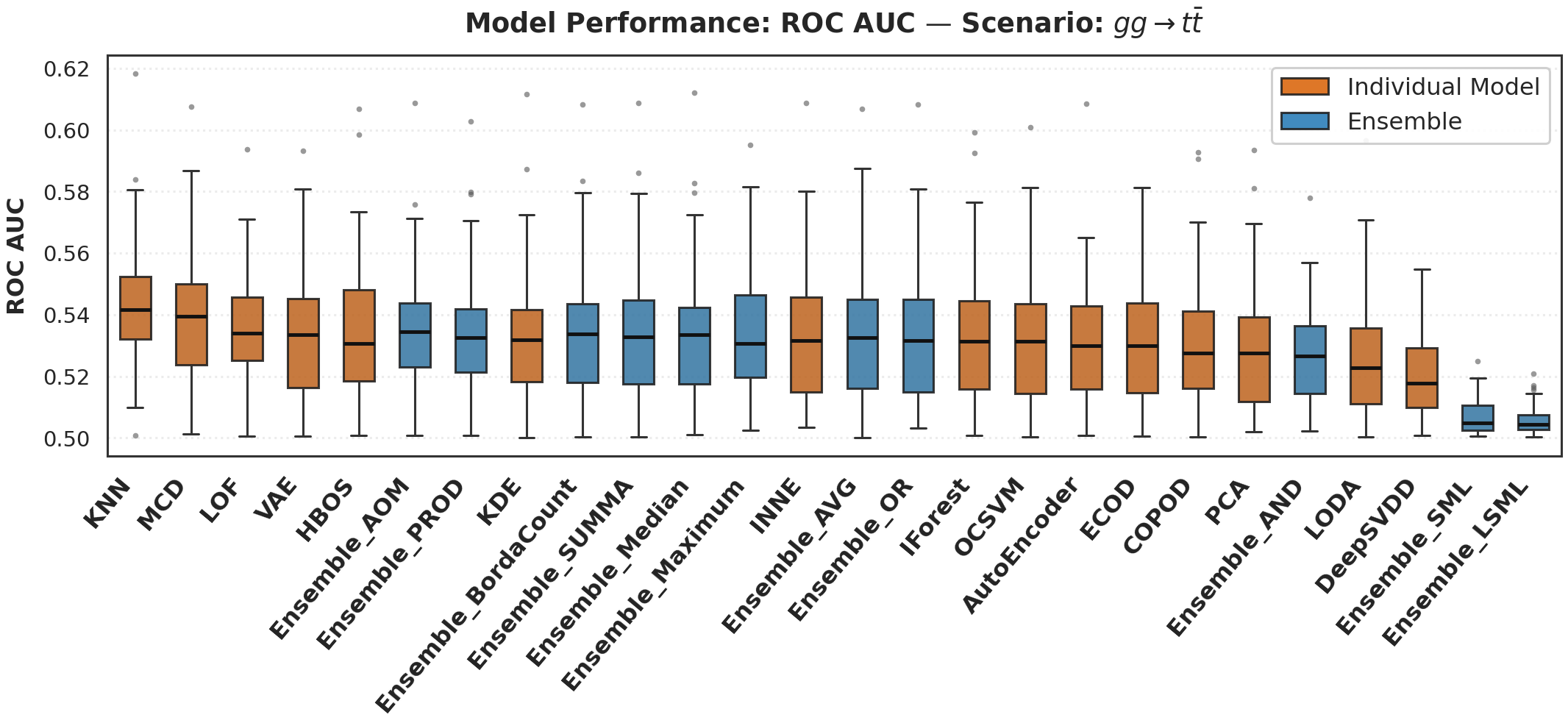}
\end{minipage}\hfill
\begin{minipage}{0.48\linewidth}
    \centering
    \includegraphics[width=\linewidth]{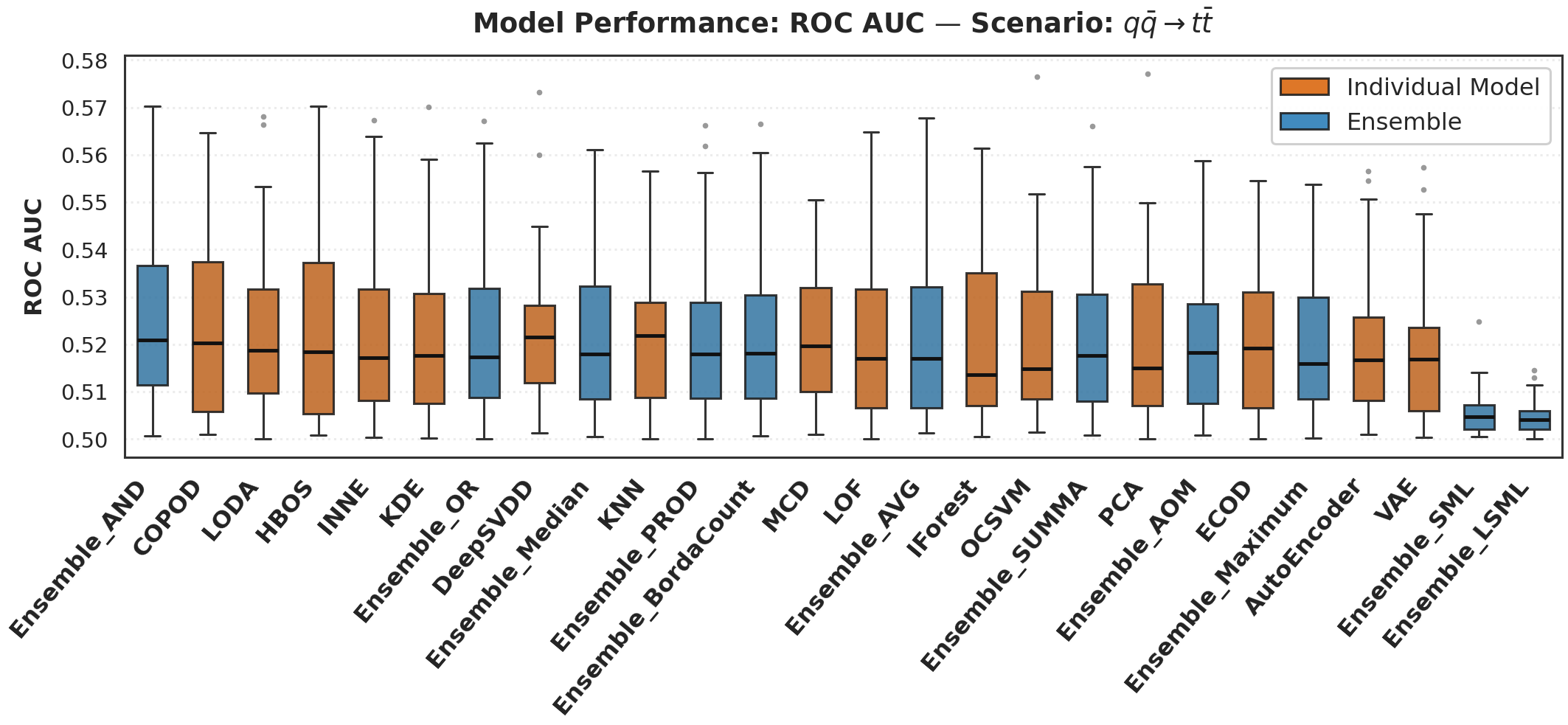}
\end{minipage}

\begin{minipage}{0.48\linewidth}
    \centering
    \includegraphics[width=\linewidth]{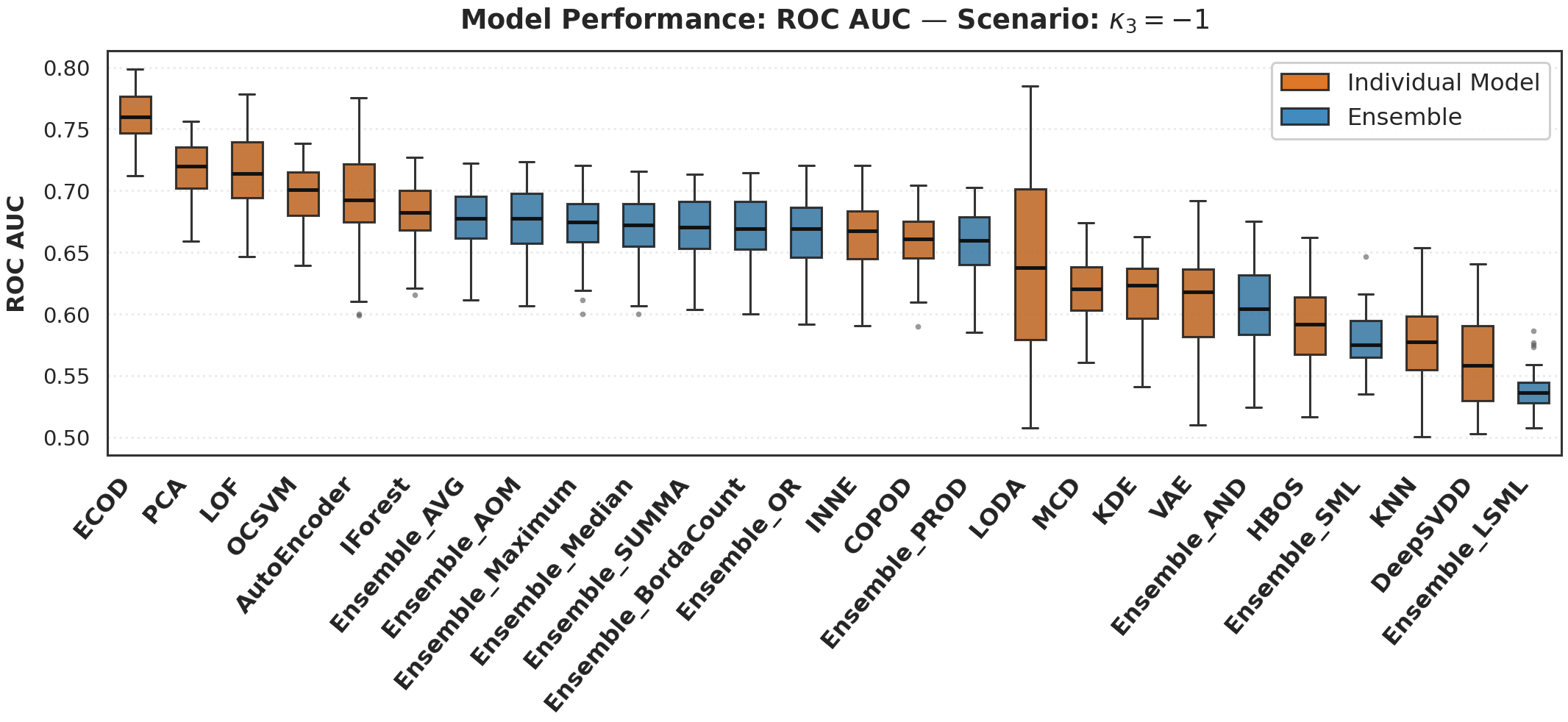}
\end{minipage}\hfill
\begin{minipage}{0.48\linewidth}
    \centering
    \includegraphics[width=\linewidth]{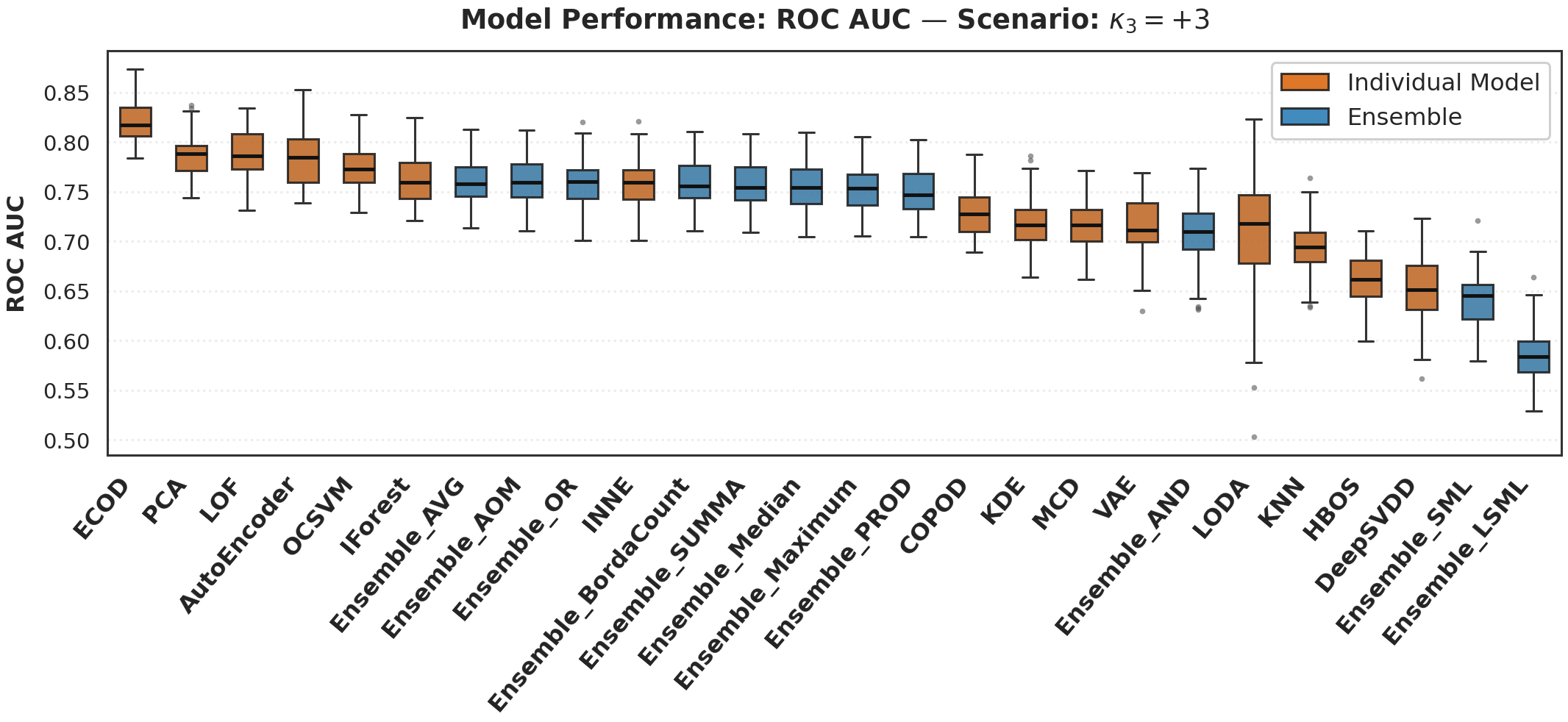}
\end{minipage}


\caption{AUC distributions obtained for the semisupervised and ensemble methods across the new physics scenarios. Each subplot corresponds to a specific physics scenario, with the distributions obtained from ten independent repeated sub-sampling runs.}
\label{fig:boxplots_semisupervised}
\end{figure}

As in the fully unsupervised setup, no single individual model consistently dominates across all physics scenarios, as shown in Figure \ref{fig:boxplots_semisupervised}. Among the individual models, the LOF exhibits a particularly interesting behavior. It is consistently among the best-ranked methods in most scenarios, but it also presents one notable deviation from this overall behavior. Its ranking changes substantially in the $H\to t\bar{t}$ scenario, where LOF becomes the individual model with the second lowest rank. This behavior is particularly noteworthy because $H\to t\bar{t}$ is comparatively less challenging for the other methods, which typically achieve AUC values around $0.9$. In contrast, LOF reaches an average AUC of approximately $0.8$, indicating a specific loss of discriminative performance in a scenario where most models maintain strong separation.

It is important to point out that an ensemble model wins in the new physics scenario of the dimension-6 EFT operator in the $q\bar{q}\to t\bar{t}$ process.

\begin{figure}[t!]
    \centering
    \includegraphics[width=0.95\linewidth]{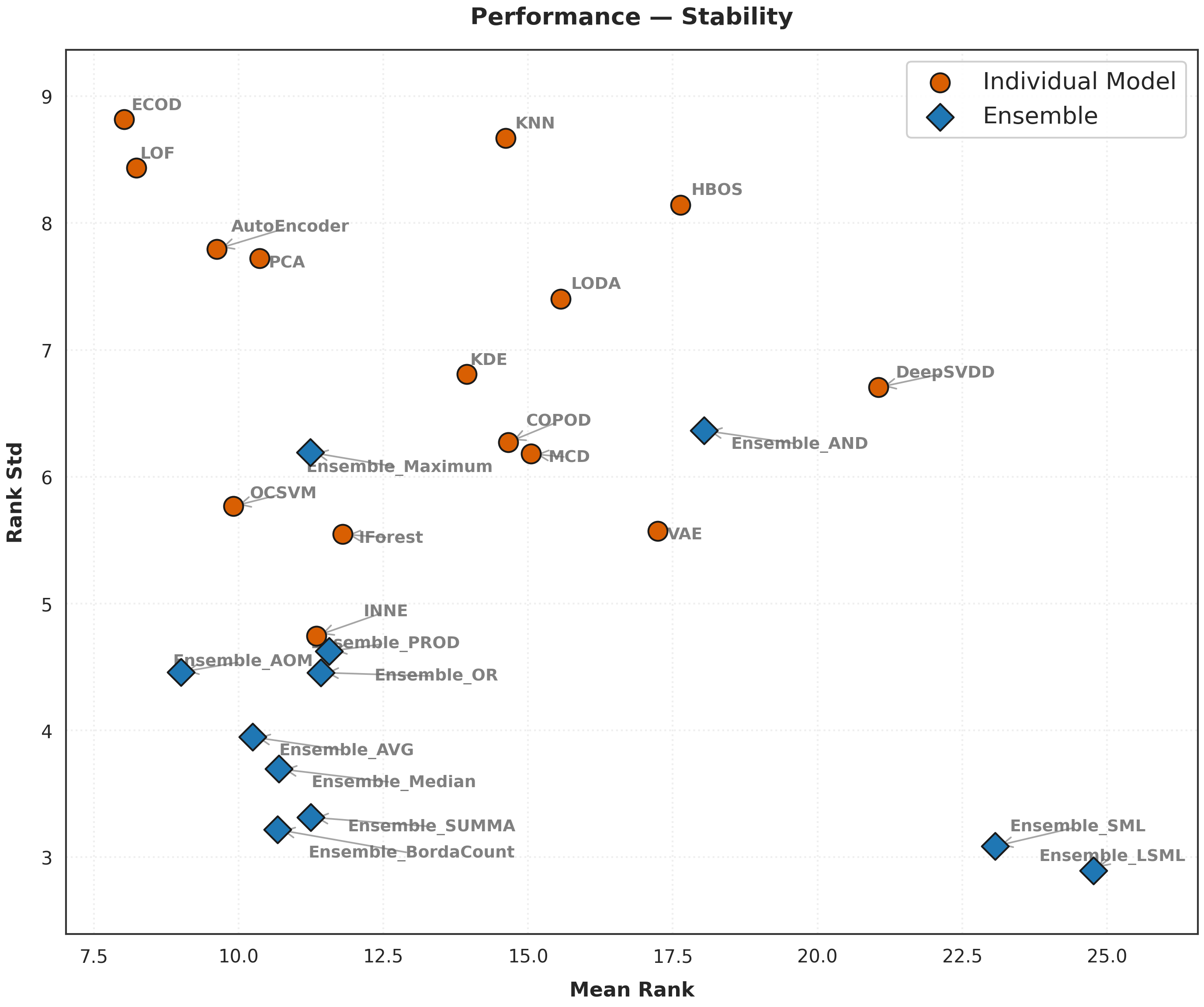}
    \caption{Performance {\it versus} stability space comparing ensemble methods and individual base models in the semi-supervised experiment. The $x$-axis shows the mean rank, $\mu_{\text{rank}}$, of each method across all runs and physics scenarios considered, while the $y$-axis represents the standard deviation of ranks, $\sigma_{\text{rank}}$, as a measure of ranking variability across runs. Orange circles represent individual models, while blue diamonds represent ensembles. The most favorable methods are located in the bottom-left region of the plot, with the ideal point at $(\mu_{\text{rank}}, \sigma_{\text{rank}}) = (1, 0)$, corresponding to the best possible mean rank and zero variability across runs.}
    \label{fig:pareto_semisupervised}
\end{figure}

The aggregate performance {\it versus} stability comparison in Figure \ref{fig:pareto_semisupervised} provides a clearer picture of the overall behavior of the methods across the complete set of experimental runs. As observed in the fully unsupervised experiment, SUMMA and Borda Count occupy the most favorable region of the performance {\it versus} stability space, combining competitive average rankings with relatively low variability in their rankings. Among the individual models, ECOD and LOF present a superior performance mean rank compared to these ensembles. However, their significantly larger rank standard deviation indicates that their relative position varies more across experimental runs than those ensembles. This distinction highlights the importance of considering both dimensions simultaneously: a method may achieve a favorable average ranking while exhibiting considerable variability in its relative position across different physics scenarios and dataset realizations.


%
The behavior of i-SML and L-SML provides a complementary example. Both ensembles exhibit the lowest average ranking standard deviations among the considered methods, indicating that their relative positions remain highly consistent across the experimental runs. However, this stability is accompanied by poor relative performance: both methods occupy the worst positions in terms of average ranking, lying at the extreme right of the performance {\it versus} stability space. No other model or ensemble reaches similar unfavorable average-rank values. Thus, although i-SML and L-SML provide highly stable rankings, this consistency reflects a persistently poor relative performance.

Table~\ref{tab:top5_semisupervised} summarizes our results, showing that ensemble methods consistently outperform the individual models in the performance {\it versus} stability space.

\begin{table}[b!]
\centering
\begin{tabular}{cccc}
\hline
\textbf{Ranking} & \textbf{Model} & \textbf{Type} & \textbf{Normalized Distance} \\
\hline
1 & BordaCount & Ensemble & 0.5466 \\
2 & SUMMA     & Ensemble & 0.5719 \\
3 & Median       & Ensemble & 0.5850 \\
4 & AVG        & Ensemble & 0.5930 \\
5 & AOM      & Ensemble & 0.6077 \\
\hline
\end{tabular}
\caption{Ranking of the top five methods in the semi-supervised experiment based on the normalized distance in the performance {\it versus} stability space.}
\label{tab:top5_semisupervised}
\end{table}

\subsection{Ensembling all models}

The third experiment combines the two preceding approaches by considering both semi-supervised and fully unsupervised models as components for the ensemble methods.

\begin{figure}
\centering
\begin{minipage}{0.48\linewidth}
\centering
\includegraphics[width=\linewidth]{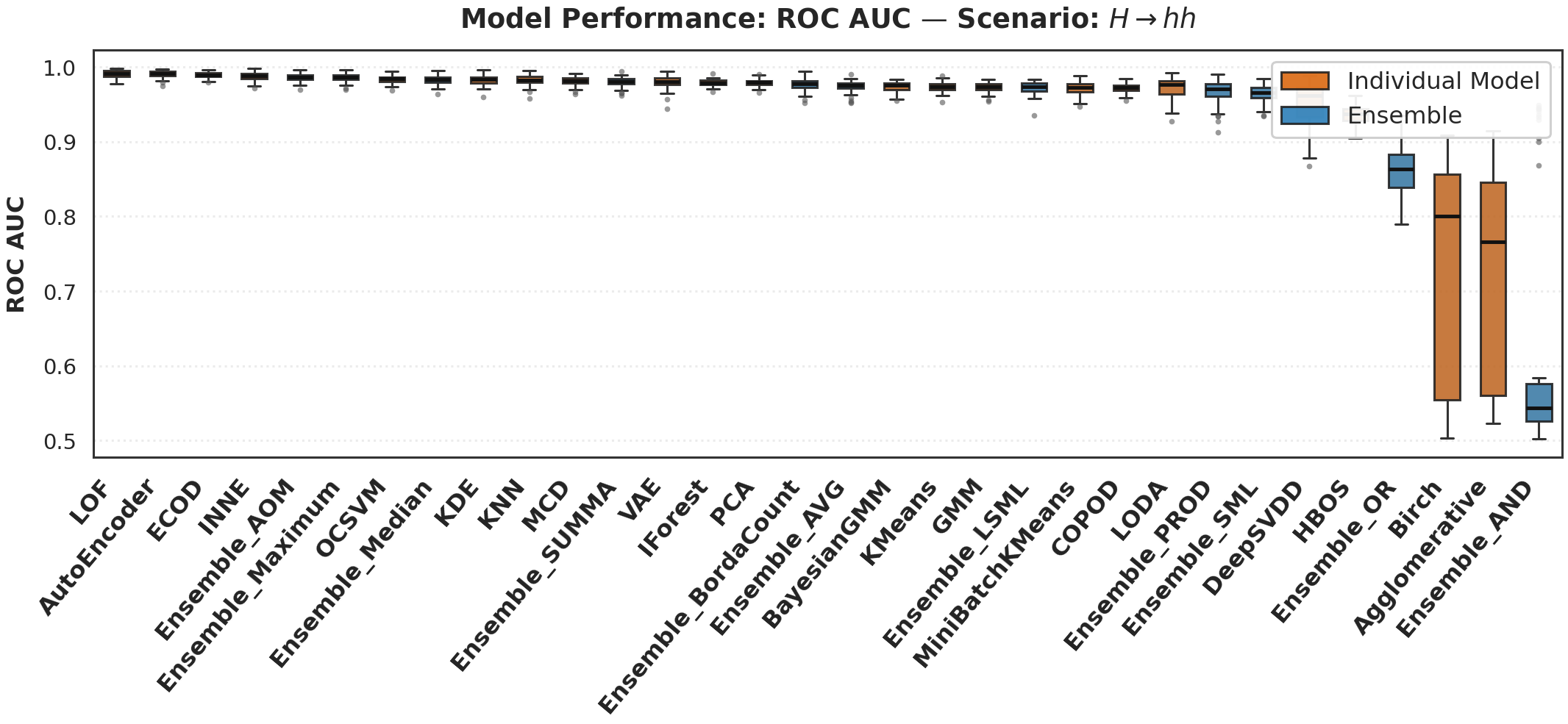}
\end{minipage}\hfill
\begin{minipage}{0.48\linewidth}
\centering
\includegraphics[width=\linewidth]{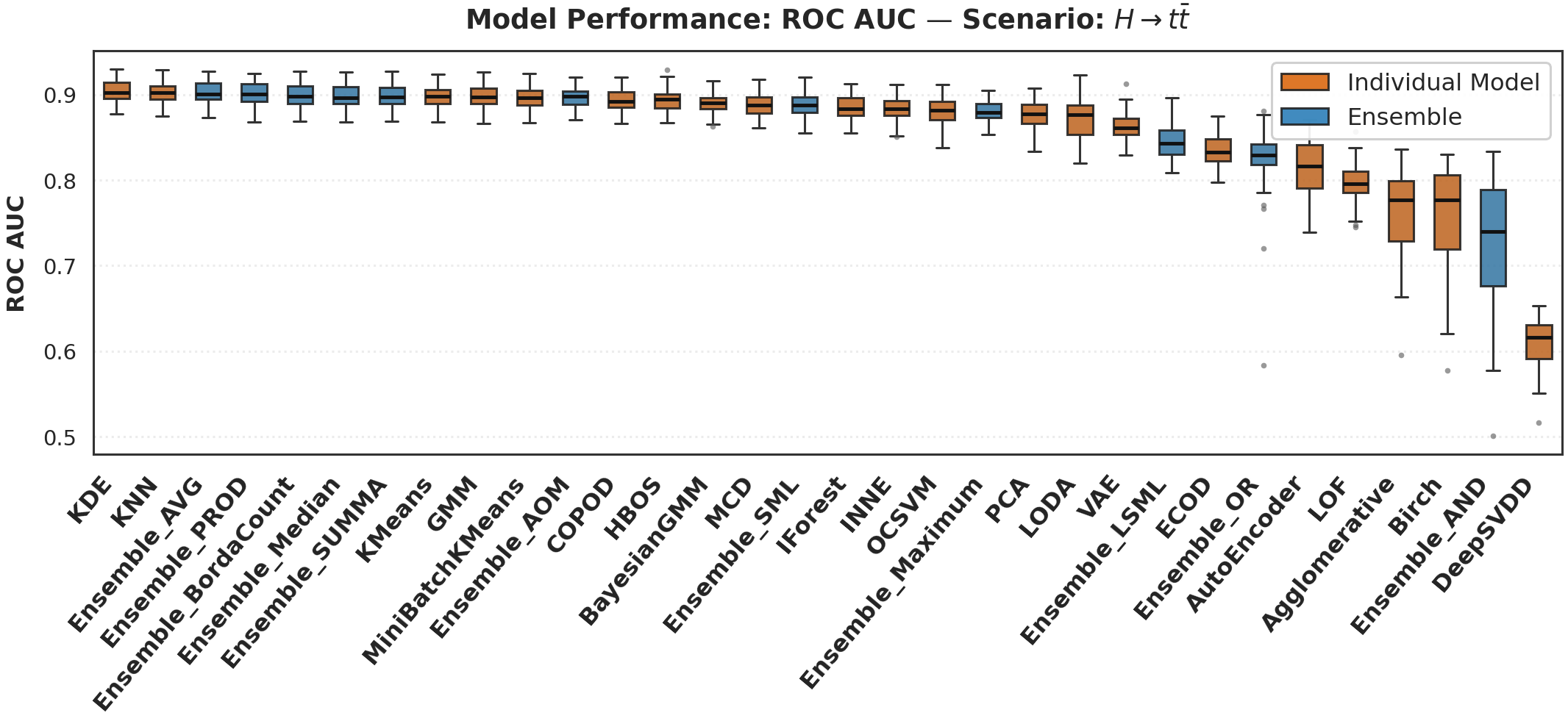}
\end{minipage}

\begin{minipage}{0.48\linewidth}
    \centering
    \includegraphics[width=\linewidth]{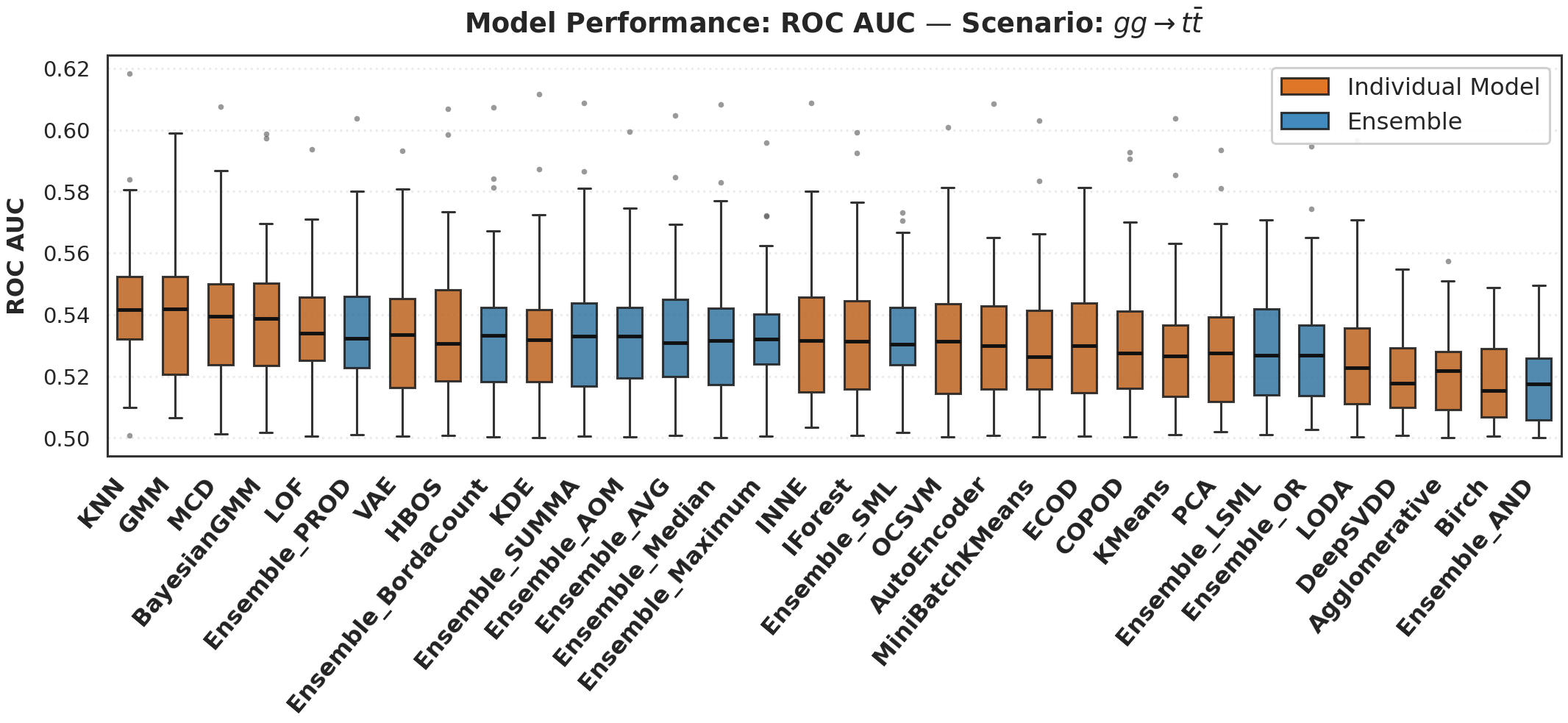}
\end{minipage}\hfill
\begin{minipage}{0.48\linewidth}
    \centering
    \includegraphics[width=\linewidth]{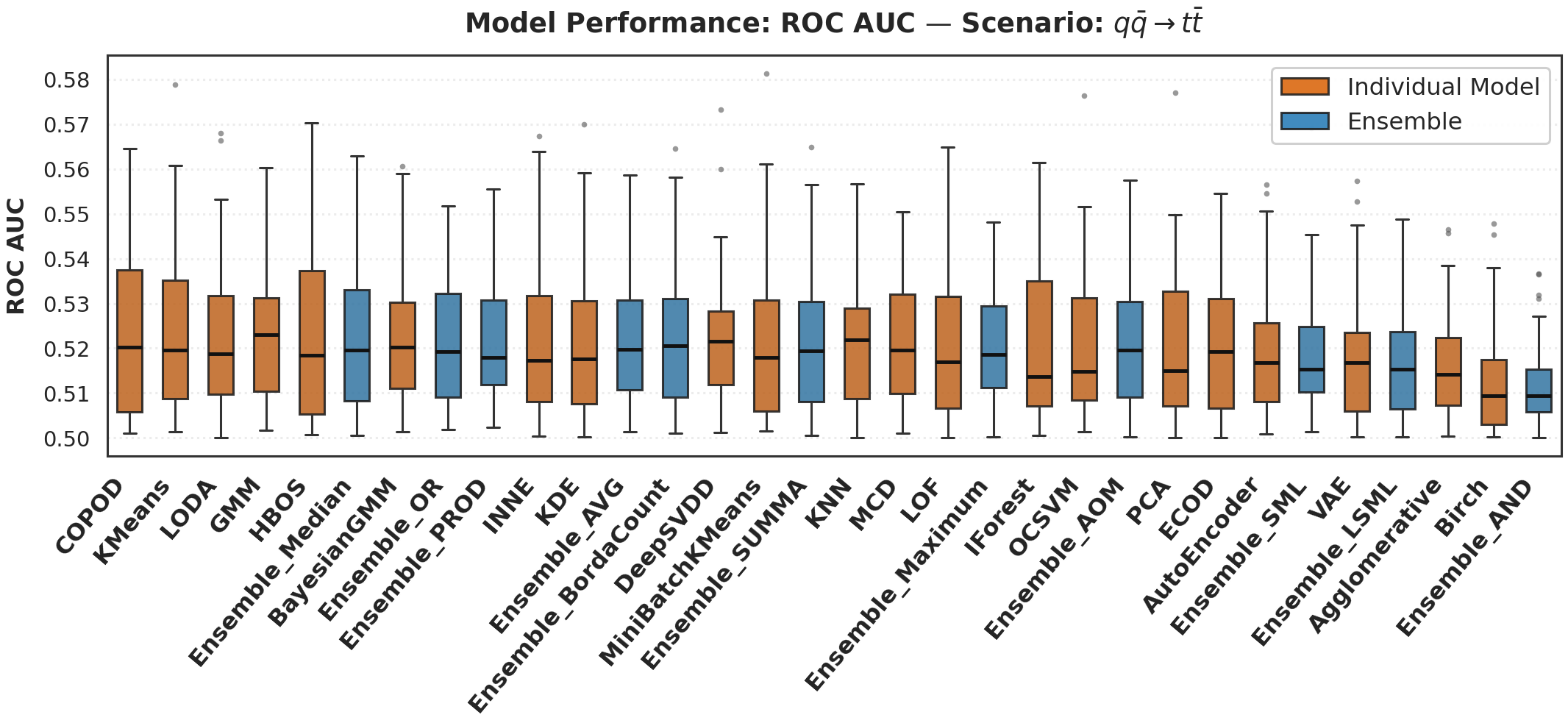}
\end{minipage}

\begin{minipage}{0.48\linewidth}
    \centering
    \includegraphics[width=\linewidth]{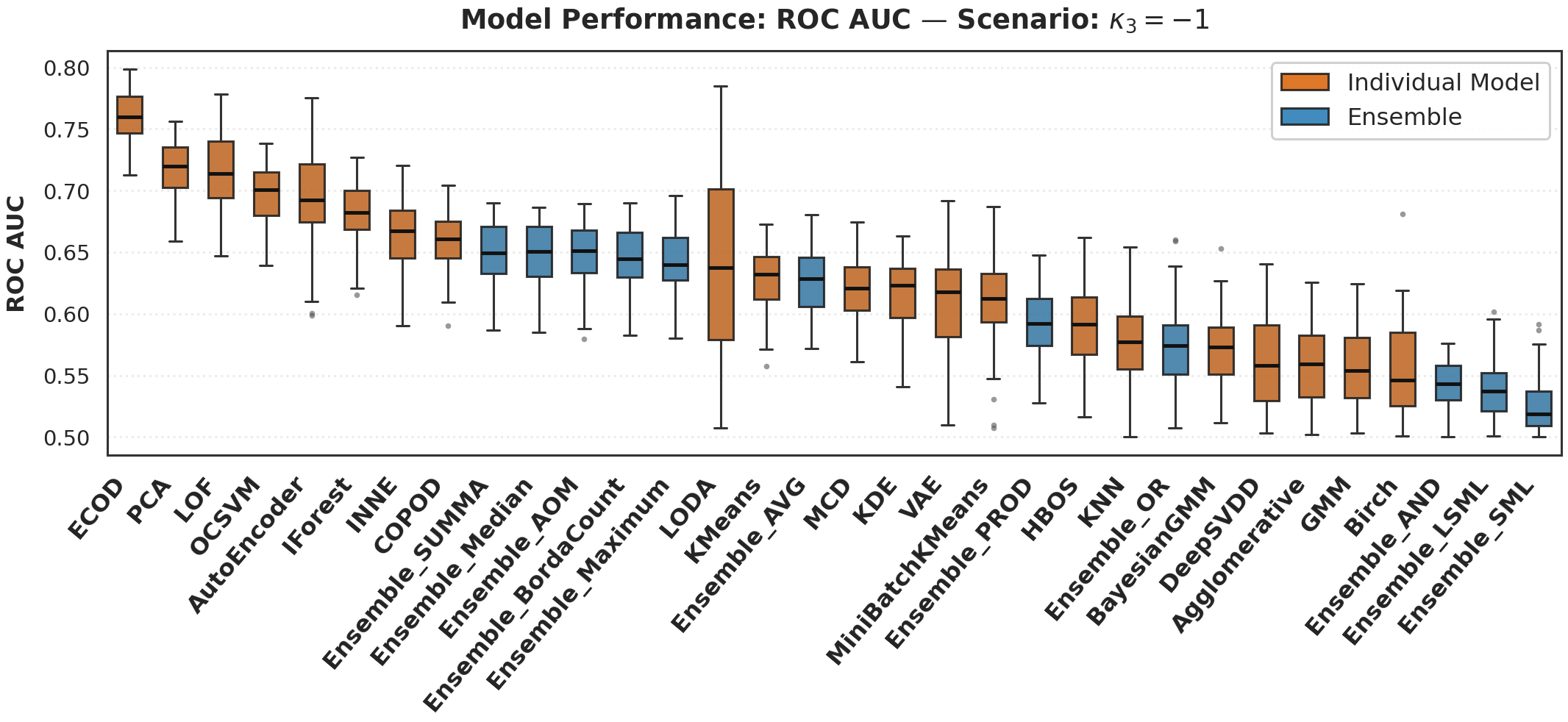}
\end{minipage}\hfill
\begin{minipage}{0.48\linewidth}
    \centering
    \includegraphics[width=\linewidth]{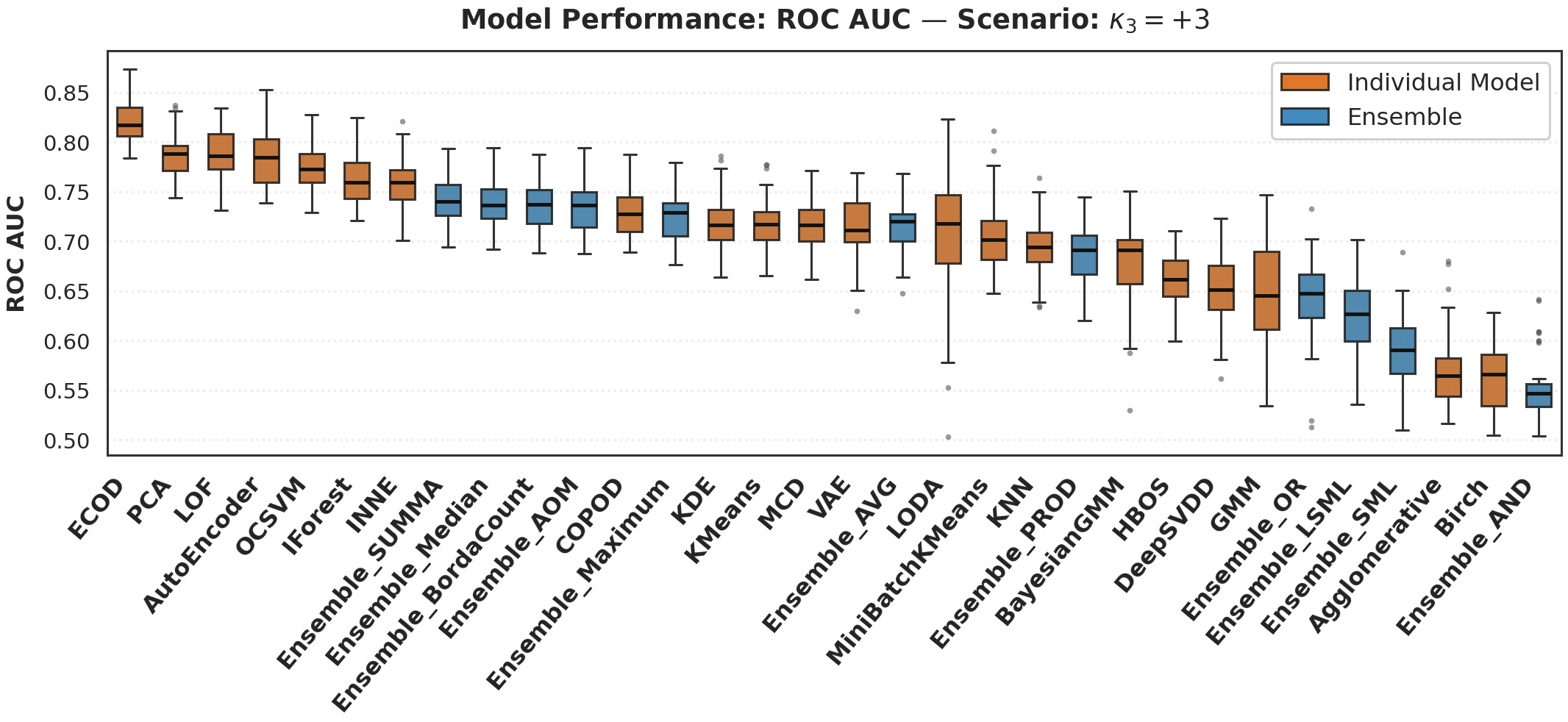}
\end{minipage}


\caption{AUC distributions obtained for the jointing semi-supervised with fully unsupervised models and ensemble methods across the new physics scenarios. Each subplot corresponds to a specific physics scenario, with the distributions obtained from ten independent repeated sub-sampling runs.}
\label{fig:boxplots_all_models}

\end{figure}

Figure \ref{fig:boxplots_all_models} presents the AUC distributions obtained for all individual models, together with the new ensembles constructed using the complete model pool. The results show a clear advantage of the semi-supervised models over the fully unsupervised ones in terms of relative performance, with the former consistently achieving higher rankings across the considered physics scenarios. In contrast, Birch and Agglomerative clustering methods exhibit particularly poor performance in the $H\to t\bar{t}$ and $H\to hh$ scenarios. In these cases, their poor rankings are accompanied by comparatively large dispersions of AUC values across the independent runs, indicating both low relative performance and substantial variability in their discriminative performance.

\begin{figure}
    \centering
    \includegraphics[width=0.95\linewidth]{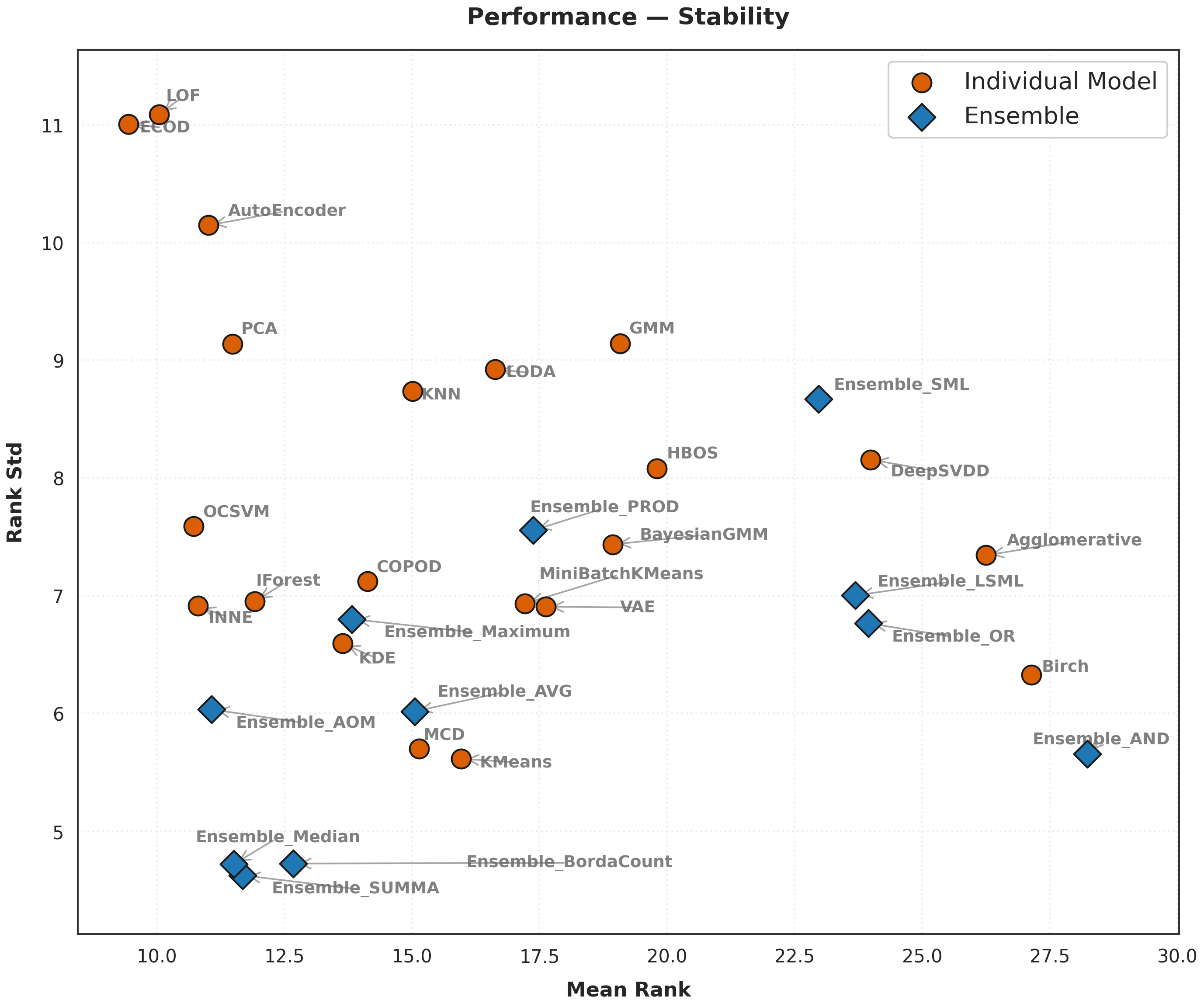}
    \caption{Performance {\it versus} stability space comparing ensemble methods and individual base models in the all models experiment. The $x$-axis shows the mean rank, $\mu_{\text{rank}}$, of each method across all runs and physics scenarios considered, while the $y$-axis represents the standard deviation of ranks, $\sigma_{\text{rank}}$, as a measure of ranking variability across runs. Orange circles represent individual models, while blue diamonds represent ensembles. The most favorable methods are located in the bottom-left region of the plot, with the ideal point at $(\mu_{\text{rank}}, \sigma_{\text{rank}}) = (1, 0)$, corresponding to the best possible mean rank and zero variability across runs.}
    \label{fig:pareto_all_models}
\end{figure}

As in previous experiments, Figure~\ref{fig:pareto_all_models} provides an aggregate assessment of relative performance and ranking stability over all scenario--dataset combinations. Once again, SUMMA, Borda Count and Median exhibit a favorable balance between these two dimensions, consistently occupying the most competitive region of the performance {\it versus} stability space. However, some individual models appear competitive, such as INNE, which figure in the top five models in Table~\ref{tab:top5_fusao}. It is worth noting the behavior of the two dimensions considered among these two groups of models within the top five. While the individual representative model (INNE) outperforms all ensemble methods in terms of average rank, the ensemble approaches are consistently superior in terms of stability.

\begin{table}[b!]
\centering
\begin{tabular}{cccc}
\hline
\textbf{Ranking} & \textbf{Model} & \textbf{Type} & \textbf{Normalized Distance} \\
\hline
1 & SUMMA      & Ensemble & 0.5725 \\
2 & Median & Ensemble & 0.5744 \\
3 & BordaCount                & Ensemble & 0.6045 \\
4 & AOM             & Ensemble & 0.6579 \\
5 & INNE               & Individual & 0.7199 \\
\hline
\end{tabular}
\caption{Ranking of the top five methods in the all models experiment based on the normalized distance in the performance {\it versus} stability space.}
\label{tab:top5_fusao}
\end{table}

\subsection{Ensembles comparison}

Following the three experimental setups, in which the ensembles were constructed from different model pools, we can now assess how their relative performance and ranking stability change with the composition of the underlying models.

\begin{figure}
    \centering
    \includegraphics[width=0.95\linewidth]{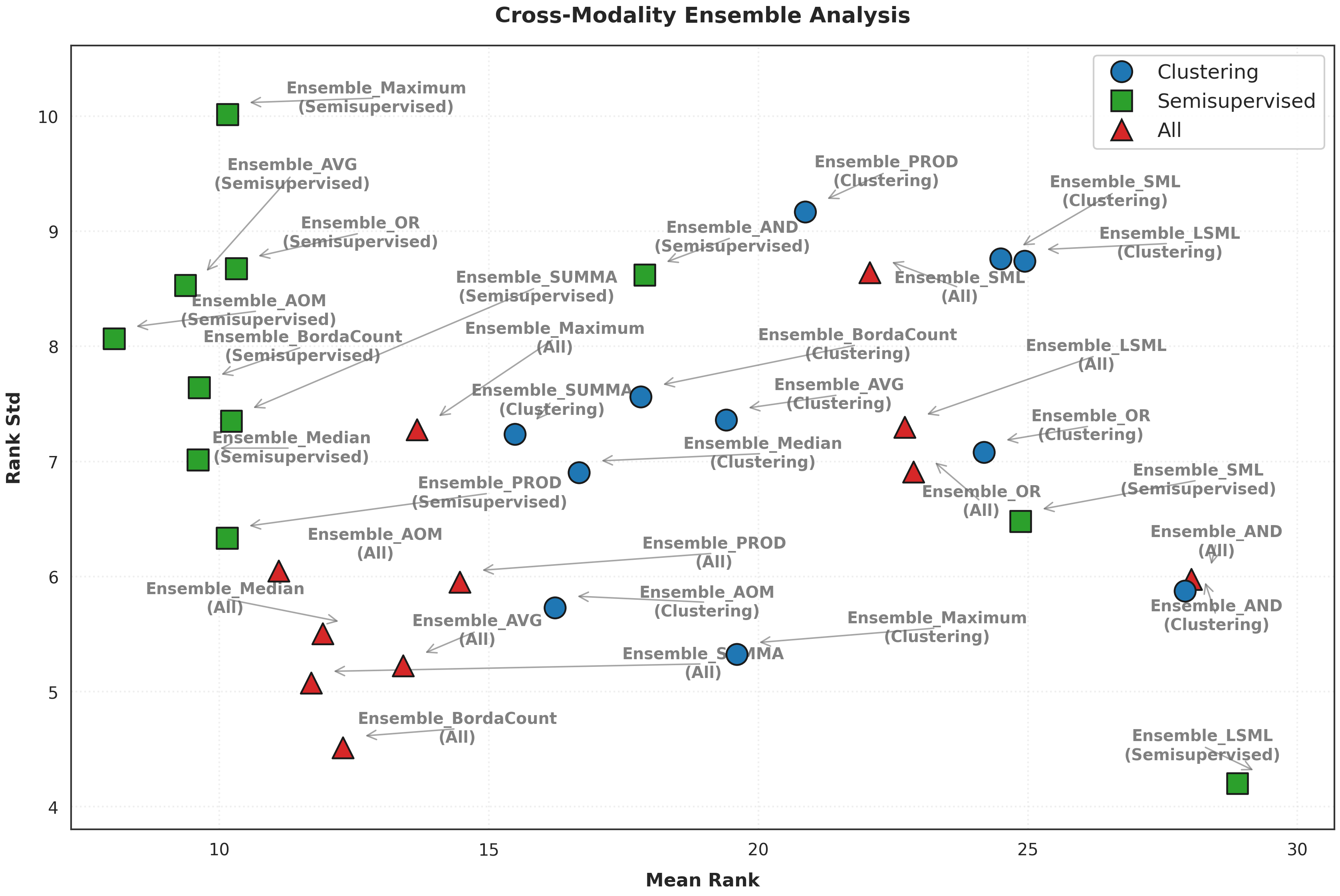}
    \caption{Performance {\it versus} stability space comparing all ensemble methods across the three experiment settings: clustering, semi-supervised and all. The $x$-axis shows the mean rank, $\mu_{\text{rank}}$, of each method across all runs and physics scenarios considered, while the $y$-axis represents the standard deviation of ranks, $\sigma_{\text{rank}}$, as a measure of ranking variability across runs. Blue circles represent the clustering setting, green squares the semi-supervised configuration, and the red triangles the fusion of all models. The most favorable methods are located in the bottom-left region of the plot, with the ideal point at $(\mu_{\text{rank}}, \sigma_{\text{rank}}) = (1, 0)$, corresponding to the best possible mean rank and zero variability across runs.}
    \label{fig:pareto_rank_ensembles}
\end{figure}

Figure \ref{fig:pareto_rank_ensembles} summarizes the ensemble methods exclusively in the performance {\it versus} stability space, with each point identified by the model pool used for its construction. An important methodological detail is that the mean rank and its standard deviation shown here are recalculated considering exclusively the ensemble methods across all three experimental setups. Therefore, the ranking values in this figure are not directly comparable to those reported in the previous figures, where different sets of methods were considered, including both individual models and ensembles. As a result, the same ensemble method may occupy a different position in the performance {\it versus} stability space, even when its underlying experimental results remain unchanged. This occurs because its relative ranking is now determined with respect to a different comparison set. The figure reveals a clear improvement in the overall behavior of the ensembles as the model pool is broadened, from fully unsupervised clustering models to semi-supervised models, and finally to the combination of both groups.

The blue circles, corresponding to ensembles constructed exclusively from fully unsupervised clustering models, are predominantly located in the central and right regions of the performance {\it versus} stability space. This indicates comparatively lower performance and, in several cases, greater variability in the rankings across independent runs. When the model pool is expanded to include the semi-supervised models, represented by the green squares, a clear improvement is observed, particularly in terms of the mean rank of the ensembles, resulting in an overall shift from right to left. 

Combining both clustering and semi-supervised models (the ``All'' configuration, represented as red triangles) systematically yields superior positioning compared to using either model family in isolation. Overall, the results indicate that broadening the pool diversity significantly enhances the ranking stability of the approaches, driving the ensembles closer to the ideal performance region. As summarized in Table~\ref{tab:top5_ensembles}, ensembles operating under the ``All'' configuration dominate the top rankings, with SUMMA (All) and Borda Count (All) securing the highest ranks.


\begin{table}[b!]
\centering
\begin{tabular}{cccc}
\hline
\textbf{Ranking} & \textbf{Model} & \textbf{Type} & \textbf{Normalized Distance} \\
\hline
1 & BordaCount      & All & 0.6058 \\
2 & SUMMA & All & 0.6357 \\
3 & Median & All & 0.6747 \\
4 & AVG      & All & 0.6857 \\
5 & AOM     & All & 0.7042 \\
\hline
\end{tabular}
\caption{Ranking of the top five methods in the ensemble comparison based on the normalized distance in the performance {\it versus} stability space.}
\label{tab:top5_ensembles}
\end{table}

\subsection{Model Performances based on AUC and signal efficiencies}

AUC is a global performance metric whose value can be strongly influenced by the classifier's behavior over a broad range of signal and background efficiencies, including regions that are less relevant for particle physics searches. To complement this metric, we also evaluate each model in terms of its signal efficiency ($\epsilon_S$) at three fixed background efficiency ($\epsilon_B$) thresholds: 0.1, 0.01 and 0.001, providing a more focused assessment of performance in the low-$\epsilon_B$ regime of particular interest in particle physics.

\begin{figure}
    \centering
    \includegraphics[width=\linewidth]{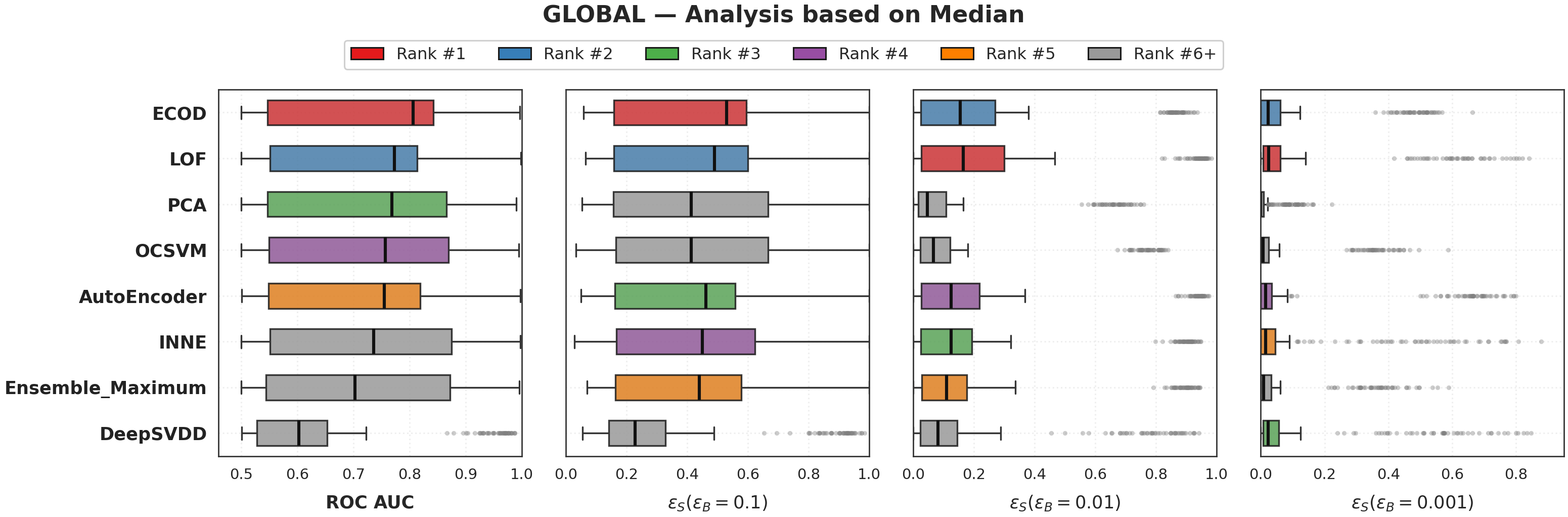}
    \includegraphics[width=\linewidth]{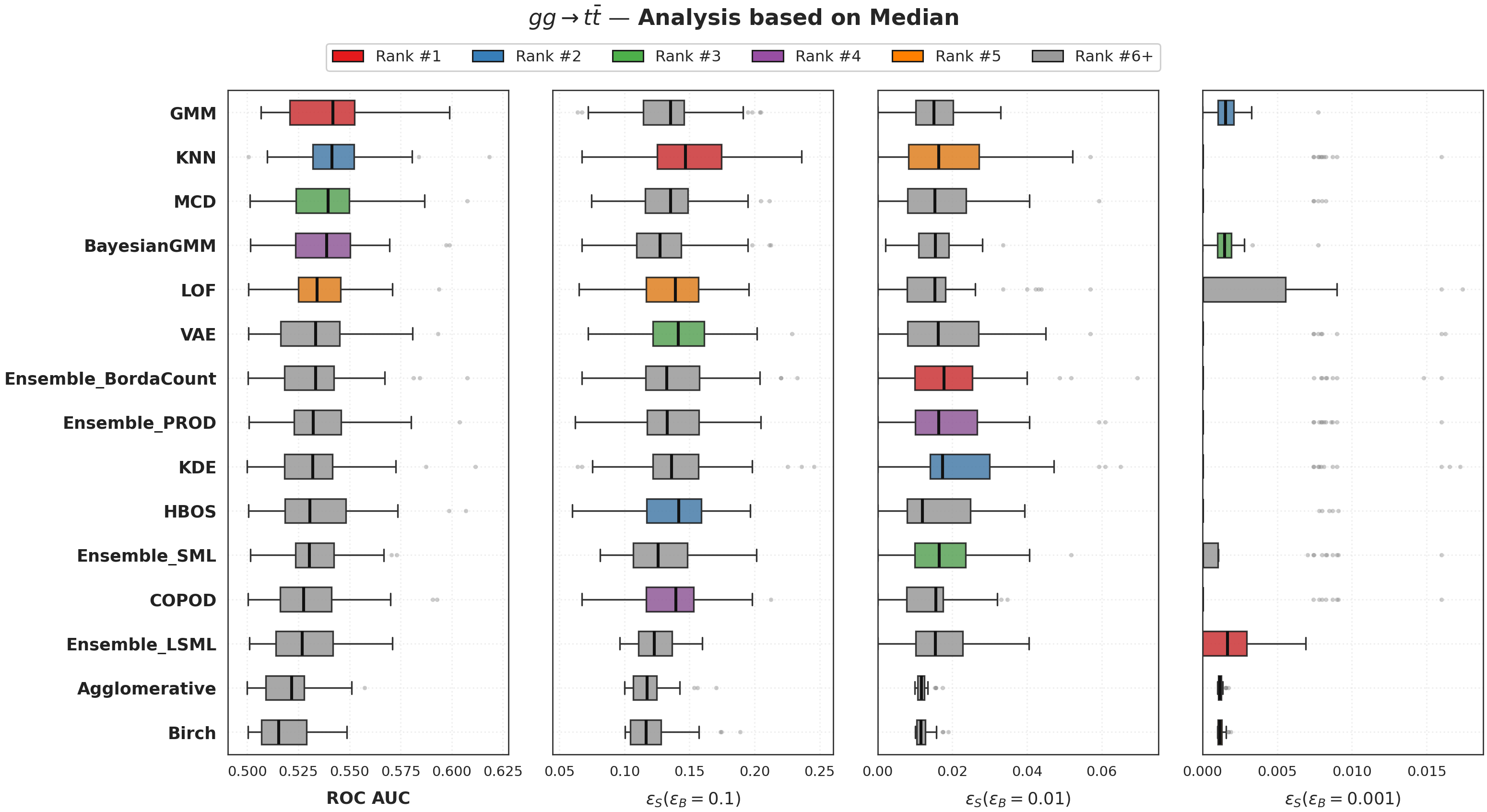}
    \caption{Discrimination performance and signal efficiency comparison between (a) the aggregate GLOBAL benchmark and (b) the specific $gg \to hh$ scenario, with models sorted by \textbf{median}. Each row displays ROC AUC (left) alongside signal efficiencies $\epsilon_S$ at background rejection thresholds $\epsilon_B \in {10^{-1}, 10^{-2}, 10^{-3}}$ (right). Boxes and central bars denote the IQR and median; whiskers denote $1.5 \times \text{IQR}$. In each metric panel, the top 5 models are uniquely highlighted (Ranks \#1–\#5), with remaining models shown in neutral gray (Rank \#6+).}
    \label{fig:auc_eff_median}
\end{figure}

Fig~\ref{fig:auc_eff_median} compares the top five approaches according to four performance metrics: AUC and $\epsilon_S$ at the three fixed $\epsilon_B$ thresholds. The ranking in this figure is based on the \textbf{median}. The top row shows the results of the GLOBAL analysis, in which all runs are considered for each approach (i.e., regardless of physical scenario). LOF and ECOD emerge as the top-performing methods, alternating between first and second place across the metrics considered. Ensemble\_Maximum provides an interesting example of an ensemble that does not achieve a competitive AUC but ranks among the top five approaches in terms of signal efficiency at $\epsilon_B = $ 0.1 and 0.01. In these metrics, it outperforms individual models that rank higher in terms of AUC, such as PCA and OCSVM.

An interesting complementary analysis is to examine the specific physics scenarios. As an example, the bottom row of Fig~\ref{fig:auc_eff_median} presents the results for the effective operator $f_{gg}$. In contrast to the GLOBAL analysis, no single model consistently dominates across all performance metrics. GMM achieves the best median AUC, whereas the top rank in terms of signal efficiency varies across approaches. Notably, the Borda Count and L-SML ensembles outperform all other methods at $\epsilon_B$ thresholds of 0.01 and 0.001, respectively.

\begin{figure}
    \centering
    \includegraphics[width=\linewidth]{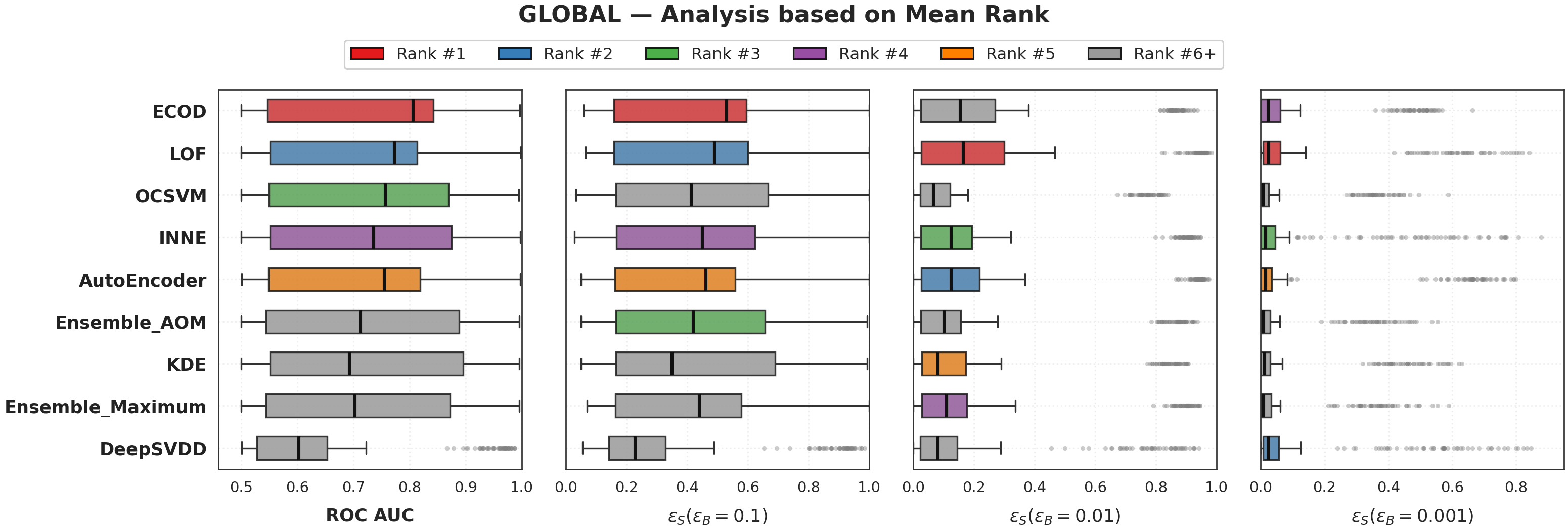}
    \includegraphics[width=\linewidth]{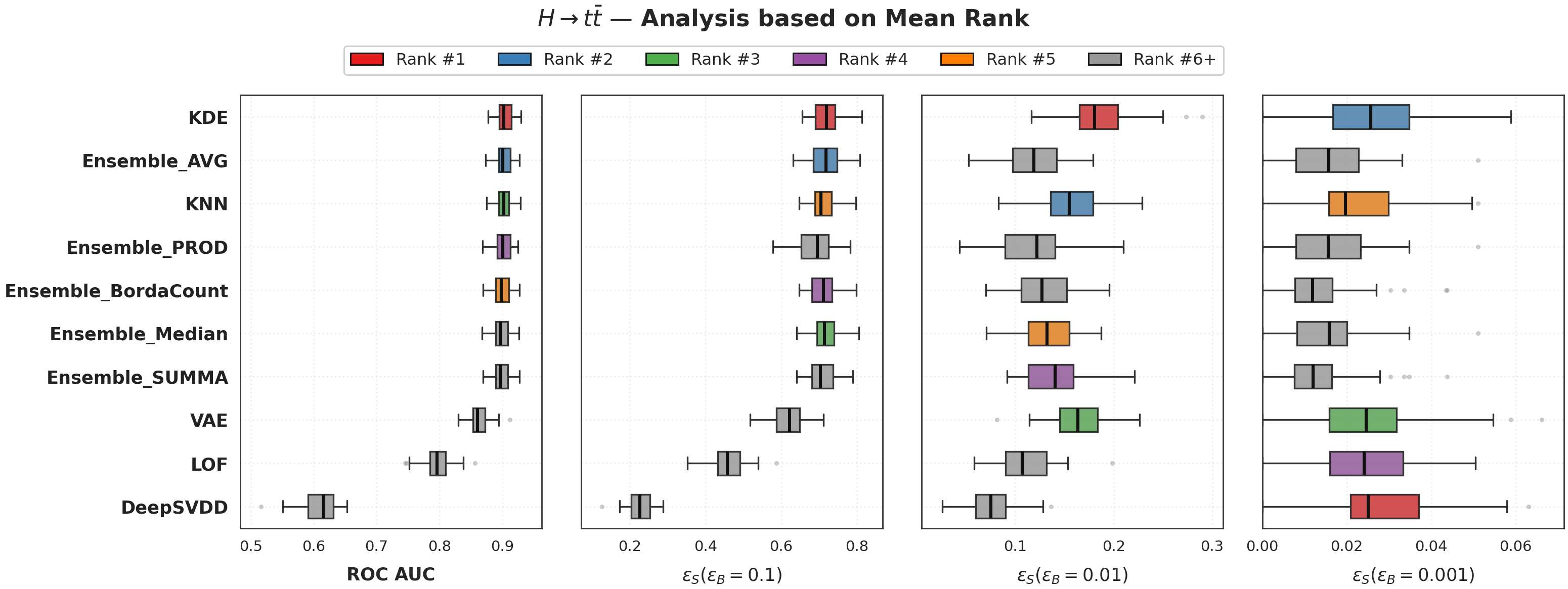}
\caption{Discrimination performance and signal efficiency comparison between (a) the aggregate GLOBAL benchmark and (b) the specific $H \to t\bar{t}$ scenario, with models sorted by \textbf{mean rank}. Each row displays ROC AUC (left) alongside signal efficiencies $\epsilon_S$ at background rejection thresholds $\epsilon_B \in {10^{-1}, 10^{-2}, 10^{-3}}$ (right). Boxes and central bars denote the IQR and median; whiskers denote $1.5 \times \text{IQR}$. In each metric panel, the top 5 models are uniquely highlighted (Ranks \#1–\#5), with remaining models shown in neutral gray (Rank \#6+).}
\label{fig:auc_eff_mean_rank}
\end{figure}

Another important comparison criterion is the \textbf{mean rank}, which provides a complementary perspective by aggregating performance across physical scenarios with different levels of difficulty and potentially different metric scales. The GLOBAL analysis based on this criterion, shown in Fig.~\ref{fig:auc_eff_mean_rank}, again identifies ECOD and LOF as the top-ranked approaches, highlighting their potential for signal detection across diverse physical scenarios. Ensembles also feature prominently in these rankings, particularly for signal efficiency at fixed $\epsilon_B$, rather than for AUC. AOM and Ensemble\_Maximum enter the top five at $\epsilon_B=0.1$ and $\epsilon_B = 0.01$, respectively, demonstrating the competitive potential of ensemble methods relative to individual models.

The bottom row of Fig~\ref{fig:auc_eff_mean_rank} presents the results for the specific $H \to t\bar{t}$ scenario. In this case, ensemble methods occupy several of the top positions, while LOF and ECOD no longer consistently dominate across the metrics. ECOD is absent from the top five, whereas LOF appears only in fourth place for signal efficiency at  $\epsilon_B = 0.001$. KDE emerges as the most consistently top-ranked individual model, while the AVG, PROD, Borda Count, Median, and SUMMA ensembles also achieve favorable rankings. In particular, AVG, Borda Count, and Median appear among the top five approaches for multiple metrics, highlighting the competitiveness of ensemble methods in this specific physical scenario.

\begin{figure}
    \centering
    \includegraphics[width=\linewidth]{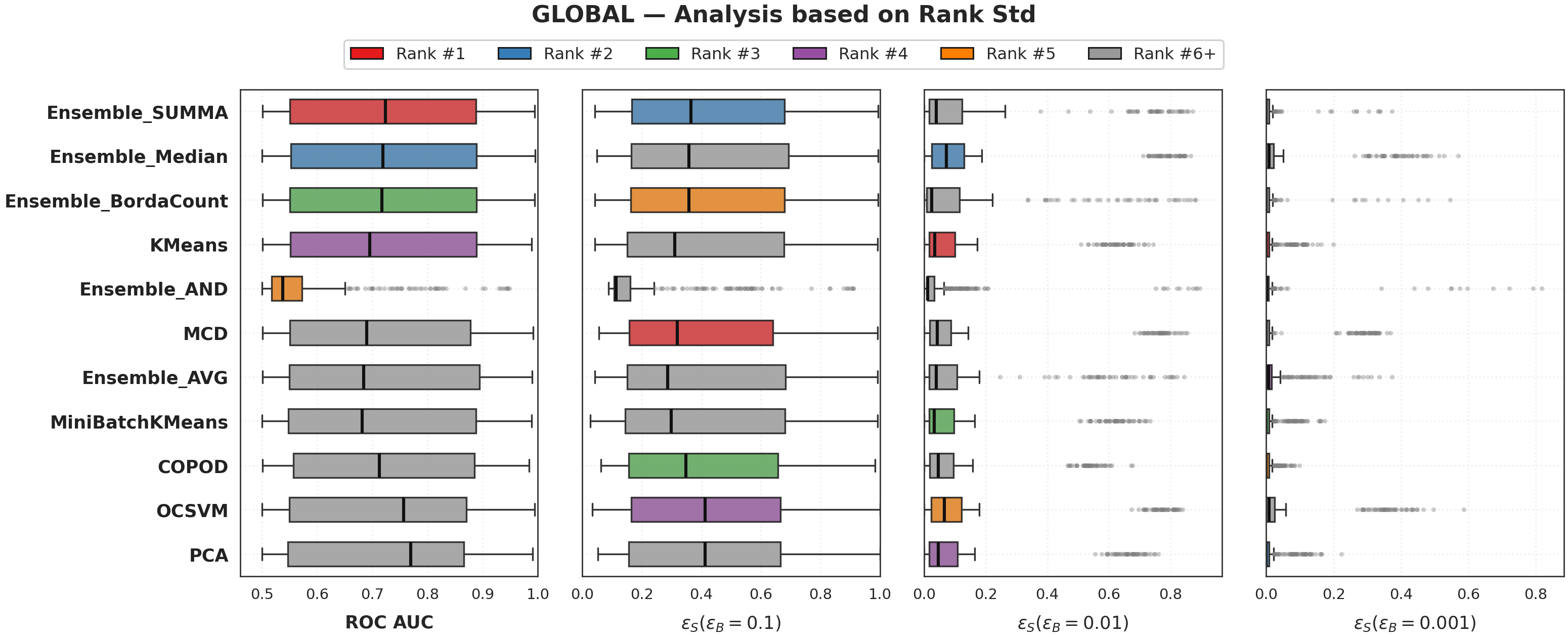}
    \includegraphics[width=\linewidth]{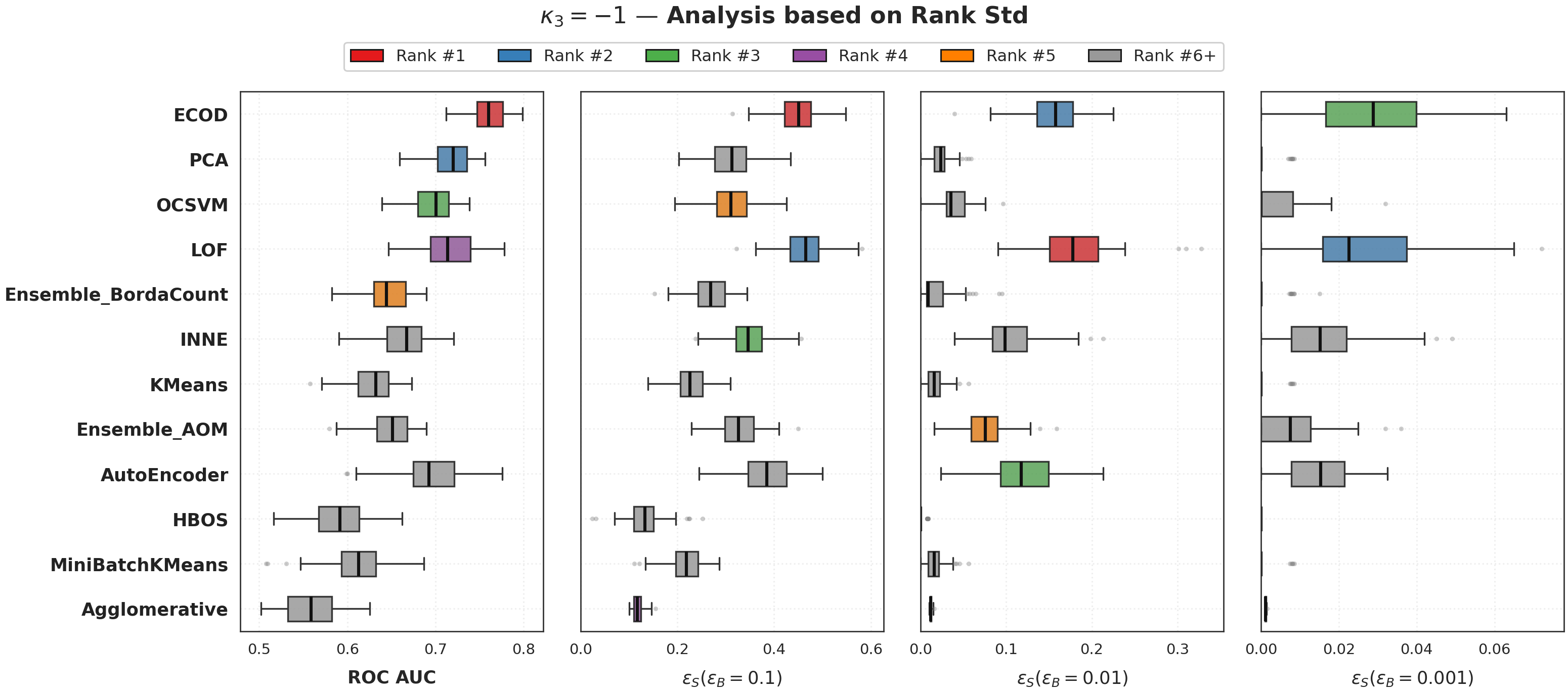}
    \caption{Discrimination performance and signal efficiency comparison between (a) the aggregate GLOBAL benchmark and (b) the specific non-resonant double Higgs production scenario with $\kappa_3 = -1$, with models sorted by \textbf{rank standard deviation}. Each row displays ROC AUC (left) alongside signal efficiencies $\epsilon_S$ at background rejection thresholds $\epsilon_B \in {10^{-1}, 10^{-2}, 10^{-3}}$ (right). Boxes and central bars denote the IQR and median; whiskers denote $1.5 \times \text{IQR}$. In each metric panel, the top 5 models are uniquely highlighted (Ranks \#1–\#5), with remaining models shown in neutral gray (Rank \#6+).}
    \label{fig:auc_eff_std_rank}
\end{figure}

The final ranking criterion considered is the \textbf{standard deviation of ranks} across independent runs, which provides a measure of the stability of each approach. The GLOBAL analysis, shown in Fig.~\ref{fig:auc_eff_std_rank}, reveals the predominance of ensemble methods. SUMMA, Median, and Borda Count occupy the top three positions in terms of AUC and also achieve favorable rankings for signal efficiency at $\epsilon_B$ thresholds of 0.1 and 0.01. These results highlight the potential of ensemble methods to provide stable performance across repeated experiments.

The bottom row of Fig.~\ref{fig:auc_eff_std_rank} presents the results for double-Higgs production with a modified trilinear Higgs self-coupling, $\kappa_3=-1$. Among the ensemble methods, Borda Count ranks among the top five for AUC, while AOM does so for signal efficiency at $\epsilon_B = 0.01$. In this specific scenario, the individual models ECOD and LOF also reappear among the top-ranked approaches, indicating that they can exhibit competitive ranking stability for this physical process.





\section{Results for the Dark Machines Challenge with Varied Final States}
\label{section:darkmachines}

The Dark Machines initiative~\cite{DarkMachines} is an effort that joins physicists and data scientists in applying machine learning to particle and astroparticle physics problems. One of the projects conducted by the initiative is the unsupervised learning of collider events. Back in 2019, a data challenge to classify a large number of new physics scenarios was organized by the initiative during the Les Houches workshop~\cite{Brooijmans:2020yij}. The outcomes and lessons learned from the data challenge can be found in Ref.~\cite{darkmachinechallenge}. 

The data analyzed in the challenge consisted of around a billion simulated collider events distributed among several new physics scenarios and their respective SM backgrounds. The new physics signals consisted of dark matter production through a new $Z^\prime$ interaction in association with SM particles, and both $R$-parity conserving and violating supersymmetry signals. The goal of the challenge was to maximize the AUC and signal acceptance for a fixed background efficiency. Contrary to the scenarios studied so far, in that challenge we encounter many distinct final states characterized by $X_{obs}+\not\!\!\! E_T$, where $X_{obs}$ consists of combinations of leptons, photons, and jets, depending on the new physics scenario.

Concerning the machine learning algorithms, versions of Kernel Density Estimators, Gaussian Mixture Models, flow models, Generative Adversarial Networks, and (Variational) Autoencoders were employed individually and also combined into ensembles. The outcomes of the study can be found in the references and links provided in Ref.~\cite{darkmachinechallenge}. In particular, output scores, AUC, and $\epsilon_S@\epsilon_B$ tables for all ML models are provided.

\begin{figure}[t!]
    \centering
    \includegraphics[width=\linewidth]{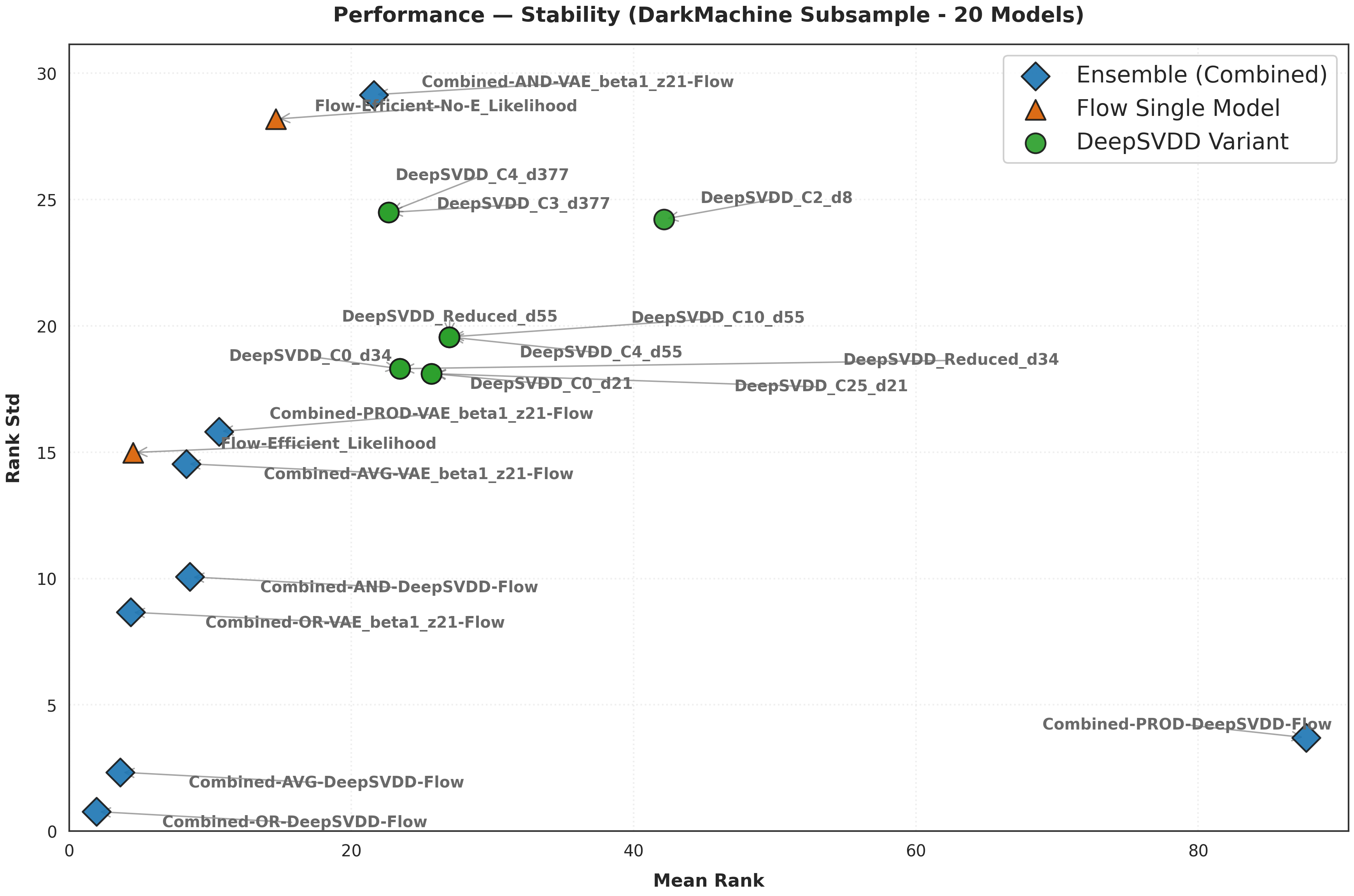}
    \caption{Performance {\it versus} stability space comparing ensemble methods and individual models evaluated in the Dark Machines Challenge \cite{darkmachinechallenge}. The $x$-axis shows the mean rank, $\mu_{\text{rank}}$, of each method across all physics scenarios considered in the original study, while the $y$-axis represents the standard deviation of ranks, $\sigma_{\text{rank}}$, as a measure of ranking variability across experiments. Blue diamonds represent ensembles, orange triangles represent flow models, and green circles represent variants of DeepSVDD evaluated in the study. The original paper considered 80 distinct DeepSVDD configurations with different hyperparameter settings. To avoid overcrowding the figure, only a subset of these configurations is shown; this selection does not affect the overall conclusions. The most favorable methods are located in the bottom-left region of the plot, with the ideal point at $(\mu_{\text{rank}}, \sigma_{\text{rank}}) = (1, 0)$, corresponding to the best possible mean rank and zero variability across experiments.}
    \label{fig:darkmachine_scatterplot}
\end{figure}

The results of the Dark Machines Unsupervised Challenge study allow us to extend the robustness analysis presented in the previous sections to the physics scenarios and algorithms addressed in that work. Figure~\ref{fig:darkmachine_scatterplot} shows the performance {\it versus} stability space for the methods evaluated in \cite{darkmachinechallenge}, including flow models, DeepSVDD variants, and ensemble methods. The green circles represent a randomly selected subset of 10 DeepSVDD variants trained in the original study. This subset is displayed to avoid overcrowding the figure.
%
The DeepSVDD variants cluster in a region characterized by relatively poor performance, with a mean rank of approximately $\mu_{\text{rank}} \sim 30$, and substantial variability across runs. The two flow models, represented by red triangles, achieve better rankings than the DeepSVDD variants. Moreover, incorporating particle energy as an input feature leads to a clear improvement in the model's performance compared with the corresponding configuration without energy.

The ensemble methods, represented by blue diamonds, predominantly occupy the most favorable region of the performance {\it versus} stability space, combining better rankings with lower variability. In particular, Combined-OR-DeepSVDD and Combined-AVG-DeepSVDD are positioned close to the ideal point, $(\mu_{\text{rank}}, \sigma_{\text{rank}}) = (1,0)$, indicating both strong relative performance and high ranking stability. Two ensembles depart from this overall pattern: Combined-AND-VAE\_beta1\_z21\_Flow exhibits the highest ranking variability among all methods, while Combined-PROD-DeepSVDD-Flow also deviates from the otherwise favorable region occupied by most ensembles. The results of the ensemble analysis are summarized in Table~\ref{tab:top5_darkmachine}.

\begin{table}[t!]
\centering
\begin{tabular}{cccc}
\hline
\textbf{Ranking} & \textbf{Model} & \textbf{Type} & \textbf{Normalized Distance} \\
\hline
1 & Combined-OR-DeepSVDD-Flow & Ensemble & 0.0288 \\
2 & Combined-AVG-DeepSVDD-Flow & Ensemble & 0.0856 \\
3 & Combined-OR-VAE\_beta1\_z21-Flow & Ensemble & 0.2998 \\
4 & Combined-AND-DeepSVDD-Flow & Ensemble & 0.3562 \\
5 & Combined-AVG-VAE\_beta1\_z21-Flow & Ensemble & 0.5060 \\
\hline
\end{tabular}
\caption{Ranking of the top five methods in the Dark Machines Challenge based on the normalized distance in the performance {\it versus} stability space.}
\label{tab:top5_darkmachine}
\end{table}

\section{Conclusions}
\label{section:conclusions}

 Identifying new physics signals in collider experiments with unsupervised and semi-supervised classifiers has become a frontier research  area in particle physics. It reflects the necessity of overcoming the difficulty of predicting what type of new physics model we should build in the face of currently available information about a possible extension of the Standard Model. In this sense, searching for specific signals based on supervised algorithms is a guessing game. 

 There are two key elements in using machine learning models to identify new signals: the physics model space and the classifier model space. Hypothesizing about the physics space reduces the scope of collider searches, but it allows us to select the best supervised classifier for signal discovery. Even better, we can use a mixture of experts to boost their individual performances. In this respect, the classifier space can be used with maximum information and efficiency for the task of identifying the hypothesized signal. 

 If we assume an agnostic posture about the physics model, an unsupervised or semi-supervised model needs to be used. However, the dependency of the classifier space still remains. The best classifier for a given new physics signal might perform poorly for a different type of signal, even while keeping the same final state topology. This dependency obliges us to construct specific classifiers to capture specific physical scenarios. As in the case of supervised algorithms, the way out is to combine the votes of the unsupervised and semi-supervised experts. Therefore, an unsupervised/semi-supervised ensemble classifier is designed to soften both the physics and machine learning space dependencies. 

 Contrary to the supervised case, unsupervised ensembles do not hold guaranties of performance enhancement. It does mean that, in some cases, a suitable combination of experts will not boost the classification metrics. Recent studies~\cite{vanBeekveld:2020txa,darkmachinechallenge} suggest that the best individual classifier changes depending on the new physics scenario and that an ensemble does not always guaranty better classification performance. Our results not only corroborate their findings by investigating a larger number of ensemble algorithms, but also complement them by addressing an important aspect of ensemble learning -- the reliability and robustness of the classification task. It is worth mentioning that our analysis factors out an important element of the study by fixing the observed final state, the totally leptonic $t\bar{t}$ production and decay, allowing for a more precise and straightforward interpretation of the relative performance of the results.

 In total, eleven ensemble algorithms were investigated, categorized into two classes of algorithms: four types of unsupervised models and eight semi-supervised models. The combinations were performed intra (with unsupervised or semi-supervised models exclusively) and inter (gathering both unsupervised and semi-supervised models) categories. We found that despite the ensembles not being capable of beating the performance of the best individual classifier in the majority of new physics scenarios (with a few exceptions, though), they always stabilize the results -- their performance varies little across all the new physics scenarios, contrary to individual models, which are highly dependent on the signal. Moreover, some ensemble algorithms showed consistent performance among intra and inter categories of ensemble models. This is the case for Borda Count, SUMMA, AOM, and Ensemble Median, which consistently presented the smallest normalized euclidean distance in the performance {\it versus} stability plane in $t\bar{t}$ final states. Interestingly, some ensemble algorithms also figured in the top five according to the mean rank and median values of AUC and $\epsilon_S@\epsilon_B$ across all physics scenarios.

 We extended the analysis of $t\bar{t}$ scenarios to physics scenarios involving supersymmetry particles and a new $Z^\prime$ boson, leading to final states with large missing energy in the context of the Dark Machines Unsupervised Challenge~\cite{darkmachinechallenge}. We evaluated the robustness of the ensemble algorithms considered in the challenge by calculating the mean and standard deviation of their ranks and then the normalized distance in the performance {\it versus} stability space. As in the other cases, the ensemble models showed much higher stability of predictions across the datasets of the challenge.

 Our findings constitute, ultimately, a complementary aspect of unsupervised and semi-supervised classifiers that is often neglected -- the variability in the performance of the algorithms as a function of the final state signal. As we demonstrated, the same algorithm can perform best for one type of signal but poorly for another. As in the case of supervised learning, though, we can combine algorithms to detect anomalies or cluster events in classes when the new physics is unknown, and benefit from their peculiarities to detect various types of signals. This work thus highlights the importance of constructing ensembles in the unsupervised search for new physics to reduce the chance of allowing potential beyond the SM signals to pass unnoticed.

\bigskip{}
	
\textbf{Acknowledgments}: This study was financed in part by Conselho Nacional de Desenvolvimento Científico e Tecnológico (CNPq), via Grant No. 300351/2026-7 (A. A.). A. A. also acknowledges support from the FAPESP (No. 2021/01089-1) Grant. 

\bibliographystyle{JHEP}
\bibliography{referencias}

\end{document}